\documentclass[useAMS,usenatbib]{mn2e}

\usepackage{epsfig}
\usepackage{amsmath}
\usepackage{amssymb}
\usepackage{ifthen}
\usepackage{txfonts}
\usepackage{rotating}
\usepackage{url}
\usepackage{varioref}
\usepackage{verbatim}
\usepackage{latexsym}
\usepackage{graphicx}

\newcommand*{\bfrac}[2]{\genfrac{}{}{0pt}{}{#1}{#2}}

\DeclareSymbolFont{cmletters}{OML}{cmm}{m}{it}

\DeclareMathSymbol{v}{\mathord}{cmletters}{"76}

\def\be{\begin{equation}}
\def\ee{\end{equation}}

\newcommand{\detg}{{\sqrt{-g}}}

\newcommand{\msun}{{\rm M_{\odot}}}

\newcommand{\rholab}{{\rho}}

\newcommand{\uvec}{{\underline{u}}}
\newcommand{\bvec}{{\underline{b}}}
\newcommand{\Bvec}{{\underline{B}}}

\newcommand{\Fvec}{{F}}

\newcommand{\xvec}{{\underline{x}}}
\newcommand{\Avpotvec}{{\underline{A}}}

\newcommand{\rvec}{{\underline{r}}}
\newcommand{\vvec}{{\underline{v}}}
\newcommand{\etavec}{{\underline{\eta}}}
\newcommand{\reluvec}{{\underline{\tilde{u}}}}

\newcommand{\alf}{Alfv\'en}

\newcommand{\gdet}{\sqrt{-g}}

\newcommand{\cut}[1]{\hbox{}}

\DeclareSymbolFont{cmletters}{OML}{cmm}{m}{it}
\DeclareMathSymbol{v}{\mathalpha}{cmletters}{"76}

\usepackage{subfigure}
\usepackage{ifthen}
\usepackage[usenames,dvipsnames]{color}
\usepackage{graphicx}

\usepackage{hyperref}

\newcommand\araa{\rmfamily{ARA\&A}}%
\newcommand\apj{\rmfamily{ApJ}}%
\newcommand\apjl{\rmfamily{ApJ}}%
\newcommand\apjs{\rmfamily{ApJS}}%
\newcommand\apss{\rmfamily{Ap\&SS}}%
\newcommand\aap{\rmfamily{A\&A}}%
\newcommand\mnras{\rmfamily{MNRAS}}%
\newcommand\prd{\rmfamily{Phys.~Rev.~D}}%
\newcommand\pasj{\rmfamily{PASJ}}%
\newcommand\nat{\rmfamily{Nature}}%
\newcommand\nar{\rmfamily{New Astronomy Reviews}}%
\newcommand\grl{\rmfamily{Geophys.~Res.~Lett.}}%
\defcitealias{bz77}{BZ77}

\newcommand{\emath}{{\rm e}}
\newcommand{\eff}{{\eta}}

\newcommand{\Qone}{Q_{\theta,\rm MRI}}
\newcommand{\Qx}{Q_{x,\rm MRI}}
\newcommand{\Qoneweak}{Q_{\theta,\rm weak,MRI}}
\newcommand{\Qtwo}{{S_{d,\rm MRI}}}
\newcommand{\Qtwoinitio}{{S_{d,t=0,\rm MRI,\{i,o\}}}}
\newcommand{\Qtwoweak}{{S_{d,\rm weak,MRI}}}
\newcommand{\Qthree}{{Q_{\phi,\rm MRI}}}
\newcommand{\Qthreeweak}{{Q_{\phi,\rm weak,MRI}}}

\newcommand{\bsqorhomax}{{{50}}}
\newcommand{\bsqoumax}{{{10^3}}}
\newcommand{\uorhomax}{{{50}}}
\newcommand{\bsqorhomaxdiaghigh}{{{30}}}
\newcommand{\bsqorhomaxdiaglow}{{{10}}}

\title[Magnetically Choked Accretion Flows]
{General Relativistic Magnetohydrodynamic Simulations of Magnetically Choked Accretion Flows around Black Holes}

\author[J.~C.~McKinney,
A.~Tchekhovskoy and
R.~D.~Blandford]
{Jonathan C. McKinney$^1$\thanks{\hbox{E-mail: jmckinne@stanford.edu~(JCM)}},
Alexander Tchekhovskoy$^2$,
Roger D. Blandford$^1$
\\
 $^1$Kavli Institute for Particle Astrophysics and Cosmology, Stanford University, P.O. Box 20450, MS 29,
Stanford, CA 94309
\\
  $^2$Center for Theoretical Science, Jadwin Hall, Princeton University, Princeton,
  NJ 08544; Princeton Center for Theoretical Science Fellow \\
}{
}

\begin{document}
\date{Accepted 2012 Jan xx.  Received 2012 Jan xx; in original form 2012 Jan xx.}
\pagerange{\pageref{firstpage}--\pageref{lastpage}} \pubyear{2012}
\maketitle

\label{firstpage}

\begin{abstract}

  Black hole (BH) accretion flows and jets are qualitatively affected
  by the presence of ordered magnetic fields.  We study fully
  three-dimensional global general relativistic magnetohydrodynamic
  (MHD) simulations of radially extended and thick (height $H$ to
  cylindrical radius $R$ ratio of $|H/R|\sim 0.2$--$1$) accretion
  flows around BHs with various dimensionless spins ($a/M$, with BH
  mass $M$) and with initially toroidally-dominated ($\phi$-directed)
  and poloidally-dominated ($R-z$ directed) magnetic fields.  Firstly,
  for toroidal field models and BHs with high enough $|a/M|$, coherent
  large-scale (i.e. $\gg H$) dipolar poloidal magnetic flux patches
  emerge, thread the BH, and generate transient relativistic
  jets. Secondly, for poloidal field models, poloidal magnetic flux
  readily accretes through the disk from large radii and builds-up to
  a natural saturation point near the BH.  While models with
  $|H/R|\sim 1$ and $|a/M|\le 0.5$ do not launch jets due to quenching
  by mass infall, for sufficiently high $|a/M|$ or low $|H/R|$ the
  polar magnetic field compresses the inflow into a geometrically thin
  highly non-axisymmetric ``magnetically choked accretion flow''
  (MCAF) within which the standard linear magneto-rotational
  instability is suppressed.  The condition of a highly-magnetized
  state over most of the horizon is optimal for the Blandford-Znajek
  mechanism that generates persistent relativistic jets with $\gtrsim
  100$\% efficiency for $|a/M|\gtrsim 0.9$.  A magnetic
  Rayleigh-Taylor and Kelvin-Helmholtz unstable magnetospheric
  interface forms between the compressed inflow and bulging jet
  magnetosphere, which drives a new jet-disk quasi-periodic
  oscillation (JD-QPO) mechanism.  The high-frequency QPO has
  spherical harmonic $|m|=1$ mode period of $\tau\sim 70GM/c^3$ for
  $a/M\sim 0.9$ with coherence quality factors $Q\gtrsim 10$.
  Overall, our models are qualitatively distinct from most prior MHD
  simulations (typically, $|H/R|\ll 1$ and poloidal flux is limited by
  initial conditions), so they should prove useful for testing
  accretion-jet theories and measuring $a/M$ in systems such as SgrA*
  and M87.

\end{abstract}


\section{Introduction}
\label{sec_intro}

Modern black hole (BH) accretion disk theory suggests that angular
momentum transport, in the past modelled using an $\alpha$-viscosity
\citep{sha73,Novikov:1973:IBH,thorne74}, is due to magnetohydrodynamic
(MHD) turbulence driven by the magneto-rotational instability (MRI)
within a differentially rotating disk \citep{bal91,bh98} and due to
large-scale magnetic torques within the plunging region of a BH
\citep{1976Natur.262..270Z,macdonald84,gammie99,krolik99,ak00,l00,lp00,li02,rgb06,omnm09,omnm10,psm11}.
However, even outside the plunging region, transport may occur via
large-scale magnetic field threading the disk (e.g.,
\citealt{blandford_accretion_disk_electrodynamics_1976,1976Natur.262..649L,2011ApJ...737...94C})
or even by ordered field within the disk far beyond the plunging
region \citep{br74,1976Ap&SS..42..401B,nia03,2007ApJ...667L.167B}.

How are strong, large-scale, or ordered magnetic fields generated in
BH accretion flows?  The MHD turbulence increases the magnetic field's
{\it strength} up to some non-linear saturation point, while the
solenoidal constraint in electrodynamics that $\nabla \cdot B=0$ for
magnetic field $B$ ensures that, within some radius $r_0$, the
magnetic field's {\it flux} (e.g. $\int_0^{r_o} B^z dA$ at the equator
for vertical field $B^z$ and equatorial area $A$) remains fixed unless
added at $r_0$.

Patches of constant polarity magnetic flux might form stochastically
out of an MHD dynamo \citep{tp96} as occurs for the toroidal
($\phi$-directed) magnetic field as seen in MHD simulations
\citep{dsp10,gg11} and for the large-scale poloidal ($R-z$-directed)
magnetic field as seen in the Sun \citep{1959ApJ...130..364B}.
However, so far no evidence exists for the development of large-scale
poloidal field from only a dominant toroidal field \citep{dhkh05}.
The MRI-driven MHD turbulence-generated poloidal field remains
small-scale on the order of the disk height (often used to estimate
jet field strength, e.g. in \citealt{meier01} and \citealt{livio03})
and does not sustain relativistic jets due to persistent mass-loading
\citep{bhk08}.  Patches of constant polarity flux may also develop by
the Poynting-Robertson drag (PRD) effect \citep{1998ApJ...508..859C},
but the PRD effect might saturate at low field strengths
\citep{2002ApJ...580..380B}.

An alternative is that ordered constant polarity magnetic flux comes
from large radii, as implicit in outflow models
\citep{blandford_accretion_disk_electrodynamics_1976,bz77,bp82,nar07,tmn09,tnm09b,tnm09},
but transport of flux from large radii may be inefficient
\citep{1994MNRAS.267..235L,su05,2008ApJ...677.1221R,bhk09trans}.  Such
magnetic fields can lead to angular momentum transport within the disk
and out of a rotating BH (the latter having some observational basis;
\citealt{2012MNRAS.419L..69N}).  Indeed, production of persistent
relativistic jets may require steady large-scale dipolar
(i.e. sub-dominant higher multipoles) fields near BHs
\citep{bhk08,mb09}.

Observations do show patches of coherent magnetic flux ($\Phi$)
surrounding astrophysical systems that can feed BHs. The
interstellar medium has $\Phi\sim 0.1 {\rm pc}^2{\rm G}$
\citep{lang99,vall11}, threads near the Galactic Nucleus have
$\Phi\gtrsim 0.01 {\rm pc}^2{\rm G}$ \citep{larosa04,ferr09,vall11},
and each O-type star with a dipolar field with strength $B\sim 1000$G
feeding SgrA* provides up to $\Phi\sim 10^{-10} {\rm pc}^2{\rm G}$.
Such a coherent flux might be available for accretion near SgrA*, M87,
and other active galactic nuclei (AGNs).  Indeed, the constancy over
several years of the sign of the circular polarization from near the
BH in SgrA* implies that the magnetic field is coherent with constant
polarity for many dynamical times \citep{2012ApJ...745..115M}.  Also,
many AGN jets show persistent signs for the transverse Faraday
rotation gradient, indicating accretion of a persistent magnetic
polarity (e.g., \citealt{mgb09,bm10}).  For BH x-ray binaries, some
portion of the donor star's surface dipolar field of order $B\sim
100$--$1000$G may be available for accretion, such that $\Phi\sim
10^{-13}$--$10^{-12} {\rm pc}^2{\rm G}$ \citep{jrp06}.  For
cosmological gamma-ray bursts (GRBs) that require $B\sim 3 \times
10^{15}{\rm G}$ for MHD-driven jets by the Blandford-Znajek (BZ)
mechanism \citep{bz77,Narayan:1992:GRB,barkov08}, one obtains a
requirement of $\Phi\sim 10^{-9} {\rm pc}^2{\rm G}$ corresponding to
about $10\%$ of the total ordered poloidal flux within a presupernova
progenitor \citep{hws05,kombar09}.  Even if $N$ random polarity
patches accrete over time, on average $\Phi\propto N^{1/2}$ in a
random walk near the BH.  Some flux can annihilate before reaching the
BH, but field reversals are unlikely to have exactly equal flux
magnitude and so magnetic flux should tend to accumulate.

Magnetic flux is conserved such that accumulation of a sufficient
amount of constant polarity poloidal magnetic flux brought in from
large radii might substantially modify the standard picture of an
MRI-driven MHD turbulent disk that applies for relatively weak
magnetic field.  Accumulation of magnetic flux can eventually lead to
the formation of a semi-permeable magnetic barrier
\citep{1976Ap&SS..42..401B}, and in such a state the inflow must
somehow accrete through the accumulated magnetosphere.

The inflow strongly interacts with the magnetospheric barrier at the
``magnetospheric radius'' $r_m$, where the force of gravity by the
accreting mass equals the magnetic force by the barrier.  $r_m$ is
readily estimated in a similar way for both neutron stars (NSs)
\citep{1973ApJ...184..271L,1977ApJ...215..897E,1978ApJ...219..617S}
and BHs \citep{br74,nia03}.  The disk surface density $\Sigma$ inside
$r_m$ is $\Sigma = \dot{M}/(2\pi r\epsilon v_{\rm ff})$, where $\dot
M$ is the mass accretion rate.  $\epsilon\equiv 10\epsilon_{-1}$ is
the ratio of mass advection velocity to free-fall velocity ($v_{\rm
  ff}$), where $\epsilon\sim 0.1$ in our simulations discussed later.
Also, let $\dot{M}=\dot{M}_{\rm H}~(r/r_g)^{n}$ for horizon accretion
rate $\dot{M}_{\rm H}$, as in \citet{Blandford:1999:FGA}.  The
condition for magnetic support against disk gravity in the equatorial
plane is $GM\Sigma/r^2 \sim 2B_r B_z/(4\pi) \sim B_z^2/(2\pi)$ for
$B_r\sim B_z$.  Then, $B_z \sim 10^5
\epsilon_{-1}^{-1/2}m_8^{-1/2}\dot{m}^{1/2}(r/r_g)^{-5/4+n/2}$~G,
where $r_g=GM/c^2$, $c$ is the speed of light, $G$ is the
gravitational constant, $M$ is the BH mass, $m_8\equiv M/(10^8\msun)$,
$\dot{m}\equiv \dot{M}/\dot{M}_{\rm Edd}$, and $\dot{M}_{\rm
  Edd}\equiv L_{\rm Edd}/(0.1 c^2)\approx 1.4\times 10^{25}m_8~{\rm
  g}~{\rm s}^{-1}$ is the Eddington mass accretion rate. If $\epsilon$
is independent of $r$, integrating $B_z$ over $r$ gives an estimate of
$\Phi$ within the ``magnetospheric radius'' $r_m$.  Inverting
$\Phi(r_m)$ gives
\begin{equation}\label{rmsimple}
  r_m \sim r_g~\left(12000\left(\frac{3}{4}+\frac{n}{2}\right)\right)^{\frac{4}{3+2n}}~\epsilon_{-1}^{\frac{2}{3+2n}}~m_8^{-\frac{6}{3+2n}}~\dot{m}_{\rm H}^{-\frac{2}{3+2n}}~\left(\frac{\Phi}{0.1{\rm pc}^2{\rm G}}\right)^{\frac{4}{3+2n}} ,
\end{equation}
such that $r_m$ varies weakly with parameters for higher $n$ and for
$n=\{0,1,2\}$ the coefficient is $\sim \{10^5,10^3,10^2\}r_g$,
respectively.  As another measure of how important the magnetic flux
is, we also consider the limit that all the surrounding flux ($\Phi$)
reaches the BH.  Then, the horizon's dimensionless magnetic
flux would be
\begin{equation}\label{upsilonsimple}
  \Upsilon_{\rm H} \approx \frac{\Phi}{5 (r_g^2 c \dot{M}_{\rm H})^{1/2}} \sim 10^{4} m_8^{-3/2}\dot{m}_{\rm H}^{-1/2} \left(\frac{\Phi}{0.1{\rm pc}^2{\rm G}}\right) ,
\end{equation}
\citep{gammie99,pmntsm10}, which measures the mass-loading of the
magnetic field lines.  An MRI-driven MHD turbulent disk has
$\Upsilon_{\rm H}\lesssim 1$ for integrals within the heavy disk
inflow \citep{gammie99,pmntsm10}.  If $\Upsilon_{\rm H}\gtrsim 1$ for
integrals over some portion of the horizon (e.g. polar regions or
heavy disk inflow), then this indicates the formation of a force-free
magnetosphere and the BZ effect can be activated there
\citep{kombar09}.

The quantities $r_m^{n=0}$, $r_m^{n=1}$, and $\Upsilon_{\rm H}$ can be
estimated for various systems (using $\Phi$ estimated in a preceding
paragraph) to check whether a magnetosphere could dominate the flow
dynamics.  For M87 with $\Phi\sim 0.1 {\rm pc}^2{\rm G}$, $M\approx
6.4\times 10^9\msun$, mass accretion rate $\dot{m}_{\rm H}\sim
10^{-4}$ \citep{dma11}, one obtains $r_m^{n=0}\sim 10^3r_g$,
$r_m^{n=1}\sim 10^2r_g$, and $\Upsilon_{\rm H}\sim 10^3$.  For SgrA*
with $\Phi\sim 0.1 {\rm pc}^2{\rm G}$, $M\approx 4.5\times
10^{6}\msun$, mass accretion rate $\dot{m}_{\rm H}\sim 1.5\times
10^{-6}$ \citep{dafm10}, one obtains $r_m^{n=0}\sim 10^{11}r_g$,
$r_m^{n=1}\sim 10^{7}r_g$, and $\Upsilon_{\rm H}\sim 10^{9}$.  A
single O-type star feeding SgrA* would give $r_m\sim r_g$ for
$n=0,1,2$ and $\Upsilon_{\rm H}\sim 2$.  For GRS1915+105 with
$\Phi\sim 10^{-12} {\rm pc}^2{\rm G}$, $M\sim 15\msun$, $\dot{m}_{\rm
  H}\sim 0.7$ \citep{greiner01}, one obtains $r_m^{n=0}\sim 10^4r_g$,
$r_m^{n=1}\sim 10^3r_g$, and $\Upsilon_{\rm H}\sim 10^4$.  For
cosmological long-duration GRBs with $\Phi\sim 10^{-9} {\rm pc}^2{\rm
  G}$, $M\sim 3\msun$ and $\dot{M}_{\rm H}\sim 0.1\msun~{\rm s}^{-1}$
\citep{Woosley:1993:EMS}, one obtains $r_m^{n=0}<r_g$, $r_m^{n=1}\sim
r_g$, and $\Upsilon_{\rm H}\sim 1$.  One obtains only smaller $r_m$
and $\Upsilon_{\rm H}$ for short-duration GRBs as modelled by NS-NS or
BH-NS collisions that reach higher $\dot{M}$ at similar $\Phi$.  For
GRBs, at late time $\dot{M}_{\rm H}$ drops, after which $r_m$ and
$\Upsilon_{\rm H}$ can be much higher \citep{pz06}.  These estimates
show that one must consider how plasma accretes through a
magnetospheric barrier.  For SgrA* (where $n\sim 1$ is plausible),
even a millionth of $\Phi$ is sufficient to lead to magnetospheric
accretion near the BH.

Such flows with accumulated magnetic flux, called ``magnetically
choked accretion flows'' (MCAFs) by us for reasons based upon our
simulation results described later, have been considered theoretically
\citep{1976Natur.262..270Z,1976Ap&SS..42..401B,nia03}, via
pseudo-Newtonian MHD simulations
\citep{igu02,ina03,pmw03,igumenshchev08,ppmgl10}, and via general
relativistic (GR) MHD (GRMHD) simulations \citep{tnm11}.  MCAFs are
qualitatively related to flows with strong magnetic field
\citep{stm90,mkm95,mnm06,omnm07,fm09}.

MCAFs effectively accrete through a magnetic flux barrier via magnetic
interchange type modes (Kruskal-Schwarzschild or magnetic Rayleigh-Taylor
instabilities)
\citep{wang84,ktl92,lovelace94,ssp95,ls95,wang96,ss01,nmf02,nia03,ln04,su05,2007ApJ...671.1726S,jl08}.
Even with flow shear that might remove some specific magnetic interchange modes,
still other non-axisymmetric magnetic modes operate \citep{tp99,ss01}.
Interchange modes may also affect jet formation and propagation
\citep{nmf02}.  One might expect that magnetic flux accumulation could
be prevented by Parker-type instabilities \citep{parker66,jl08}.
However, the Parker instability cannot operate if the magnetic
pressure dominates the gas pressure due to strong magnetic tension
\citep{stm90}.  The MCAF-type state is also seen in other systems that
develop a magnetosphere, including young-stellar objects, star
formation regions \citep{2012ApJ...744..185C}, and neutron stars
\citep{1967MNRAS.137...95M,1976ApJ...207..914A,1980ApJ...235.1016A,1983ApJ...266..175B,1990A&A...227..473A,1994ApJ...429..781S,rkl08,kr08,2011arXiv1111.3068R}.
Unlike these systems that can have a closed dipolar magnetosphere, BHs
harbor a split-monopolar magnetosphere \citep{igumenshchev08}.

Most 3D pseudo-Newtonian MHD simulations (except in
\citealt{ina03,igumenshchev08}) and most 3D GRMHD simulations (except
in \citealt{tnm11,tmn12,2012arXiv1201.4385T}) of accretion disks have
only considered relatively weak magnetic fields leading to little
accumulated poloidal flux.  Such simulations do not reach the MCAF
state (for GRMHD simulations, e.g., see
\citealt{dev03,mg04,mck05,dhkh05,hk06,mck06jf,km07,kom07,mck07a,mck07b,fragile07,bhk08,mb09,nkh10,sdgn11}),
typically because their initial conditions used relatively small-sized
hydro-equilibrium tori within which weak magnetic fields are inserted
(so only little poloidal flux can accumulate, see section 4.3 in
\citealt{ina03}).  \citet{tnm11} also used an initial torus, but the
torus is radially extended and they ensured the initial conditions
make available a sufficient amount of magnetic flux to reach flux
saturation near the BH.  Note that 3D is required for simulations to
avoid decaying turbulence \citep{Cowling:1934:SGS} and to properly
resolve a MCAF.  For example, the suspended inflow seen in 2D
axisymmetric MHD simulations \citep{mg04,pz06,kombar09} is due to
axisymmetry \citep{igumenshchev08,igu09}.  Other 2D MHD simulations
have constant poloidal flux due to using a purely radial field
\citep{pb03}, which gives results similar to torus simulations with
limited poloidal flux.

Also, MHD simulations have typically studied relatively thin disks
compared to expected for radiatively inefficient accretion flows
(RIAFs), as applicable to (e.g.) SgrA*, M87, and some x-ray binary
states.  Quasi-analytical RIAF models include advection-dominated
accretion flows (ADAFs)
\citep{Ichimaru:1977:BBA,nar94,nar95b,acgl96,Abramowicz:1995:TEA,pg98},
convection-dominated accretion flows (CDAFs) \citep{nia00,qg00}, and
advection-dominated inflow-outflow solutions (ADIOSs)
\citep{Blandford:1999:FGA,beg11}.  These models suggest that RIAFs
should have disk height ($H$) to cylindrical radius ($R$) ratio of
$|H/R|\sim 0.5$--$0.9$. Simulations have studied $|H/R|\sim 0.05$--$0.1$
\citep{shafee08,rm09,noble09,nkh10,sorathia10,bas11}, $|H/R|\sim
0.1$--$0.15$ \citep{hk01,dev03,beckwith08b,bhk08,bhk09trans}, $|H/R|\sim
0.2$ \citep{hb02,mhm00,mm03,mg04,fragile07,mb09}, $|H/R|\sim
0.3$--$0.4$ \citep{ina03,pmntsm10}, $|H/R|\sim 0.6$ (Run F type in
\citealt{sto01,2001ApJ...554L..49H}), and rarely $|H/R|\sim 1$
\citep{igu02,pmw03,ppmgl10}.

In this work, we use fully 3D GRMHD simulations to study radially
extended and thick (including $|H/R|\sim 1$) radiatively inefficient
BH accretion flows with poloidal and toroidal magnetic field
geometries and various $a/M$.  Our initially poloidally-dominated
field models are designed so that near the horizon the poloidal
magnetic flux reaches a saturation point independent of the initial
poloidal magnetic flux.  Most prior MHD simulations would have only
required a few times more poloidal magnetic flux to reach this natural
saturation point.  The only other natural limit of poloidal magnetic
flux is it's negligible.  So, for comparison, we also consider models
with initially toroidally-dominated magnetic field.

The equations solved are presented in \S\ref{sec:goveqns}, diagnostics
are described in \S\ref{sec:diagnostics}, and models are described in
\S\ref{sec:physnummodels}.  Results for our fiducial model of a thick
disk around a rapidly rotating BH are in \S\ref{sec:fiducial}.
Results for various field geometries, BH spins, and disk thicknesses
are in \S\ref{sec:tables}.  We discuss our results in
\S\ref{sec:discussion} and conclude in \S\ref{sec:conclusions}.  Some
numerical method details are given in Appendix~\ref{sec:nummethods}.

\section{Governing Equations}
\label{sec:goveqns}

We solve the GRMHD equations for a radiatively inefficient magnetized
accretion flow around a rotating black hole defined by the Kerr metric
in Kerr-Schild coordinates with internal coordinates
$\xvec^\alpha\equiv (t,\xvec^{(1)},\xvec^{(2)},\xvec^{(3)})$ mapped to
the spherical polar coordinates $\rvec^\alpha\equiv
(t,r,\theta,\phi)$.  We write orthonormal vectors as $u_i$,
contravariant (covariant) vectors as $\uvec^i$ ($\uvec_i$), and
higher-ranked coordinate basis tensors with no underbar.  We work with
Heaviside-Lorentz units, often set $c=GM=1$, and let the horizon radius
be $r_{\rm H}$.

Mass conservation gives
\begin{equation}
\nabla_\mu (\rho_0 \uvec^\mu) = 0 ,
\end{equation}
where $\rho_0$ is the rest-mass density, $\uvec^\mu$ is the
contravariant 4-velocity, and $\rho=\rho_0 \uvec^t$ is the lab-frame
mass density.

Energy-momentum conservation gives
\begin{equation}\label{emomeq}
\nabla_\mu  T^\mu_\nu = 0 ,
\end{equation}
where the stress energy tensor $T^\mu_\nu$ includes both matter (MA)
and electromagnetic (EM) terms:
\begin{eqnarray}\label{MAEM}
{T^{\rm MA}}^\mu_\nu &=& (\rho_0 + u_g + p_g ) \uvec^\mu \uvec_\nu + p_g \delta^\mu_\nu \nonumber ,\\
{T^{\rm EM}}^\mu_\nu &=& b^2 \uvec^\mu \uvec_\nu + p_b\delta^\mu_\nu - \bvec^\mu \bvec_\nu \nonumber ,\\
T^\mu_\nu &=& {T^{\rm MA}}^\mu_\nu + {T^{\rm EM}}^\mu_\nu .
\end{eqnarray}
The MA term can be decomposed into a particle (PA) term: ${T^{\rm
    PA}}^\mu_\nu = \rho_0 \uvec_\nu \uvec^\mu$ and an enthalpy (EN)
term.  The MA term can be reduced to a free thermo-kinetic energy
(MAKE) term, which is composed of free particle (PAKE) and enthalpy
(EN) terms:
\begin{eqnarray}\label{MAKEs}
{T^{\rm MAKE}}^\mu_\nu &=& {T^{\rm MA}}^\mu_\nu - \rho_0 \uvec^\mu \etavec_\nu/\alpha ,\\
{T^{\rm PAKE}}^\mu_\nu &=& (\uvec_\nu - \etavec_\nu/\alpha)\rho_0 \uvec^\mu  \nonumber ,\\
{T^{\rm EN}}^\mu_\nu &=&  (u_g + p_g)\uvec^\mu \uvec_\nu + p_g \delta^\mu_\nu \nonumber ,
\end{eqnarray}
such that ${T^{\rm MAKE}}^\mu_\nu = {T^{\rm PAKE}}^\mu_\nu+{T^{\rm
    EN}}^\mu_\nu$.  Here, $u_g$ is the internal energy density and
$p_g=(\Gamma-1)u_g$ is the ideal gas pressure with adiabatic index
$\Gamma=4/3$ ($\Gamma=5/3$ may lead to somewhat different results ;
\citealt{mg04,mm07}). The contravariant fluid-frame magnetic 4-field
is given by $\bvec^\mu$, which is related to the lab-frame 3-field via
$\bvec^\mu = \Bvec^\nu h^\mu_\nu/\uvec^t$ where $h^\mu_\nu = \uvec^\mu
\uvec_\nu + \delta^\mu_\nu$ is a projection tensor, and
$\delta^\mu_\nu$ is the Kronecker delta function.  The magnetic energy
density ($u_b$) and pressure ($p_b$) are $u_b=p_b=\bvec^\mu
\bvec_\mu/2 = b^2/2$.  The total pressure is $p_{\rm tot} = p_g +
p_b$, and plasma $\beta\equiv p_g/p_b$.  The 4-velocity of a zero
angular momentum observer (ZAMO) is $\etavec_\mu=\{-\alpha,0,0,0\}$
where $\alpha=1/\sqrt{-g^{tt}}$ is the lapse.  The 4-velocity relative
to this ZAMO is $\reluvec^\mu = \uvec^\mu - \gamma \etavec^\mu$ where
$\gamma=-\uvec^\alpha \etavec_\alpha$.  For any 3-vector
(e.g. $\Bvec^i$), the ``quasi-orthonormal'' vector is $B_i\equiv
\Bvec^i \sqrt{g_{ii}}$ computed in spherical polar coordinates.

Magnetic flux conservation is given by the induction equation
\begin{equation}
\partial_t(\detg \Bvec^i) = -\partial_j[\detg(\Bvec^i \vvec^j - \Bvec^j \vvec^i)] ,
\end{equation}
where $g={\rm Det}(g_{\mu\nu})$ is the metric's determinant, and the
lab-frame 3-velocity is $\vvec^i = \uvec^i/\uvec^t$.  No explicit
viscosity or resistivity are included, but we use the energy
conserving HARM scheme so all dissipation is captured
\citep{gam03,mck06ffcode}.

The energy-momentum conservation equations are only modified due to
so-called numerical density floors that keep the numerical code stable
as described in detail in Appendix~\ref{sec:nummethods}.  The injected
densities are tracked and removed from all calculations.

\section{Diagnostics}
\label{sec:diagnostics}

Diagnostics are computed from snapshots produced every $\sim 2r_g/c$.
For quantities $Q$, averages over space ($\langle Q \rangle$) and time
($[Q]_t$) are performed directly on $Q$ (e.g. on $v_\phi$ rather than
on any intermediate values).  Any flux ratio vs. time with numerator
$F_N$ and denominator $F_D$ ($F_D$ often being mass or magnetic flux)
is computed as $R(t)=\langle F_N(t) \rangle / [\langle F_D
\rangle]_t$.  Time-averages are then computed as $[R]_t$.

\subsection{Fluxes and Averages vs. Radius}
\label{sec:avgradquant}

For flux density $F_d$, the flux integral is
\begin{equation}
F(r) \equiv \int dA_{23} F_d ,
\end{equation}
where $dA_{23}=\gdet d\xvec^{(2)} d\xvec^{(3)}$ ($dA_{\theta\phi}$ is
the spherical polar version).  For example, $F_d=\rho_0 \uvec^{(1)}$
gives $F=\dot{M}$, the rest-mass accretion rate.  For weight $w$, the
average of $Q$ is
\begin{equation}
Q_w(r) \equiv \langle Q \rangle_w \equiv \frac{\int dA_{\theta\phi} w Q}{\int dA_{\theta\phi} w} ,
\end{equation}
All $\theta,\phi$ angles are integrated over.

\subsection{Fluxes and Averages vs. $\theta$}
\label{sec:avgthetaquant}

The flux angular distribution, at any given radius, is
\begin{equation}\label{fluxtheta}
F(\theta) = \int_{\theta'=0}^{\pi/2-|\theta-\pi/2|} dA_{\theta'\phi} F_d + \int^{\pi}_{\theta'=\pi/2+|\theta-\pi/2|} dA_{\theta'\phi} F_d  ,
\end{equation}
which just integrates up from both poles towards the equator, is
symmetric about the equator, and gives the total flux value at
$\theta=\pi/2$.  The average of $Q$ vs. $\theta$ using weight $w$ is
given by
\begin{equation}
Q_w(\theta) = \frac{\int dA_{\theta\phi} w Q}{\int dA_{\theta\phi} w} .
\end{equation}
All $\phi$-angles are integrated over.

\subsection{Disk Thickness Measurements}
\label{sec:diskthick1}

The disk's geometric half-angular thickness is given by
\begin{equation}\label{thicknesseq}
\theta^d \equiv \left(\left\langle \left(\theta-\theta_0\right)^2 \right\rangle_{\rholab}\right)^{1/2} ,
\end{equation}
where we integrate over all $\theta$ for each $r,\phi$, and $\theta_0
\equiv \pi/2 + \left\langle \left(\theta-\pi/2\right)
\right\rangle_{\rholab}$ is also integrated over all $\theta$ for each
$r,\phi$, and the final $\theta^d(r)$ is from $\phi$-averaging with no
additional weight or $\detg$ factor.  This way of forming
$\theta^d(r)$ works for slightly tilted thin disks or disordered thick disks.
For a Gaussian distribution in density, this satisfies
$\rholab/(\rholab[\theta=0]) \sim \exp(-\theta^2/(2(\theta^d)^2))$.
For sound speed $c_s=\sqrt{\Gamma p_g/(\rho_0 + u_g + p_g)}$, the
thermal half-angular thickness is
\begin{equation}\label{thetateq}
\theta^t_w \equiv \arctan{\left(\frac{\langle c_s\rangle_w}{\langle v_{\rm rot} \rangle_w}\right)} ,
\end{equation}
where $v_{\rm rot}^2 = v_\phi^2 + v_\theta^2$.  For a thin hydrostatic
non-relativistic Keplerian (i.e. $v_{\rm rot}=|v_{\rm K}|$ with
$v_{\rm K}\approx R/(a + R^{3/2})$) Gaussian disk,
$\theta^d=c_s/v_{\rm rot}$ for $c_s$ and $v_{\rm rot}$ at the disk
plane.  Also, $\theta^t_{\rholab}\approx 0.93\theta^d$ for
$\Gamma=4/3$.  Ram pressure forces \citep{2005A&A...433..619B} and
magnetic forces \citep{2011ApJ...737...94C} can cause $\theta^d\ll
\theta^t$. See Tables~\ref{tbl1},~\ref{tbl3}.  Note that ADAFs have
$\theta^t\gtrsim 1$ \citep{nar94}.

\subsection{BH, Disk, Jet, Magnetized Wind, and Entire Wind}
\label{integrations}

Many quantities ($Q$) vs. $r$ or vs. $\theta$ or vs. $\phi$ are
considered for various weights and conditions.  We define the
superscript ``f'' (full flow) case as applies for weight $w=1$ with no
conditions, ``fdc'' (full flow except avoids highly magnetized jet
where numerical floors are activated), ``dc'' (disk plus corona but no
jet) case as applies for $w=1$ with condition $b^2/\rho_0<1$,
``dcden'' (density-weighted average) with $w=\rholab$ and no
conditions, ``$\theta^d$'' (within $1$ disk half-angular thickness)
case with $w=1$ and condition of $|\theta-\theta_0|<\theta^d$, ``eq''
(within 3 cells around the equator) case with $w=1$, and ``jet'' or
``j'' case (jet only) with $w=1$ and the condition that density floors
are activated (see Appendix~\ref{sec:nummethods}).  For quantities
vs. $\theta$ or vs. $\phi$, we radially average within $\pm 0.1r$ at
radius $r$.

Fluxes, described in the next section, have integrals computed for a
variety of (somewhat arbitrary) conditions.  The subscript ``BH'' or
``H'' is for all angles on the horizon.  The subscript ``j'' or
``jet'' is for the ``jet'' with condition $b^2/\rho_0\ge 1$.  When the
jet is measured at a single radius, we use $r=50r_g$ (except the MB09Q
model that uses $r=30r_g$ due to its limited radial range).  The
subscript ``mw'' is for the ``magnetized wind'' with conditions
$b^2/\rho_0<1$ and $\beta<2$ for all fluxes, except for the rest-mass
flux that also has $-(\rho_0+u_g+p_g)\uvec_t/\rho_0>1$ (i.e.
thermo-kinetically unbound).  The ``w'' or ``wind'' subscript is for
the ``entire wind'' with the condition of $b^2/\rho_0<1$ that includes
all of the flow except the jet.  The subscript is ``in'' (``out'') for
the condition $u_r<0$ ($u_r>0$).

We also compute (as shown in Table~\ref{tbl3}) the number of grid
cells that cover the disk half-angular thickness ($\theta^d$) computed
by Eq.~(\ref{thicknesseq}) at the horizon ($\theta^d_{\rm H}$) and
other radii $r$ ($\theta^d_{r/r_g}$).  For comparison, we also compute
the thermal half-angular thickness ($\theta^t=\theta^t_{\rholab}$).
Also computed are the disk-corona interface angular location at a
given radius $r$ (denoted $\theta^{dc}_{r/r_g}$) defined by where
$\beta=1$ and the corona-jet interface angular location at a given
radii $r$ (denoted $\theta^{cj}_{r/r_g}$) defined by where
$b^2/\rho_0=1$.  In practice, these interface locations are defined
similarly to Eq.~(\ref{thicknesseq}), except $\theta_0=\pi/2$ and
weight $w=\uvec^t (\rho_0 + u_g + p_g + b^2)$ are chosen with the
following conditions.  The disk-corona interface calculation uses the
condition $1/2<\beta<1$, unless that condition is not met by any grid
cells at that radius -- in which case the condition $1/10<\beta<1$ is
used.  The corona-jet interface calculation uses the condition
$1<b^2/\rho_0<2$, unless (very rarely) that condition is not met by
any grid cells at that radius -- in which case the condition
$\bsqorhomaxdiaghigh>b^2/\rho_0>1$ is used.

\subsection{Fluxes of Mass, Energy, and Angular Momentum}
\label{fluxes}

The rest-mass flux, specific energy flux, and specific angular
momentum flux are respectively given by
\begin{eqnarray}\label{Dotsmej}
\dot{M} &=&  \left|\int\rho_0 \uvec^r dA_{\theta\phi}\right| , \\
\emath \equiv \frac{\dot{E}}{[\dot{M}]_t} &=& -\frac{\int T^r_t dA_{\theta\phi}}{[\dot{M}]_t} , \\
\jmath \equiv \frac{\dot{J}}{[\dot{M}]_t} &=& \frac{\int T^r_\phi dA_{\theta\phi}}{[\dot{M}]_t} ,
\end{eqnarray}
and are computed in Tables~\ref{tbl6}--\ref{tbl7}.

The net flow efficiency is given by
\begin{equation}\label{eff}
  \eff = \frac{\dot{E}-\dot{M}}{[\dot{M}]_t} = \frac{\dot{E}^{\rm EM}(r) + \dot{E}^{\rm MAKE}(r)}{[\dot{M_{\rm H}}]_t} . \\
\end{equation}
Positive values correspond to an extraction of positive energy from
the system at some radius.  These $\eff$'s are computed in
Tables~\ref{tbl6},~\ref{tbl7}.

The BH's dimensionless spin-up parameter is
\begin{equation}\label{spinevolve}
s \equiv \frac{d(a/M)}{dt}\frac{M}{[\dot{M}]_t}  =  -\jmath - 2\frac{a}{M}(1-\eff) ,
\end{equation}
(computed in Table~\ref{tbl15}).  All $\theta$ and $\phi$ angles are
integrated over.  The BH is in ``spin equilibrium'' for $s=0$
\citep{gammie_bh_spin_evolution_2004}.

\subsection{Magnetic Flux}
\label{magneticfluxdiag}

The radial magnetic flux vs. $\theta$ at any radius is
\begin{equation}
\Psi_r(r,\theta) = \int dA_{\theta\phi} \Bvec^r .
\end{equation}
The signed value of the maximum absolute value over all $\theta$
angles (${\rm smaxa}_\theta$) of the magnetic flux is
\begin{equation}
\Psi_{\rm t}(r) \equiv {\rm smaxa}_\theta \Psi_r ,
\end{equation}
and $\Psi_{\rm tH} \equiv \Psi_{\rm t}(r=r_{\rm H})$ is the horizon's
magnetic flux.  The half-hemisphere horizon flux is
\begin{equation}
\Psi_{\rm H} \equiv \Psi_r(r=r_{\rm H}) ,
\end{equation}
as integrated from $\theta=\pi/2$ to $\pi$ (negative compared to the
integral from $\theta=0$ to $\pi/2$). The $\theta$ magnetic flux
vs. radius at angle $\theta$ is
\begin{equation}
\Psi_\theta(r,\theta) = \int_{r_{\rm H}}^{r} \detg d\xvec^{(1)} d\xvec^{(3)} \Bvec^{\xvec^{(2)}} ,
\end{equation}
where the vertical magnetic flux threading the equator is
\begin{equation}
\Psi_{\rm eq}(r) \equiv \Psi_\theta(r,\theta=\pi/2) .
\end{equation}
The total magnetic flux along the equator is
\begin{equation}
\Psi(r) \equiv \Psi_{\rm H} + \Psi_{\rm eq}(r) .
\end{equation}
For all forms of $\Psi$, all $\phi$-angles are integrated over.

The magnetic flux can be normalized in various ways (as computed in
Table~\ref{tbl11}).  Normalization by the initial flux at $r_0$ gives
$\Psi(r)/\Psi(r_0)$.  One type of field geometry we will use has
multiple field loops of alternating polarity as a function of radius.
So another normalization is by the initial $i$-th extrema vs. radius,
which gives $\Psi/\Psi_i$ that picks up the extrema in the magnetic
flux over each field loop.  Normalization by the initial value of an
extrema gives $\Psi/\Psi_i(t=0)$.  We also need to form a measure that
indicates how much flux is available to the BH.  So we
consider the normalization by the flux in the disk that is immediately
available to the horizon of the same polarity.  This measure is given
by $\Psi_{\rm H}/\Psi_a$, where $\Psi_a$ is the value where $\Psi(r)$
goes through its first extremum of the same sign of magnetic flux
(i.e. out to the radius with the same polarity of dipolar-like field)
as on the horizon.  If the horizon value is itself an extremum, then
$\Psi_{\rm H}/\Psi_a=1$ implying that the region immediately beyond
the horizon only has opposite polarity field.

The absolute magnetic flux ($\Phi$) is computed similarly to $\Psi$,
except one 1) inserts absolute values around the field (e.g. $\Bvec^r$
and $\Bvec^\theta$ in the integrals); 2) puts absolute values around
the integral ; and 3) divides by $2$ so that a dipolar field has
$|\Psi_{\rm t}|=\Phi$.  For example, $\Phi_r(r,\theta) =
(1/2)\left|\int dA_{\theta\phi} |\Bvec^r| \right|$.  The quantity
$\Phi/\Psi_{\rm t}$ (computed in Table~\ref{tbl11}, and which is the
only flux ratio directly time-averaged as $[\Phi/\Psi_{\rm t}]_t$) is
roughly the vector spherical harmonic multipole $l$ of the
$\phi$-component of the magnetic vector potential:
\begin{equation}\label{vpot}
\Avpotvec_\phi=\int_{\theta'=0}^\theta \detg \Bvec^r d\theta'
\end{equation}
as integrated over all $\phi$.  For example, for $l=\{1\ldots 8\}$ one
gets $|\Phi/\Psi_{\rm t}| = 1$, $2$, $2.6$, $3.5$, $4$, $5.6$, $5.7$,
and $6.7$.

The \citet{gammie99} model normalization gives
\begin{equation}\label{equpsilon}
\Upsilon \approx 0.7\frac{\Phi_r}{\sqrt{[\dot{M}]_t}} ,
\end{equation}
which accounts for $\Phi_r$ being in Heaviside-Lorentz units
\citep{pmntsm10}.  Compared to Gaussian units version of $\phi_{\rm
  H}\equiv \Phi_{\rm H}/\sqrt{\dot{M} r_g^2 c}$ defined in
\citet{tnm11}, $\Upsilon\approx 0.2\phi_{\rm H}$.  $\Upsilon_{\rm H}$
and $\Upsilon_j$ are normalized by $\dot{M}_{\rm H}$, $\Upsilon_{\rm
  in}$ by $\dot{M}_{\rm in}$, and $\Upsilon_{\rm mw}$ and
$\Upsilon_{\rm w}$ respectively by $\dot{M}_{\rm mw}$ and
$\dot{M}_{\rm w}$.  $\Upsilon$ is computed in Table~\ref{tbl11}.

The field line rotation frequency with respect to the BH spin ($z$)
axis is computed various ways.  We consider $\Omega^a_{\rm F} \equiv
\Fvec_{tr}/\Fvec_{r\phi}$, $\Omega^b_{\rm F} \equiv
\Fvec_{t\theta}/\Fvec_{\theta\phi}$, $\Omega^c_{\rm F} \equiv
|\vvec^\phi| + {\rm sign}[\uvec^r] (v_p/B_p)|\Bvec^\phi|$ with
$v_p=\sqrt{v_r^2 + v_\theta^2}$ and $B_p=\sqrt{B_r^2 + B_\theta^2}$,
and
\begin{equation}\label{omegaeq}
\Omega_{\rm F}\equiv \Omega^d_{\rm F} \equiv \vvec^\phi - \Bvec^\phi \left(\frac{v_r B_r + v_\theta B_\theta}{B_r^2 + B_\theta^2}\right) .
\end{equation}
We also consider $\Omega^e_{\rm
  F}=[|\Fvec_{t\theta}|]_t/[|\Fvec_{\theta\phi}|]_t$.  These
$\Omega_{\rm F}$ are normalized by the BH rotation angular
frequency $\Omega_{\rm H}=a/(2Mr_{\rm H})$.

\subsection{Inflow Equilibrium and $\alpha$ Viscosity}
\label{sec_infloweq}

Inflow equilibrium is defined as when the flow is in a complete
quasi-steady-state and the accretion fluxes are constant (apart from
noise) vs. radius and time.  The inflow equilibrium timescale is
\begin{equation}\label{tieofrie}
t_{\rm ie} = N \int_{r_i}^{r_{\rm ie}} dr\left(\frac{-1}{[\langle v_r\rangle_{\rholab}]_t}\right) ,
\end{equation}
for $N$ inflow times from $r=r_{\rm ie}$ and $r_i=12r_g$ to focus on
the more self-similar flow.  $t_{\rm ie}$ is used in
Table~\ref{tbl14}, where $r^{\rm dcden}_{\rm i} = r_i$, $r^{\rm
  dcden}_{\rm f}=r_{\rm ie}$ with $N=1$, and $r^{\rm
  dcden}_{\rm o}$ uses $r_{\rm ie}$ with $N=3$.

Viscous theory gives a GR $\alpha$-viscosity estimate for $v_r$ of
$v_{\rm visc}\sim -G\alpha(\theta^d)^2 |v_{\rm rot}|$
\citep{pt74,pmntsm10}, with GR correction $G$ ($\lesssim 1.5$ for
$r\gtrsim 58r_g$) and (not the lapse)
\begin{eqnarray}\label{alphaeq}
\alpha &=& \alpha_{\rm PA} + \alpha_{\rm EN} + \alpha_{\rm M1} + \alpha_{\rm M2} , \\
\alpha_{\rm PA} &\approx& \frac{\rho_0 \delta u_r (\delta \uvec_\phi \sqrt{g^{\phi\phi}})}{p_{\rm tot}} , \nonumber\\
\alpha_{\rm M2} &\approx& -\frac{b_r (\bvec_\phi \sqrt{g^{\phi\phi}})}{p_{\rm tot}} , \nonumber\\
\alpha_{\rm mag} &\approx& -\frac{b_r (\bvec_\phi \sqrt{g^{\phi\phi}})}{p_b} ,\nonumber \\
\alpha_{\rm eff} &\equiv& \frac{v_r}{v_{\rm visc}/\alpha} ,\nonumber
\end{eqnarray}
$\alpha_{\rm eff2} \equiv \alpha_{\rm eff}(|v_{\rm rot}|/|v_{\rm K}|)$,
and (small) $\alpha_{\rm EN} \approx (u_g+p_g)
\delta u_r (\delta \uvec_\phi \sqrt{g^{\phi\phi}})/p_{\rm tot}$ and
$\alpha_{\rm M1} \approx b^2 \delta u_r (\delta\uvec_\phi
\sqrt{g^{\phi\phi}})/p_{\rm tot}$.  Here, $\delta u$ is the deviation
of the velocity from its average (taken over all $\phi$ and over the
time-averaging period). The $\alpha$ (e.g. in Table~\ref{tbl4}) is
averaged as follows.  The numerator and denominator are separately
volume averaged in $\theta,\phi$ for each $r$.  Weight $w=1$ with
condition $b^2/\rho_0<1$ gives $\alpha_a$ for the disk+corona, while
$w=\rholab$ gives $\alpha_b$ for the heavy disk.  Notice $\alpha_{\rm
  M2} = \alpha_{\rm mag}/(1+\beta_{\rm mag})$ for some $\beta$ denoted
$\beta_{\rm mag}$, and $\sin(2\theta_b)=\alpha_{\rm mag}$ for tilt
angle $\theta_b$ \citep{2011arXiv1106.4019S}.  These $\alpha$'s are
accurate for $|v|\ll c$ as true for $r\gtrsim 2r_g$ in our models, while
$\alpha_{\rm eff}$ is accurate far outside the inner-most stable
circular orbit (ISCO).

\subsection{Modes and Correlation Lengths}
\label{sec_cor}

The flow structure is studied via the discrete Fourier transform of
$dq$ (related to quantity $Q$) along $x=r,\theta,\phi$ giving
amplitude $a_p$ for $p=n,l,m$, respectively.  The averaged amplitude
is
\begin{equation}\label{am}
\left\langle \left|a_p\right| \right\rangle \equiv \left\langle \left|\mathcal{F}_p\left(dq\right)\right| \right\rangle \equiv \int_{\rm not\ x} \left| \sum_{k=0}^{N-1}dq~{\rm e}^{\frac{-2\pi i p k}{N-1} } \right| ,
\end{equation}
computed at $r=r_{\rm H},4r_g,8r_g,30r_g$.  The $x$ is one of
$r,\theta,\phi$ and ``not x'' are others (e.g. $\theta,\phi$ for
$x=r$).  The $dq$ is (generally) a function of $x$ on a uniform grid
indexed by $k$ of $N$ cells that span: $\delta r$ equal to $0.75r$
around $r$ for $x=r$, $\pi$ for $x=\theta$, and $2\pi$ for
$x=\phi$. The $N$ is chosen so all structure from the original grid is
resolved, while the span covered allows many modes to be resolved.

For all $x$, $dq \equiv \detg d\xvec^{(1)} d\xvec^{(2)} d\xvec^{(3)}
\delta Q/q_N$.  For $x=r,\theta$, we let $q_N=\int_{\rm\ not\ x} \detg
d\xvec^{(1)} d\xvec^{(2)} d\xvec^{(3)} \langle [Q]_t \rangle$,
$\langle [Q]_t \rangle$ as the time-$\phi$ averaged $Q$, and $\delta Q
= Q - \langle[Q]_t\rangle$.  Using $dq$ removes gradients with
$r,\theta$ so the Fourier transform acts on something closer to
periodic with constant amplitude (see also \citealt{bas11}).  For
$x=\phi$, we let $q_N=1$ and $\delta Q = Q$ because the equations of
motion are $\phi$-ignorable.  For $x=\theta,\phi$, the radial integral
is computed within $\pm 0.1r$.  For $x=r,\theta$, the $\phi$ integral
is over all $2\pi$.  For $x=r,\phi$, the $\theta$ integral is over all
$\pi$.  For all $x$ cases, the $\theta$ range of values uses the
``fdc'' or ``jet'' conditions (respectively called ``Disk'' and
``Jet'' in
sections~\ref{sec:phidep},\ref{sec:alphamri},\ref{sec:resolution}),
where these conditional regions are defined via $\phi$-averaged
quantities at each time. Notice we average the mode's absolute
amplitude, because the amplitude of $\langle \delta Q\rangle$
de-resolves power (e.g. $m=1$ out of phase at different $\theta$ gives
$\langle \delta Q\rangle \to 0$ and $a_m\to 0$) and is found to
underestimate small-scale structure.

We also compute the correlation length: $\lambda_{x,\rm cor}=x_{\rm
  cor}-x_0$, where $x_0=0$ for $x=\theta,\phi$ and $x_0$ is the inner
radius of the above given radial span for $x=r$, where $n_{\rm cor} =
\delta r/\lambda_{r,\rm cor}$, $l_{\rm cor}=\pi/\lambda_{\theta,\rm
  cor}$, and $m_{\rm cor} =(2\pi)/\lambda_{\phi,\rm cor}$.  The
Wiener-Khinchin theorem for the auto-correlation gives
\begin{equation}\label{mcor}
\exp(-1) = \frac{\mathcal{F}^{-1}_{x=x_{\rm cor}}[\langle|a_{p>0}|\rangle^2]}{\mathcal{F}^{-1}_{x=x_0}[|\langle a_{p>0}|\rangle^2]} ,
\end{equation}
where $\mathcal{F}^{-1}[\langle |a_{p>0}|\rangle ^2]$ is the inverse
discrete Fourier transform of $\langle |a_p|\rangle^2$ but with
$\langle a_0\rangle$ reset to $0$ (i.e. mean value is excluded).

\subsection{Suppression of the MRI}
\label{sec_mri}

The MRI is a linear instability with fastest growing wavelength of
\begin{equation}\label{lambdamri}
\lambda_{x,\rm MRI} \approx  2\pi \frac{|v_{x,\rm A}|}{|\Omega_{\rm rot}|} , \\
\end{equation}
for $x=\theta,\phi$, where $|v_{x,\rm A}|=\sqrt{\bvec_x
  \bvec^x/\epsilon}$ is the $x$-directed~\alf~speed, $\epsilon\equiv
b^2 + \rho_0 + u_g + p_g$, and $r\Omega_{\rm rot} = v_{\rm rot}$.
$\lambda_{\rm MRI}$ is accurate for $\Omega_{\rm rot}\propto r^{-5/2}$
to $r^{-1}$.  $\Omega_{\rm rot},v_{\rm A}$ are separately
angle-volume-averaged at each $r,t$.

The MRI suppression factor corresponds to the number of MRI
wavelengths across the full disk:
\begin{equation}\label{q2mri}
\Qtwo \equiv \frac{2 r \theta^d}{\lambda_{\theta,\rm MRI}} .
\end{equation}
Wavelengths $\lambda<0.5\lambda_{\theta,\rm MRI}$ are stable, so the
linear MRI is suppressed for $\Qtwo<1/2$ when no unstable wavelengths
fit within the full disk \citep{bh98,pp05}.  $\Qtwo$ (or $\Qtwoweak$)
uses averaging weight $w=(b^2\rholab)^{1/2}$ (or $w=\rholab$),
condition $\beta>1$, and excludes regions where density floors are
activated. Weight $w=(b^2\rholab)^{1/2}$ is preferred, because much
mass flows in current sheets where the magnetic field vanishes and yet
the MRI is irrelevant.  When computing the averaged $\Qtwo$, $v_{\rm
  A}$ and $|\Omega_{\rm rot}|$ are separately
$\theta,\phi$-volume-averaged within $\pm 0.2r$ for each $t,r$.  The
averaged $\Qtwo$ is at most $30\%$ smaller than $\Qtwoweak$.

$\Qtwoinitio$ (in Table~\ref{tbl1}) gives $\Qtwo$ at $t=0$ at
$r=r_i=30r_g$ and $r=r_o=50r_g$ averaged within $\pm 0.2r$, except
models MB09D/Q use $r_i=10r_g$ and $r_0=15r_g$ due to their disk's
limited radial extent.  Time-averages (see Table~\ref{tbl4}) are
obtained for $r=r^{\rm dcden}_{\rm i}$ and $r=r^{\rm dcden}_{\rm o}$
(see Table~\ref{tbl14}) averaged within $\pm 0.2r$.  Table~\ref{tbl4}
also gives $r=r_{\Qtwo=1/2}$, within which the linear MRI is
suppressed.

\section{Physical and Numerical Models}
\label{sec:physnummodels}

This section describes our models with parameters shown in
Tables~\ref{tbl1},~\ref{tbl2}.  The model names are in the form
AxByNz, where x is the approximate value of the BH spin, y identifies
the field geometry (p=poloidal, f=flipping poloidal, t=toroidal), and
z identifies the normalization of the magnetic field.  For instance,
our fiducial model A0.94BfN40 has a spinning BH ($a/M=0.9375$), a
poloidal field that flips polarity with radius, and $\beta_{\rm
  min}\approx 40$ (i.e. smallest value of $\beta$ is $40$).  The label
$c?$ (with number $?$) is appended for convergence tests, $r$ is
appended to the model name if it is another realization of an
identical model, and $HR$ is appended if it is a high-resolution
continuation of some model.  2D axisymmetric models are marked with
$^*$.  Models from \citet{mb09} are denoted MB09D and MB09Q for their
dipolar and large-scale quadrupolar models, respectively.  The
remaining TMN11 models are similar to \citet{tnm11}.  Primary models
(fiducial thick poloidal: A0.94BfN40, thick retrograde poloidal:
A-0.94BfN40HR, thick toroidal: 0.94BtN10HR, thinner poloidal:
A0.99N100) have bold font labels.

Table~\ref{tbl2} gives the time-averaging period (from $T^a_i$ to
$T^a_f$ in $r_g/c$).  Models have $T_i=0$, except A0.94BtN10HR starts
as a super-sampled A0.94BtN10 at $T_i=57400r_g/c$, A-0.94BfN40HR
starts as A0.94BfN40 at $T_i=7995r_g/c$ with $a/M=-0.9375$, and
A0.99N100 doubles $N_\phi$ (to shown $N_\phi$) at $t\approx 14675r_g/c$.
For A0.94BpN100, $T^a_i,T^a_f=14000,27048$ and
$T^a_i,T^a_f=8000,13000$ give similar results (validating
$T^a_f=13000$ for similar models).

\subsection{Physical Models}
\label{sec:modelsetup}

\begin{table*}
\caption{Physical Model Parameters}
\begin{center}
\begin{tabular}[h]{|l|r|r|r|r|r|r|r|r|r|r|}
\hline
ModelName         &               $a/M$    &        FieldType     &             $\beta_{\rm{}min}$  &    $\beta_{\rm{}rat-of-avg}$  &     $\beta_{\rm{}rat-of-max}$  &    $\theta^d_{r_{\rm{}max}}$  &     $\theta^t_{r_{\rm{}max}}$  &    $S_{d,t=0,\rm{}MRI,\{i,  o\}}$  &    $T_f$   \\
\hline
{\bf              A0.94BfN40}     &        0.9375   &             PoloidalFlip  &                   40   &                          230   &                          40   &                          0.61  &                          1.1  &                        1.6,   2.6  &       26548  \\  
A0.94BfN100c1     &               0.9375   &        PoloidalFlip  &             120                 &    490                        &     130                        &    0.61                       &     1.1                        &    2.4,                     4.4    &    13000   \\     
A0.94BfN100c2     &               0.9375   &        PoloidalFlip  &             130                 &    480                        &     140                        &    0.61                       &     1.1                        &    2.3,                     4.4    &    13000   \\     
A0.94BfN100c3     &               0.9375   &        PoloidalFlip  &             120                 &    490                        &     130                        &    0.61                       &     1.1                        &    2.3,                     4.5    &    13000   \\     
A0.94BfN100c4     &               0.9375   &        PoloidalFlip  &             130                 &    490                        &     130                        &    0.61                       &     1.1                        &    2.3,                     4.4    &    13000   \\     
A0.94BfN40c5$^*$  &               0.9375   &        PoloidalFlip  &             40                  &    230                        &     43                         &    0.61                       &     1.1                        &    1.6,                     2.6    &    13000   \\     
\\
{\bf              A-0.94BfN40HR}  &        -0.9375  &             PoloidalFlip  &                   40   &                          230   &                          40   &                          0.61  &                          1.1  &                        1.6,   2.6  &       18416  \\  
A-0.94BfN30       &               -0.9375  &        PoloidalFlip  &             30                  &    120                        &     31                         &    0.6                        &     1.1                        &    1.2,                     2.2    &    13000   \\     
A-0.5BfN30        &               -0.5     &        PoloidalFlip  &             34                  &    130                        &     35                         &    0.61                       &     1.1                        &    1.3,                     2.2    &    13000   \\     
A0.0BfN10         &               0        &        PoloidalFlip  &             10                  &    42                         &     11                         &    0.61                       &     1.1                        &    0.68,                    1.2    &    13832   \\     
A0.5BfN30         &               0.5      &        PoloidalFlip  &             33                  &    120                        &     36                         &    0.61                       &     1.1                        &    1.3,                     2.2    &    13000   \\     
A0.94BfN30        &               0.9375   &        PoloidalFlip  &             30                  &    120                        &     32                         &    0.61                       &     1.1                        &    1.2,                     2.2    &    13000   \\     
A0.94BfN30r       &               0.9375   &        PoloidalFlip  &             32                  &    120                        &     33                         &    0.61                       &     1.1                        &    1.2,                     2.2    &    13000   \\     
\\
A0.94BpN100       &               0.9375   &        Poloidal      &             93                  &    240                        &     110                        &    0.61                       &     1.1                        &    1.8,                     1.8    &    27048   \\     
\\
A-0.94BtN10       &               -0.9375  &        Toroidal      &             10                  &    1600                       &     25                         &    0.6                        &     1.1                        &    370,                     350    &    93944   \\     
A-0.5BtN10        &               -0.5     &        Toroidal      &             10                  &    980                        &     17                         &    0.61                       &     1.1                        &    300,                     260    &    129392  \\     
A0.0BtN10         &               0        &        Toroidal      &             10                  &    1200                       &     28                         &    0.61                       &     1.1                        &    320,                     300    &    96796   \\     
A0.5BtN10         &               0.5      &        Toroidal      &             10                  &    1000                       &     17                         &    0.61                       &     1.1                        &    290,                     250    &    135572  \\     
A0.94BtN10        &               0.9375   &        Toroidal      &             10                  &    1300                       &     20                         &    0.61                       &     1.1                        &    300,                     290    &    93648   \\     
{\bf              A0.94BtN10HR}   &        0.9375   &             Toroidal      &                   10   &                          1300  &                          20   &                          0.61  &                          1.1  &                        300,   290  &       80184  \\  
\\
MB09D             &               0.92     &        PoloidalOld   &             14                  &    530                        &     100                        &    0.18                       &     0.24                       &    3.4,                     16     &    5662    \\     
MB09Q             &               0.9375   &        LSQuad        &             0.0067              &    540                        &     140                        &    0.19                       &     0.25                       &    41,                      60     &    5688    \\     
\\
A-0.9N100         &               -0.9     &        Poloidal2     &             100                 &    150                        &     220                        &    0.21                       &     0.27                       &    3.6,                     4.4    &    20060   \\     
A-0.5N100         &               -0.5     &        Poloidal2     &             100                 &    140                        &     210                        &    0.22                       &     0.28                       &    3.5,                     4.3    &    16335   \\     
A-0.2N100         &               -0.2     &        Poloidal2     &             100                 &    140                        &     200                        &    0.22                       &     0.29                       &    3.5,                     4.3    &    15175   \\     
A0.0N100          &               0        &        Poloidal2     &             100                 &    140                        &     210                        &    0.22                       &     0.29                       &    3.6,                     4.3    &    18585   \\     
A0.1N100          &               0.1      &        Poloidal2     &             100                 &    150                        &     210                        &    0.22                       &     0.29                       &    3.6,                     4.4    &    20000   \\     
A0.2N100          &               0.2      &        Poloidal2     &             100                 &    140                        &     210                        &    0.23                       &     0.29                       &    3.6,                     4.3    &    19350   \\     
A0.5N100          &               0.5      &        Poloidal2     &             100                 &    150                        &     210                        &    0.23                       &     0.29                       &    3.6,                     4.4    &    19540   \\     
A0.9N25           &               0.9      &        Poloidal2     &             25                  &    36                         &     51                         &    0.23                       &     0.3                        &    1.8,                     2.2    &    17350   \\     
A0.9N50           &               0.9      &        Poloidal2     &             50                  &    72                         &     100                        &    0.23                       &     0.3                        &    2.5,                     3.1    &    14385   \\     
A0.9N100          &               0.9      &        Poloidal2     &             100                 &    140                        &     210                        &    0.23                       &     0.3                        &    3.6,                     4.4    &    19895   \\     
A0.9N200          &               0.9      &        Poloidal2     &             200                 &    290                        &     420                        &    0.23                       &     0.3                        &    5.1,                     6.2    &    28620   \\     
{\bf              A0.99N100}      &        0.99     &             Poloidal2     &                   100  &                          160   &                          230  &                          0.24  &                          0.3  &                        3.7,   4.6  &       31400  \\  
\hline
\hline
\end{tabular}
\end{center}
\label{tbl1}
\end{table*}

This study considers BH accretion disk systems.  Our ``thick disk''
models have initial geometric half-angular thickness $\theta^d\sim
0.6$, a range of BH spins ($a/M= ~-0.9375, ~-0.5, ~0, ~0.5, ~0.9375$),
a range of field geometries (constant polarity poloidally-dominated,
flipping polarity poloidally-dominated, and toroidally-dominated), and
a range of initial disk magnetic field strengths ($\beta_{\rm min}=
~10, ~30, ~40, ~100$).  Our ``thinner disk'' TNM11 models have initial
$\theta^d\sim 0.2$, $a/M= ~-0.9, ~-0.5,~-0.2, ~0, ~0.1, ~0.2, ~0.5,
~0.9, ~0.99$, a poloidally-dominated field geometry, and $\beta_{\rm
  min}= ~25, ~50, ~100, ~200$ for $a/M=0.9$.  MB09 models are like
most prior MHD simulations with limited poloidal flux.

The initial mass for the thick disk models is an isentropic
hydro-equilibrium torus \citep{fis76,gam03} with inner edge at $r_{\rm
  in}=10r_g$ and pressure maximum at $r_{\rm max}=100r_g$.  The torus
is marginally unbound by tens of percent, as similar to ADAFs that we
want to model \citep{1995ApJ...444..231N}.  The magnetic field
inserted (described later) makes negligible changes to the torus'
boundedness.  Table~\ref{tbl3} shows the disk's geometric half-angular
thickness ($\theta^d$) at $r_{\rm max}$.  We set $\rho_0=\rho_{\rm
  max}=1$ at the maximum rest-mass density.  To seed the MRI, $u_g$ is
perturbed by a factor $1 + F_R(E-0.5)$, where $F_R=0.1$ and $E$ is a
random number from $0$ to $1$.  The torus is surrounded by an
atmosphere with $\rho_0=10^{-4} (r/r_g)^{-2}$, $u_g=10^{-6}
(r/r_g)^{-5/2}$, $\reluvec^i=0$, and $B^i=0$.

We consider an initial poloidal field geometry to seek poloidal
magnetic flux saturation near the BH.  Field polarity flips are
inserted by modulating the poloidal polarity vs. radius in order to
generate multiple loops of alternating polarity for studying magnetic
field inversion/annihilation.  For this poloidal field geometry, the
$\phi$-component of the magnetic vector potential is
\begin{eqnarray}
\Avpotvec_\phi &\propto& f_1 f_2 ,\\
f_1 &=& |q|^p |r\sin{(\theta)}|^\nu ,\nonumber \\
q &=& u_g/u_{g,\rm max} - f_c ,\nonumber \\
f_2 &=& \sin{(\log{(r/S)}/T)} ,\nonumber
\end{eqnarray}
where $f_1$ has $p=1$ and $\nu=2$, $q$ has $f_c=0.2$, $u_{g,\rm max}$
is the maximum $u_g$, $q=0$ is set if $q<0$, and $f_2$ has
$S=0.5r_{\rm in}$ and $T=0.28$ for the flipping field and $f_2=1$ for
the non-flipping field.

We also consider an initial toroidal field geometry, as the limit of
negligible coherent poloidal magnetic flux, where the
$\theta$-component of the magnetic vector potential is
\begin{equation}
\Avpotvec_\theta \propto r^2 (u_g/u_{\rm max} - f_c) ,
\end{equation}
with $f_c=0.2$.  If $\Avpotvec_\theta<0$, then $\Avpotvec_\theta=0$ is
set.  Then $\detg \Bvec^\phi = \Avpotvec_{\theta,r}$ is computed.  The
random perturbations of $u_g$ also lead to a small radial field via
$\detg \Bvec^r = -\Avpotvec_{\theta,\phi}$, which corresponds to
radial wiggles in the toroidal field.  Within $r\sim 100r_g$ there is
about $10$ times less energy in this radial field compared to the
toroidal field.  Very small truncation-level $\Bvec^\theta$ is also
present.

Table~\ref{tbl1} shows each model's initial field, marked as
``Poloidal'' for the non-flipping poloidal field, ``PoloidalFlip'' for
the flipping poloidal field, and ``Toroidal'' for the toroidal field.
The models from \citet{mb09} use the ``PoloidalOld'' poloidal field
geometry (i.e. $\Avpotvec_\phi\propto (\rho_0 - \rho_{\rm cut})$ and, e.g.,
$\rho_{\rm cut}\sim 0.25 \rho_{\rm max}=0.25$) or use the ``LSQuad''
large-scale quadrupolar field geometry.  The models marked as
``Poloidal2'' use the magnetic field geometry described in
\citet{tnm11}.

The magnetic field strength is set via the plasma $\beta=p_g/p_b \sim
(2/\Gamma)(c_s/v_a)^2$ where $v_a^2 = b^2/(\rho_0 + u_g + p_g + b^2)$
gives the~\alf~speed $v_a$.  Our thick disk models have $\beta_{\rm
  min}$, the smallest value of $\beta$ (within the resolved disk
region, e.g., $r\sim 1000r_g\ll R_{\rm out}$) of $\beta_{\rm
  min}\approx 10$ to $200$. An alternative measure is $\beta_{\rm
  rat-of-maxes}\equiv p_{g,\rm max}/p_{b,\rm max}$, where $p_{g,\rm
  max}$ is the maximum thermal pressure on the domain and $p_{b,\rm
  max}$ is the maximum magnetic pressure on the domain.  Another
alternative is $\beta_{\rm rat-of-avg}\equiv p_{g,\rm avg}/p_{b,\rm
  avg} = \langle p_g \rangle/\langle p_b \rangle$.  These $\beta$ (see
Table~\ref{tbl1}) are computed with condition $b^2/\rho_0<1$.  For poloidal field models,
our choices for $\beta$ ensure that $\Qtwo>1$ so the MRI operates, while
we push close to $\Qtwo\sim 1$ as reached even in toroidal field
models.

\subsection{Numerical Models}
\label{sec:numsetup}

\begin{table*}
\caption{Numerical Model Parameters}
\begin{center}
\begin{tabular}[h]{|l|r|r|r|r|r|r|r|r|r|r|r|r|r|}
\hline
ModelName         &               GridType  &       $N_r$  &    $N_\theta$  &    $N_\phi$  &    $R_{\rm{}in}/r_{\rm{}H}$  &      $R_{\rm{}out}$  &       $\Delta\phi$  &       $A_{r_{\rm{}H}}$  &          $A_{r_i}$  &          $A_{r_o}$  &          $\bfrac{Q_{\theta,\rm{}MRI}}{\rm{}t=0,\{i,  o\}}$  &              $T^a_i$--$T^a_f$  \\
\hline
{\bf              A0.94BfN40}     &         HypExp  &      272  &           128  &         256  &                         0.855  &               26000   &             $2\pi$  &                 4.6:1:4.3  &          1.1:1.1:1  &          1.2:2.1:1  &                                           17,    11             &                 8000--17000   \\               
A0.94BfN100c1     &               HypExp    &       136    &    64          &    128       &    0.727                     &      26000           &       $2\pi$        &       4.6:1:4.2         &          1.2:1.1:1  &          1.2:2.1:1  &          5.6,                                        3.4    &              8000--13000       \\            
A0.94BfN100c2     &               HypExp    &       136    &    64          &    64        &    0.727                     &      26000           &       $2\pi$        &       4.7:1:8.5         &          1:1:1.8    &          1:1.8:1.7  &          5.8,                                        3.4    &              8000--12000       \\            
A0.94BfN100c3     &               HypExp    &       136    &    64          &    32        &    0.727                     &      26000           &       $2\pi$        &       4.6:1:17          &          1:1:3.6    &          1:1.8:3.4  &          5.6,                                        3.4    &              8000--13000       \\            
A0.94BfN100c4     &               HypExp    &       136    &    64          &    16        &    0.727                     &      26000           &       $2\pi$        &       4.5:1:33          &          1:1:7.2    &          1:1.8:6.7  &          5.7,                                        3.4    &              9000--13000       \\            
A0.94BfN40c5$^*$  &               HypExp    &       272    &    128         &    1         &    0.855                     &      26000           &       $2\pi$        &       4.7:1:1100        &          1:1:230    &          1:1.8:220  &          16,                                         11     &              8000--13000       \\            
\\
{\bf              A-0.94BfN40HR}  &         HypExp  &      272  &           128  &         256  &                         0.855  &               26000   &             $2\pi$  &                 4.6:1:4.3  &          1.1:1.1:1  &          1.2:2.1:1  &                                           17,    11             &                 10000--15000  \\               
A-0.94BfN30       &               HypExp    &       136    &    64          &    128       &    0.727                     &      26000           &       $2\pi$        &       4.4:1:4           &          1.2:1.1:1  &          1.2:2.1:1  &          11,                                         6.5    &              8000--13000       \\            
A-0.5BfN30        &               HypExp    &       136    &    64          &    128       &    0.735                     &      26000           &       $2\pi$        &       4.2:1:4           &          1.1:1.1:1  &          1.1:2.1:1  &          9.6,                                        6.7    &              8000--13000       \\            
A0.0BfN10         &               Exp       &       128    &    64          &    128       &    0.808                     &      1000            &       $2\pi$        &       2.2:1:2           &          1.1:1.1:1  &          1.1:2.2:1  &          19,                                         12     &              8000--13832       \\            
A0.5BfN30         &               HypExp    &       136    &    64          &    128       &    0.735                     &      26000           &       $2\pi$        &       4.2:1:4           &          1.1:1.1:1  &          1.1:2.1:1  &          10,                                         7      &              8000--13000       \\            
A0.94BfN30        &               HypExp    &       136    &    64          &    128       &    0.727                     &      26000           &       $2\pi$        &       4.6:1:4.2         &          1.2:1.1:1  &          1.2:2.1:1  &          11,                                         6.8    &              8000--13000       \\            
A0.94BfN30r       &               HypExp    &       136    &    64          &    128       &    0.727                     &      26000           &       $2\pi$        &       4.6:1:4.2         &          1.2:1.1:1  &          1.2:2.1:1  &          11,                                         6.8    &              8000--13000       \\            
\\
A0.94BpN100       &               HypExp    &       136    &    64          &    128       &    0.727                     &      26000           &       $2\pi$        &       4.5:1:4.1         &          1.2:1.1:1  &          1.2:2.1:1  &          7.5,                                        8.4    &              14000--27048      \\            
\\
A-0.94BtN10       &               Exp       &       128    &    64          &    128       &    0.797                     &      1000            &       $2\pi$        &       2.2:1:2           &          1.2:1.1:1  &          1.2:2.1:1  &          $<1$                                        &      58000--93944   \\                
A-0.5BtN10        &               Exp       &       128    &    64          &    128       &    0.806                     &      1000            &       $2\pi$        &       2:1:1.9           &          1.1:1.1:1  &          1.1:2.1:1  &          $<1$                                        &      80000--129392  \\                
A0.0BtN10         &               Exp       &       128    &    64          &    128       &    0.808                     &      1000            &       $2\pi$        &       2:1:1.9           &          1.1:1.2:1  &          1.1:2.2:1  &          $<1$                                        &      58000--96796   \\                
A0.5BtN10         &               Exp       &       128    &    64          &    128       &    0.806                     &      1000            &       $2\pi$        &       2:1:1.9           &          1.1:1.1:1  &          1.1:2.1:1  &          $<1$                                        &      80000--135572  \\                
A0.94BtN10        &               Exp       &       128    &    64          &    128       &    0.797                     &      1000            &       $2\pi$        &       2.4:1:2.1         &          1.2:1.1:1  &          1.2:2.1:1  &          $<1$                                        &      58000--93648   \\                
{\bf              A0.94BtN10HR}   &         Exp     &      256  &           128  &         256  &                         0.894  &               1000    &             $2\pi$  &                 2.2:1:2    &          1.2:1.1:1  &          1.2:2.1:1  &                                           $<1$   &              58000--80184      \\            
\\
MB09D             &               ExpOld    &       256    &    128         &    32        &    0.79                      &      1000            &       $2\pi$        &       1.6:1:12          &          1.4:1:11   &          1.4:1:10   &          4.7,                                        1.3    &              2000--3000        \\            
MB09Q             &               ExpOld    &       128    &    128         &    32        &    0.816                     &      40              &       $2\pi$        &       1.6:1:11          &          1.5:1:10   &          1.4:1:9.7  &          $<1$                                        &      1500--5688     \\                
\\
A-0.9N100         &               TNM11     &       288    &    128         &    64        &    0.7                       &      100000          &       $2\pi$        &       2.3:1:7.9         &          2:1:6.8    &          1.6:1:5.4  &          7,                                          7.2    &              11000--20060      \\            
A-0.5N100         &               TNM11     &       288    &    128         &    32        &    0.81                      &      100000          &       $\pi$         &       1.6:1:8.1         &          1.4:1:6.8  &          1.1:1:5.3  &          7.4,                                        7.5    &              11000--16335      \\            
A-0.2N100         &               TNM11     &       288    &    128         &    32        &    0.81                      &      100000          &       $\pi$         &       1.6:1:8.2         &          1.4:1:6.9  &          1.1:1:5.2  &          7.6,                                        7.5    &              11000--15175      \\            
A0.0N100          &               TNM11     &       288    &    128         &    32        &    0.81                      &      100000          &       $\pi$         &       1.6:1:8           &          1.4:1:6.8  &          1.1:1:5.2  &          7.6,                                        7.6    &              11000--18585      \\            
A0.1N100          &               TNM11     &       288    &    128         &    32        &    0.81                      &      100000          &       $\pi$         &       1.6:1:8           &          1.4:1:6.8  &          1.1:1:5.2  &          7.5,                                        7.4    &              11000--20000      \\            
A0.2N100          &               TNM11     &       288    &    128         &    32        &    0.81                      &      100000          &       $\pi$         &       1.6:1:8.2         &          1.4:1:6.8  &          1.1:1:5.2  &          7.7,                                        7.6    &              11000--19350      \\            
A0.5N100          &               TNM11     &       288    &    128         &    32        &    0.81                      &      100000          &       $\pi$         &       1.6:1:8.1         &          1.4:1:6.9  &          1.1:1:5.2  &          7.7,                                        7.6    &              11000--19540      \\            
A0.9N25           &               TNM11     &       288    &    128         &    32        &    0.8                       &      100000          &       $\pi$         &       1.7:1:8.2         &          1.4:1:6.7  &          1.2:1:5.3  &          16,                                         15     &              11000--17350      \\            
A0.9N50           &               TNM11     &       288    &    128         &    32        &    0.8                       &      100000          &       $\pi$         &       1.7:1:8.2         &          1.5:1:6.8  &          1.2:1:5.3  &          11,                                         11     &              11000--14385      \\            
A0.9N100          &               TNM11     &       288    &    128         &    64        &    0.8                       &      100000          &       $2\pi$        &       1.7:1:8.2         &          1.5:1:6.8  &          1.2:1:5.3  &          7.9,                                        7.7    &              11000--19895      \\            
A0.9N200          &               TNM11     &       288    &    128         &    32        &    0.8                       &      100000          &       $\pi$         &       1.7:1:8.2         &          1.5:1:6.8  &          1.2:1:5.4  &          5.5,                                        5.4    &              16000--28620      \\            
{\bf              A0.99N100}      &         TNM11   &      288  &           128  &         128  &                         0.83   &               100000  &             $2\pi$  &                 1.8:1:4    &          2:1:3.3    &          1.7:1:2.7  &                                           7.5,   7.3            &                 15000--31400  \\               
\hline
\hline
\end{tabular}
\end{center}
\label{tbl2}
\end{table*}

The uniform spatial coordinates $\xvec^{(i)}$ have resolution
$N_r\times N_\theta \times N_\phi$ active grid cells and $4$ boundary
cells for each of the $6$ boundaries in 3D.  The radial grid of $N_r$
cells spans from $R_{\rm in}$ to $R_{\rm out}$ with mapping
\begin{equation}
r(x^{(1)}) = R_0 + \exp{f[\xvec^{(1)}]}
\end{equation}
where $R_0=0$ is chosen in this paper.  For $\xvec^{(1)}<x_{\rm
  break}$
\begin{equation}
f[\xvec^{(1)}] = n_0 \xvec^{(1)} ,
\end{equation}
where $x_{\rm break}=\log(r_{\rm break}-R_0)/n_0$ (with $n_0=1$), and
otherwise
\begin{equation}
f[\xvec^{(1)}] = n_0 \xvec^{(1)} + c_2 (\xvec^{(1)} - x_{\rm break})^{n_2} ,
\end{equation}
where $c_2=1$, and $n_2=10$.  The $\xvec^{(1)}$ grid ranges from
$\xvec^{(1)}_s = (\log(R_{\rm in}-R_0))/n_0$ to $\xvec^{(1)}_f$, which
is $\xvec^{(1)}_f=(\log(R_{\rm out}-R_0))/n_0$ if $R_{\rm out}<r_{\rm
  break}$ and otherwise determined iteratively from $R_{\rm out} =
r[\xvec^{(1)}_f]$.  The value of $R_{\rm in}$ is chosen so that there
are $6$ active grid cells inside the outer horizon, while $R_{\rm in}$
is outside the inner horizon.  So the boundary cells only connect to
stencils (each $\pm 4$ cells) that are inside the horizon, which
avoids causal connection between the inner boundary and the flow
outside the horizon.  For models where no persistent jet is launched,
we set $R_{\rm out}=10^3r_g$ and $r_{\rm break}\gg R_{\rm out}$.  For
models where a jet is launched, we set $R_{\rm out}=26000r_g$ and
$r_{\rm break}=5\times 10^2r_g$.  The radius $r_{\rm break}$ is where
the grid changes from exponential to hyperexponential, which allows
the grid to focus on the dynamics at small radii while avoiding
numerical reflections off the outer grid.  Radial boundaries use
absorbing conditions.

The $\theta$-grid of $N_\theta$ cells spans from $0$ to $\pi$ with
mapping
\begin{equation}
\theta(\xvec^{(2)}) = \Theta_2 + W (\Theta_1 - \Theta_2)
\end{equation}
where $\xvec^{(2)}$ ranges from $0$ to $1$ (i.e. no polar cut-out ; but
see Appendix~\ref{sec:nummethods}).  The first grid mapping function
is given by
\begin{eqnarray}
\Theta_1 &=& T_0 S_2 + T_2 S_0 ,\\
T_2 &=& (\pi/2)(1 + \arctan{(h_2(\xvec^{(m2)}-(1/2)))}/\arctan{(h_2/2)} ) ,\nonumber \\
h_2 &=& h_3 + ((r-r_{sj3})/r_{0j3})^{n_{j1}} ,\nonumber \\
T_0 &=& \pi  \xvec^{(2)} + ((1- h_0)/2)  \sin{(2 \pi  \xvec^{(2)})} ,\nonumber \\
h_0 &=& 2-Q_j (r/r_{1j})^{-n_{j2}( (1/2) + (1/\pi)   \arctan{(r/r_{0j}-r_{sj}/r_{0j})})} ,\nonumber \\
S_2 &=& (1/2) - (1/\pi) \arctan{((r-r_s)/r_0)} ,\nonumber \\
S_0 &=& (1/2) + (1/\pi) \arctan{((r-r_s)/r_0)} \nonumber ,
\end{eqnarray}
where $r_s=40$ and $r_0=20$.  For $h_2$, we set $h_3=0.3$,
$r_{0j3}=20$, $r_{sj3}=0$, and $n_{j1}=1$ so the jet is resolved with
grid lines following $\theta_j\propto r^{-n_{j1}}$.  For $h_0$, we set
$r_{1j}=2.8$, $n_{j2}=1$, $r_{0j}=15$, $r_{sj}=40$, and $Q_j=1.3$.  We
set $\xvec^{(m2)}=\xvec^{(2)}$ unless $\xvec^{(2)}>1$ then
$\xvec^{(m2)}=2-\xvec^{(2)}$ and unless $\xvec^{(2)}<0$ then
$\xvec^{(m2)}=-\xvec^{(2)}$.  $\Theta_1$ focuses on the disk at small
radii and the jet at large radii.  The second mapping function is
\begin{equation}
\Theta_2 =  (\pi/2) (h_\theta (2 \xvec^{(2)}-1)+(1-h_\theta) (2 \xvec^{(2)}-1)^{n_\theta}+1) ,
\end{equation}
where $n_\theta=5$ and $h_\theta=0.15$.  $\Theta_2$ focuses on the
thin inflow near the horizon in poloidal field models, while it also
avoids small $\phi$ polar cells that would limit the time step.  The
interpolation factor is
\begin{equation}
W = (1/2) + (1/\pi)(\arctan{( (r-r_{sj2})/r_{0j2})} ) ,
\end{equation}
where $r_{sj2}=5$ and $r_{0j2}=2$.  The polar axis boundary condition
is transmissive as described in Appendix~\ref{sec:nummethods}.

The $\phi$-grid of $N_\phi$ cells spans from $0$ to $2\pi$ with
mapping $\phi(\xvec^{(3)}) = 2\pi \xvec^{(3)}$.  Many of our simulations have
$\xvec^{(3)}$ vary from $0$ to $1$ such that $\Delta\phi = 2\pi$.  This is
a fully 3D (no assumed symmetries) domain.  Periodic boundary
conditions are used in the $\phi$-direction.  Our TNM11 type models
use various $\Delta\phi$, and the spatial integrals are renormalized
to account for the full $2\pi$ range in $\phi$.

Table~\ref{tbl2} marks the grid as ``Exp'' if exponential, ``HypExp''
if hyperexponential, ``ExpOld'' for MB09's exponential grid, and
``TNM11'' for TNM11's hyperexponential grid.

We choose a resolution $N_r\times N_\theta\times N_\phi$ that has a
grid aspect ratio of 1:1:1 for most of the inner-radial domain.  This
allows the $\phi$ dimension to be treated equally to the $r-\theta$
dimensions.  The aspect ratio (as volume-averaged within
$|\theta-\pi/2|\le [\theta^d]_t$) is given as $A_r$ in
Table~\ref{tbl2}, where $r_{\rm H}$ is the horizon (focusing on the
geometrically thinning disk), $r_i=20r_g$ (but applies for
$50r_g\gtrsim r\gtrsim 5r_g$ such that $A$ is uniform for much of the
flow), and $r_o=100r_g$ (showing the aspect ratio changes at large
radii due to focusing on the jet).  The MB09D/Q models use $r_i=10r_g$
and $r_o=20r_g$.

The MRI is resolved for grid cells per wavelength
(Eq.~(\ref{lambdamri})),
\begin{equation}\label{q1mri}
\Qx \equiv \frac{\lambda_{x,\rm MRI}}{\Delta_{x}} ,
\end{equation}
of $\Qx\ge 6$, for $x=\theta,\phi$, where $\Delta_{r} \approx
d\xvec^{(1)} (dr/d\xvec^{(1)})$, $\Delta_{\theta} \approx r
d\xvec^{(2)} (d\theta/d\xvec^{(2)})$, and $\Delta_{\phi} \approx r
\sin\theta d\xvec^{(3)} (d\phi/d\xvec^{(3)})$.  Volume-averaging is
done as with $\Qtwo$, except $v_{x,\rm A}/\Delta_{x}$ and
$|\Omega_{\rm rot}|$ are separately $\theta,\phi$-volume-averaged
before forming $\Qx$.  Averaged $\Qone,\Qthree$ are typically $30\%$
larger than $\Qoneweak,\Qthreeweak$ for toroidal field and MB09
models.  The $t=0$ values (see Table~\ref{tbl1}) and time-averaged
values (see Table~\ref{tbl4}) are measured at same radii as $\Qtwo$.

Turbulence is resolved for grid cells per correlation length
(Eq.~(\ref{mcor})),
\begin{equation}\label{Qmcor}
Q_{p,\rm{}cor} \equiv \frac{\lambda_{x,\rm cor}}{\Delta_{x}} ,
\end{equation}
of $Q_{p,\rm{}cor}\ge 6$, for $x=r,\theta,\phi$ and $p=n,l,m$,
respectively.  Otherwise, modes are numerically damped on a dynamical
timescale (even $Q=5$ would not indicate the mode is marginally
resolved, because numerical noise can keep $Q\approx 5$ at increasing
resolution until finally the mode is actually resolved -- finally
leading to an increasing $Q\ge 6$ with increasing resolution ; as seen
by \citealt{sdgn11}).  Reported $Q_{p,\rm{}cor}$ take $1/\Delta_{x}$
as the number of grid cells covering the span of $\lambda_{x,\rm cor}$
as centered on: middle of $x^{(1)}$ within the used radial span for $x=r$, $\theta=\pi/2$ for
$x=\theta$ for the ``Disk'' and $\theta=0$ for $x=\theta$ for the
``Jet'', and anywhere for $x=\phi$.  For $\Delta\phi<2\pi$,
$\Qthree,Q_{m,\rm{}cor}\ll N_\phi$ is required to avoid truncating the
mode (as happens in model A0.9N25).

\section{Fiducial Thick Disk Model}
\label{sec:fiducial}

Our fiducial model, A0.94BfN40, consists of an initially weakly
magnetized thick accretion disk around a rotating ($a/M=0.9375$) BH.
The initial magnetic field consists of field loops that alternate
polarity with radius (the ``flipping'' field geometry).  Model
parameters given in Table~\ref{tbl1} and Table~\ref{tbl2}.

\subsection{Initial and Evolved Disk Structure}

\begin{figure}
\centering
\includegraphics[width=3.2in,clip]{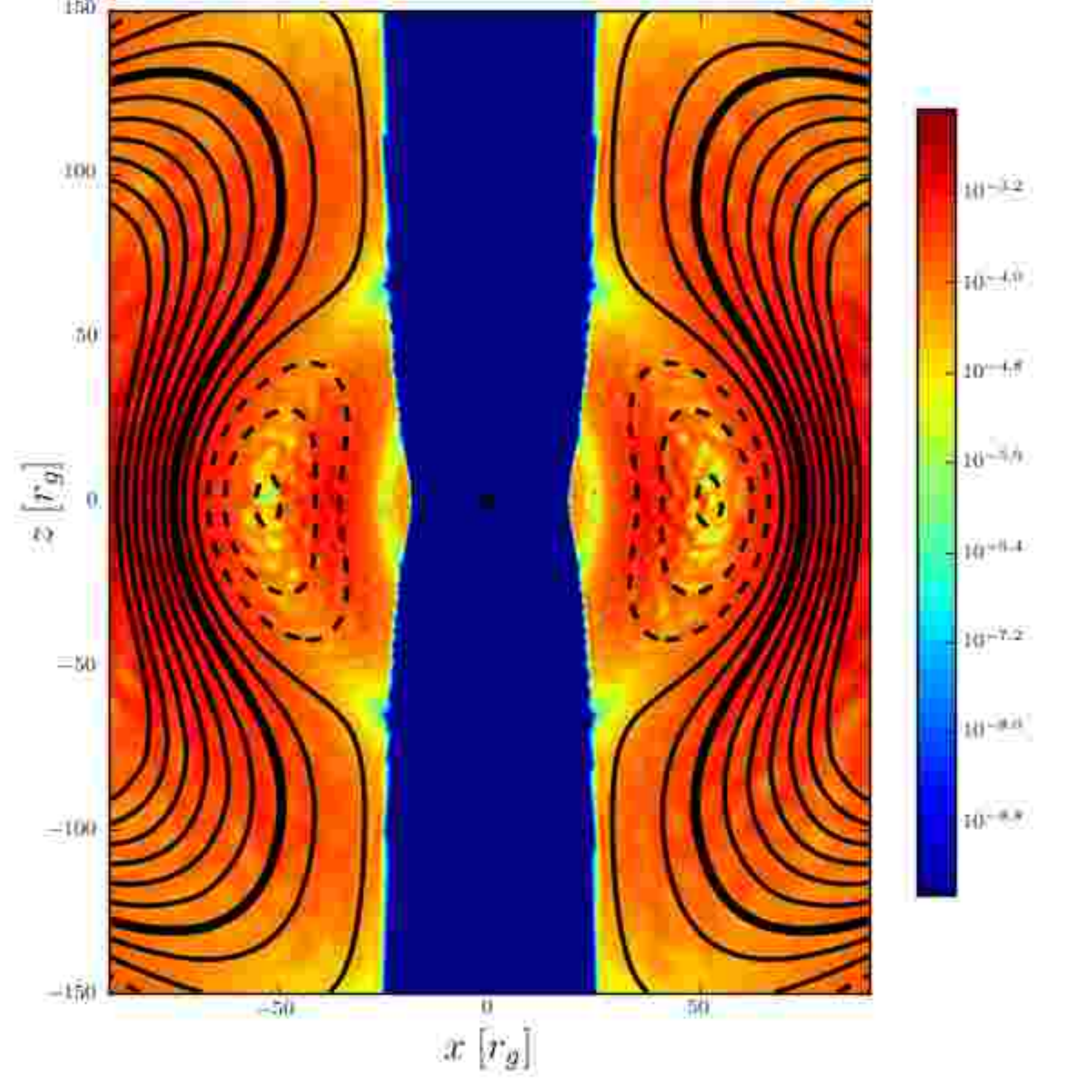}
\caption{The fiducial model's initial ($t=0$) state consists of a
  weakly magnetized geometrically thick torus around a spinning
  ($a/M=0.9375$) BH. $b^2$ is shown as color with legend.  Black lines
  show $\Avpotvec_\phi$ (integrated over all $\phi$).  The
  inner-most-radial field loop is small and shows up as a high
  magnetic energy density with no field lines shown, the next field
  lines (black dashed lines) show the 2nd field loop, and the next
  field lines (black solid lines) show the 3rd field loop.  The
  thickest black solid line is the same field line in
  Figure~\ref{middle3plot} that tracks the outer-most-angular part of
  the BH-driven jet.}
\label{initial3plot}
\end{figure}

\begin{figure}
\centering

\includegraphics[width=3.2in,clip]{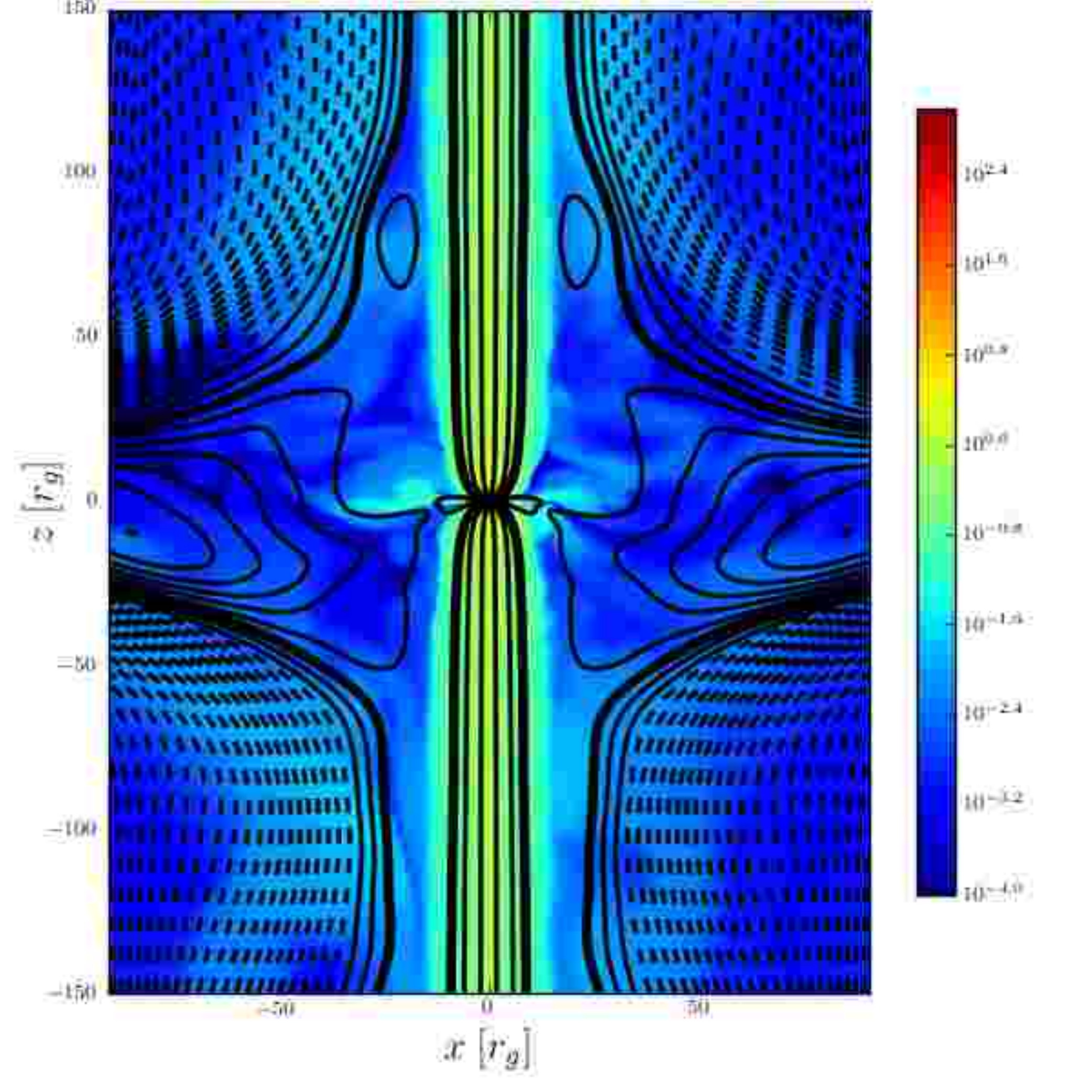}
\caption{The evolved ($t\approx 10412r_g/c$) state of the fiducial
  model (otherwise similar to Figure~\ref{initial3plot}) consists of
  strongly magnetized gas near the BH that launches a jet that
  is visible as the vertical beam of large electromagnetic energy
  density.  The jet is nearly cylindrical at small radii due to the
  pressure support from the thick accretion flow.  All field lines in
  the 1st and 2nd field loops (shown in Figure~\ref{initial3plot})
  have been accreted or ejected in an outflow.  The 3rd (solid black
  lines) and 4th (dashed black lines) evolved field loops are shown,
  where the thickest solid black line traces the field line that
  connects to the outer angular part of the jet from the BH.}
\label{middle3plot}
\end{figure}

Figure~\ref{initial3plot} and Figure~\ref{middle3plot} show color
plots of $b^2$ and field line contours (contours of $\Avpotvec_\phi$
integrated over $\phi$, so is axially symmetric) for the initial and
quasi-steady-state evolved solution, respectively.  The initial
solution consists of a radially extended thick torus within which
several weak field loops (of alternating poloidal polarity) are
embedded.  The disk is geometrically thick with $\theta^d\sim 1$ and
also quite thermally thick with $c_s/v_{\rm rot}\gtrsim 1$ and
$c_s/v_{\rm K}\gtrsim 1$ through-out the solution both initially and
at late times.

The evolved solution, shown in Figure~\ref{middle3plot}, shows that
the first two field loops have been completely accreted or ejected.
During accretion of the 3rd field loop, the magnetic flux reaches a
maximum saturated value over a long period of inflow equilibrium.

\begin{figure}
\centering

\includegraphics[width=3.0in,clip]{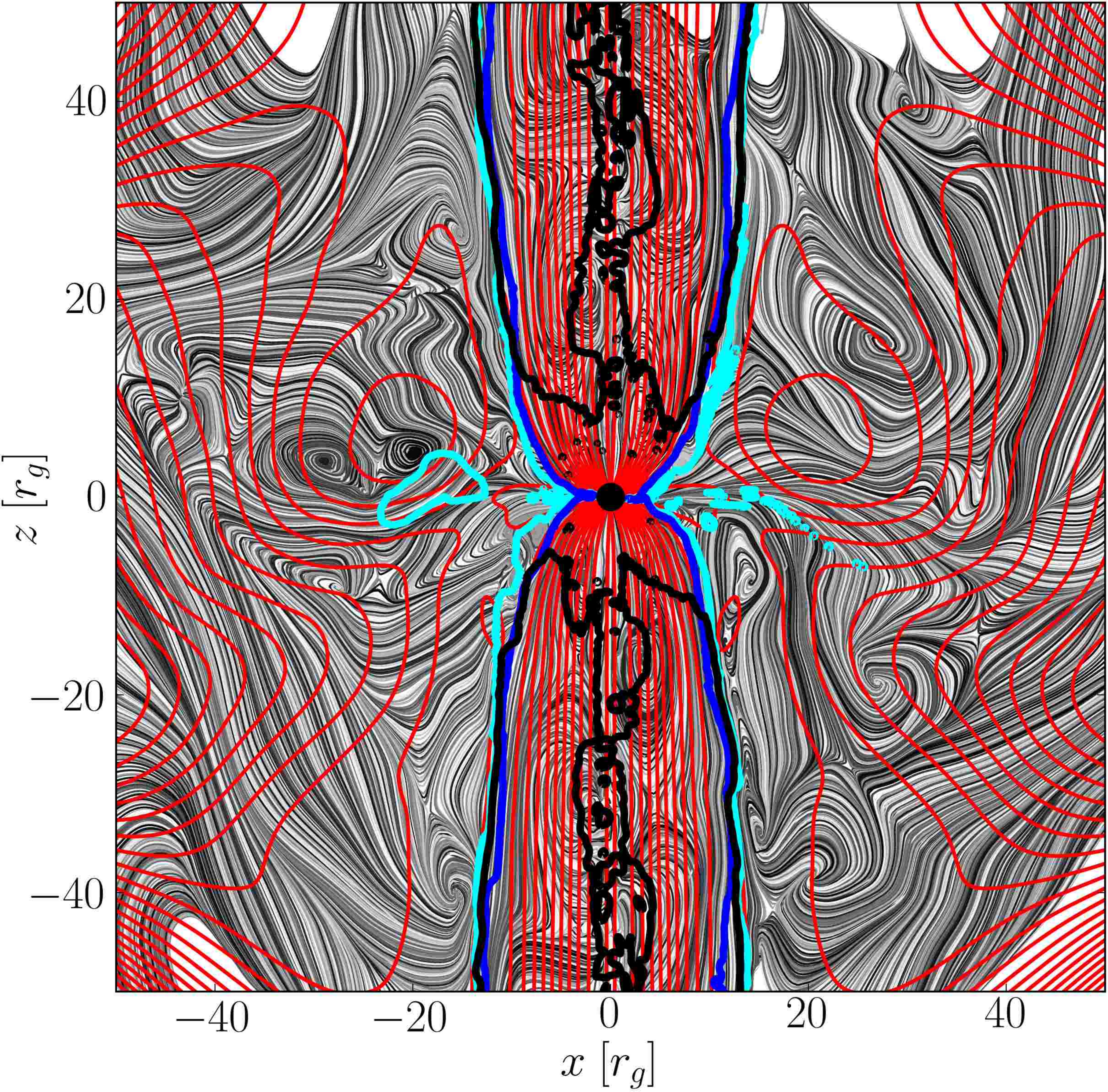}
\caption{The evolved ($t\approx 10412r_g/c$) state of the fiducial
  model showing velocity flow lines (traces of $v_i$ shown as
  grayscale) and field lines ($\Avpotvec_\phi$ integrated over all
  $\phi$, shown as red lines).  Also shown are contour lines, where
  $-\uvec_t=1$ (most particles are unbound in the jet and in the wind
  at larger radii) is a thick black line, $\beta=2$ (beyond where
  corona and jet exist) is a thick cyan line, and $b^2/\rho_0=1$
  (beyond where the relativistic jet exists) is a thick blue line.
  The accretion flow is turbulent with circulating eddies.}
\label{figmiddleflowfield}
\end{figure}

Figure~\ref{figmiddleflowfield} shows an instantaneous meridional
snapshot of the flow-field, which shows significant circulation.
The field lines threading the BH are a dipolar split-monopolar
magnetosphere.

\subsection{Overall Time Dependence}\label{sec:timedep}

\begin{figure*}
\centering
\includegraphics[width=6.6in,clip]{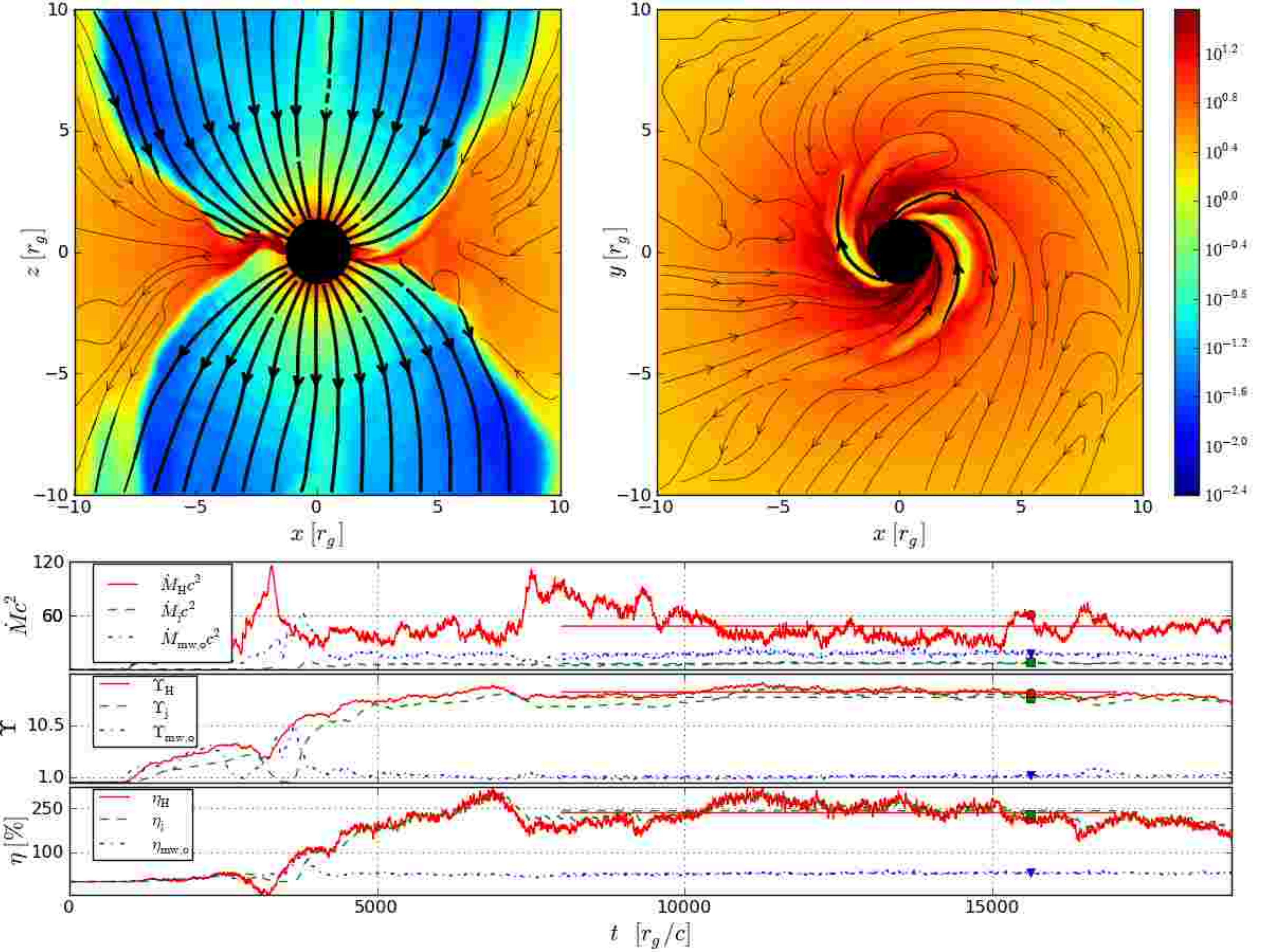}
\caption{Evolved snapshot (see Supporting Information for the movie)
  of the fiducial model at $t\approx 15612r_g/c$ showing log of
  rest-mass density in color (see legend on right) in both the $z-x$
  plane at $y=0$ (top-left panel) and $y-x$ plane at $z=0$ (top-right
  panel).  Black lines trace field lines, where thicker black lines
  show where field is lightly mass-loaded.  The bottom panel has 3
  subpanels.  The top subpanel shows $\dot{M}$ through the BH
  ($\dot{M}_{\rm H}$), out in the jet ($\dot{M}_{\rm j}$, at
  $r=50r_g$), and out in the magnetized wind ($\dot{M}_{\rm mw,o}$, at
  $r=50r_g$) with legend.  The middle subpanel shows $\Upsilon$ for
  similar conditions.  The bottom subpanel shows the efficiency
  ($\eta$) for similar conditions.  Horizontal lines of the same
  colors show the averages over the averaging period, while
  square/triangle/circle tickers are placed at the given time and
  values.  In summary, the efficiency is high at $\eta\sim 200\%$.
  Also, despite plenty (up to $10\times$ around $t\sim
  8500r_g/c$) of same-signed polarity magnetic flux surrounding the
  BH, the magnetic flux reaches a stable saturated value of $\Upsilon_{\rm
    H}\approx 17$ as managed by magnetic
  Rayleigh-Taylor modes.  This suggests that the simulation has
  reached a force balance between the magnetic flux in the disk and
  the hot heavy inflow.}
\label{evolvedmovie}
\end{figure*}

Figure~\ref{evolvedmovie} shows a typical snapshot for the rest-mass
density, field lines, and fluxes ($\dot{M}$, $\Upsilon$, and $\eta$)
on the BH, through $r=50r_g$ in the jet, and at $r=50r_g$ in
the magnetized wind.

The flow consists of a magnetized polar jet and a turbulent equatorial
disk inflow.  Near the BH, the rest-mass density in the inflow rises
substantially due to vertical compression by the accumulated polar
magnetic flux that surrounds the BH.  The large-scale polar magnetic flux forms
a semi-permeable magnetic barrier to the massive inflow, which is
forced to undergo non-axisymmetric accretion through magnetic
Rayleigh-Taylor instabilities.

The BH's magnetic flux dominates the mass influx with $\Upsilon_{\rm
  H}\approx 17$ during the quasi-steady-state period.  Any additional
magnetic flux that temporarily accretes onto the hole is ejected in
magnetic Rayleigh-Taylor modes that push the flux back into the disk.
This suggests that the magnetic flux near the BH has reached a maximum
saturation point via some force balance condition.  Because
$\Upsilon\gg 1$, one expects the BZ effect to be activated, and indeed
the energy extraction efficiency is high at $\eta\sim 200\%$.  Most of
the energy extracted from the BH reaches the jet at large radii
(i.e. $\eta_{\rm H}\sim \eta_{\rm j}$).  The shown temporal behavior
tracks the fact that $\eta\propto \Upsilon^2$.  At the latest times
the efficiency drops as the magnetic flux (from the 3rd field loop)
begins to be destroyed by an incoming polarity field reversal (outer
part of 3rd field loop and inner part of 4th field loop).  The obvious
complete polarity reversal (destruction of inner part of 2nd field
loop by outer 2nd and inner 3rd) occurs at $t\sim 2700r_g/c$ when
$\dot{M}$ doubles.  At $t\sim 7000r_g/c$, magnetic flux is ejected in
a large magnetic Rayleigh-Taylor disruption allowing $\dot{M}$ to
double.

\begin{figure}
\centering
\includegraphics[width=3.1in,clip]{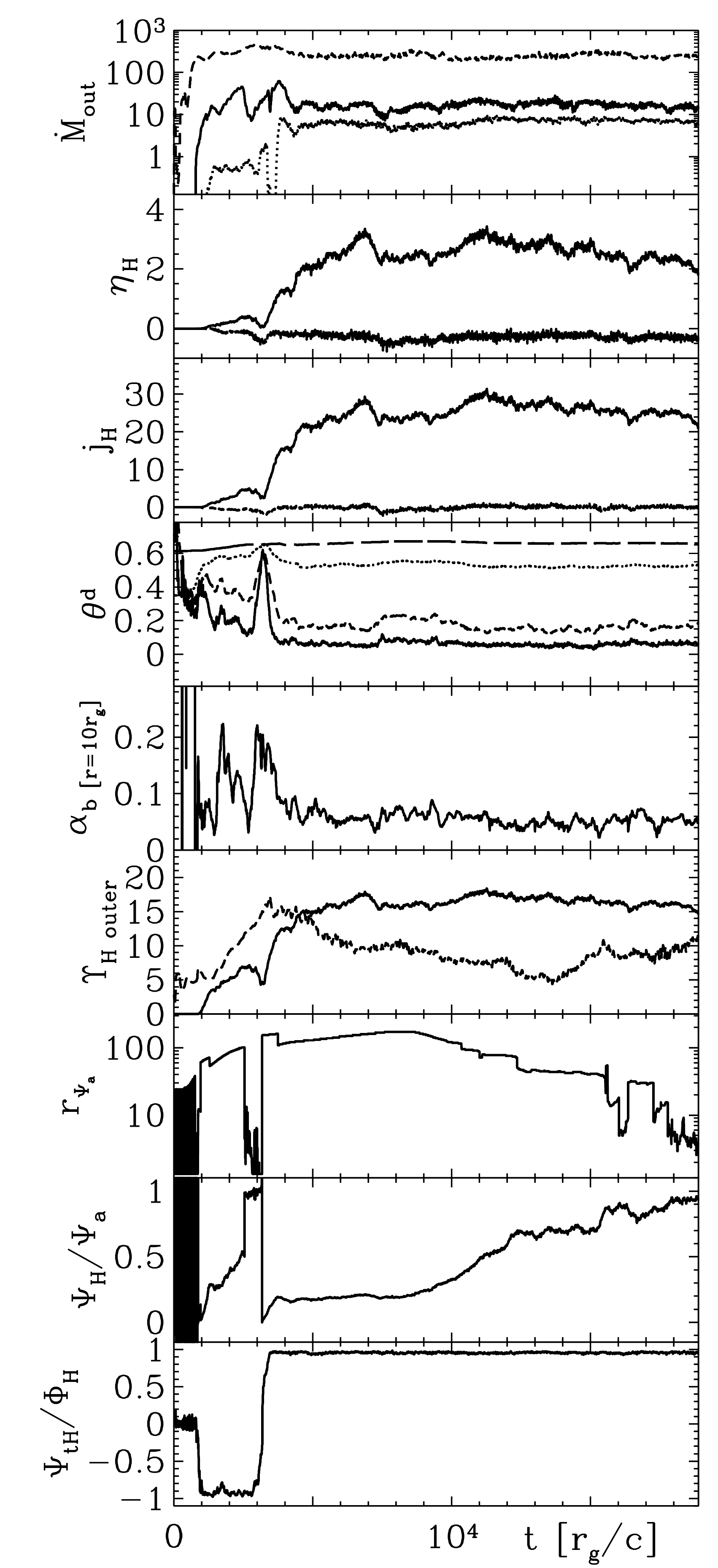}
\caption{Quantities vs. time.  Top Panel: $\dot{M}_{\rm out}$ for
  magnetized wind ($\dot{M}_{\rm mw,o}$, solid line, and mostly middle
  line), entire wind ($\dot{M}_{\rm w,o}$, short-dashed line, and
  upper-most line), and jet ($\dot{M}_{\rm j}$, dotted line, and
  lowest line).  Next Panel: Efficiency ($\eta_{\rm H}$) for EM (solid
  line, upper line) and MAKE (short-dashed line, lower line)
  terms. Next Panel: Specific angular momentum ($\jmath_{\rm H}$) for
  EM (solid line, upper line) and MAKE (short-dashed line, lower line)
  terms. Next Panel: $\theta^d$ at $r=\{r_{\rm H}/r_g,5,20,100\}r_g$
  with, respectively, lines: \{solid, short-dashed, dotted, long
  dashed\} corresponding to the lowest to upper-most lines. Next
  panel: $\alpha_b$ at $r=10r_g$.  Next Panel: $\Upsilon$ on the
  horizon (solid line, mostly upper line) and in the disk at $r=50r_g$
  for only the ingoing flow (short-dashed line, mostly lower line).
  Next Panel: $r_{\Psi_a}$ for the radius out to where there is the
  same magnetic polarity as on the hole (solid line). Next panel:
  Magnetic flux on the BH per unit flux available in the flow with the
  same polarity: $\Psi_{\rm H}(t)/\Psi_a(t)$.  Bottom panel:
  $\Psi_{\rm tH}(t)/\Phi_{\rm H}(t)\sim 1/l$, for $l$ mode of vector
  spherical harmonic multipole expansion of $\Avpotvec_\phi$.
  In summary, the flow has reached a quasi-steady-state at late
  times. The magnetic flux on the horizon has saturated to a large
  value leading to a high efficiency for energy and angular momentum
  extraction from the BH.}
\label{fluxvst}
\end{figure}

Figure~\ref{fluxvst} shows various quantities vs. time.  All
quantities are in a quasi-steady-state for $t\gtrsim 8000r_g/c$ far
after the last field polarity inversion at $t\sim 2700r_g/c$ and after
magnetic flux has accumulated near the BH.  The mass ejected in the
circulating wind ($\dot{M}_{\rm w,o}$, seen as eddies in
Fig.~\ref{figmiddleflowfield}) dominates the magnetized wind
($\dot{M}_{\rm mw,o}$) and jet ($\dot{M}_{\rm j}$) at large radii
($r_{\rm o}=50r_g$ here); see \S\ref{integrations} for definitions of
various outflow components.  The EM term dominates the MAKE term in
$\eta_{\rm H}$ and ${\jmath}_{\rm H}$, and the flow has a high
efficiency of $\eta_{\rm H}\sim 200\%$.  The MAKE term is composed of
a particle term (i.e. $\eta^{\rm PAKE}=1+\uvec_t$) and an enthalpy
term (i.e. $\eta^{\rm EN}=\uvec_t(u_g+p_g)/\rho_0$).  We find that
$\eta^{\rm MAKE}\sim -30\%$ as composed of $\eta^{\rm PAKE}\sim 63\%$
and $\eta^{\rm EN}\sim -93\%$ ($\eta^{\rm MAKE}\approx -26\%$ in the
initial torus, so $\eta^{\rm EN}\sim -67\%$ can be chosen for
marginally bound inflow).  So the inflow is bound as particles but
thermo-kinetically unbound due to high enthalpy.

Accumulation of magnetic flux near the BH leads to geometric
compression of the inflow as it approaches the horizon.  The geometric
thickness drops before the field inversion at $t\sim 2700r_g/c$ as
magnetic flux accumulates and compresses the disk inflow.  However,
during the field inversion, the geometric thickness restores to the
prior geometric thickness ($\theta^d\simeq 0.7$) at all radii, which
indicates that the field (lost during the field annihilation) is
responsible for the thinning of the dense part of the disk.  After the
field polarity inversion, the magnetic flux re-accumulates near the
BH, which leads again to the vertical compression of the disk flow.
The $\alpha$-viscosity parameter holds steady at about $\alpha_b\sim
0.05$.  $\Upsilon$ in the pure inflow ($u_r<0$ only) available at
large radii (here $r=50r_g$, giving $\Upsilon_{\rm outer}$ in the
plot) is large (the BH and ``outer'' values are similar for this
chosen ``outer'' radius).

The value of $r_{\Psi_a}$ shows the radius out to which the magnetic
polarity is the same as on the horizon.  As expected, $r_{\Psi_a}$
drops to the horizon during the field inversion (destruction of inner
part of 2nd field loop) at $t\sim 2700r_g/c$.  It also gradually drops
as the next polarity inversion (outer part of 3rd field loop) eats
away at the magnetic flux outside the BH.  The process of field
inversion is also evident by looking at $\Psi_{\rm H}(t)/\Psi_a(t)$
(i.e. ratio of time-dependent fluxes) corresponding to [the flux on
the hole] per unit [flux on the hole plus available of the same
polarity just beyond the hole].  $\Psi_{\rm H}(t)/\Psi_a(t)\sim 1$ is
reached during the field polarity inversion, and at late times
$\Psi_{\rm H}(t)/\Psi_a(t)\sim 1$ is approached.  However, while
$\Upsilon$ holds steady, the value of $|\Psi_{\rm H}(t)/\Psi_a(t)|\ll
1$, which indicates that much more same-polarity flux is available.
This shows that the saturated value of $\Upsilon$ (and so $\eta$) is
controlled by some force balance condition and not simply limited by
initial conditions.  Finally, $|\Psi_{\rm tH}(t)/\Phi_{\rm H}(t)|\sim
1$ shows that the horizon's field is dipolar ($l\approx 1$).

\subsection{Time-Averaged Poloidal ($r-\theta$) Dependence}\label{sec:poldep}

\begin{figure}
\centering
\includegraphics[width=3.0in,clip]{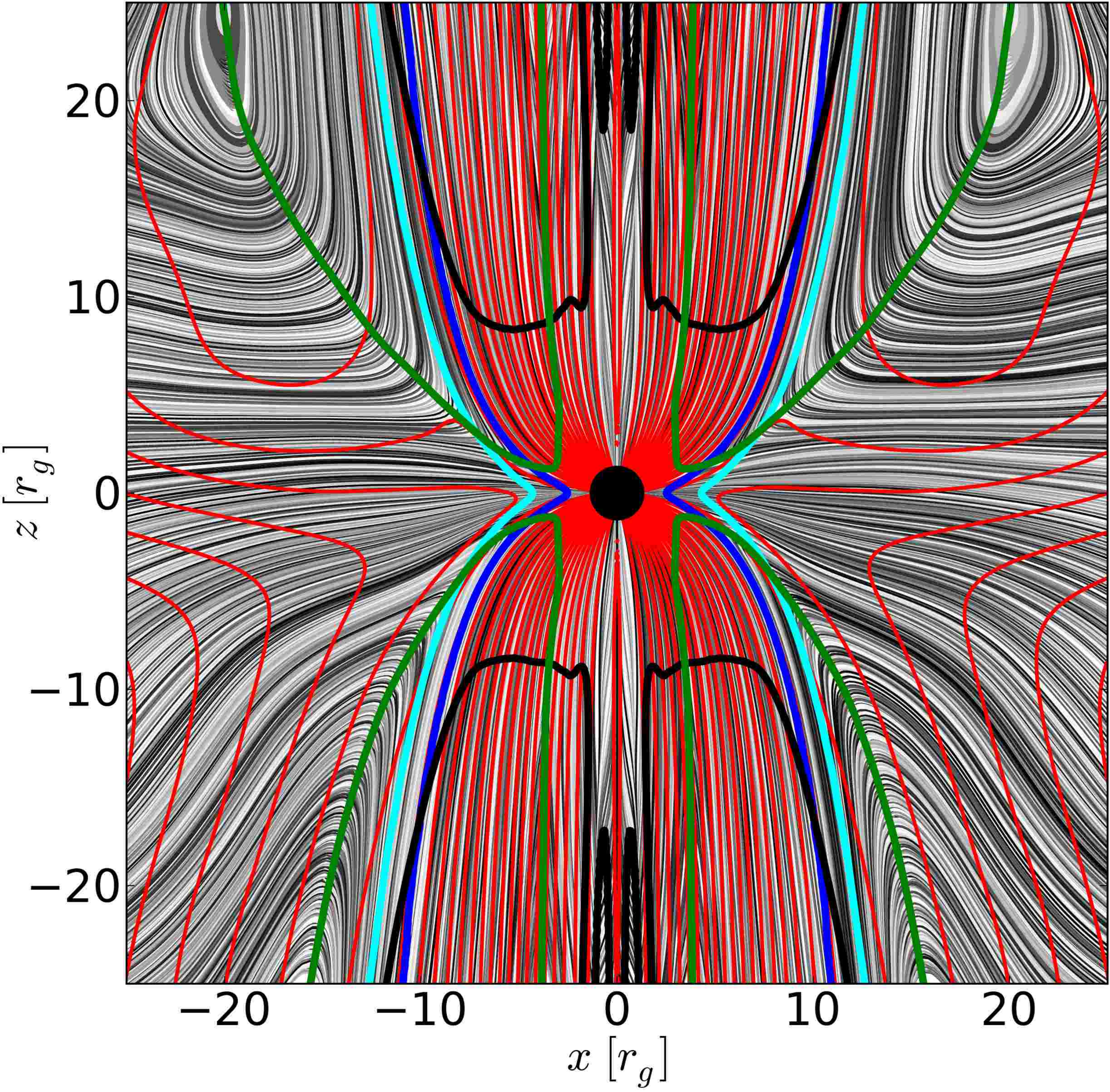}
\caption{Flow-field as in Figure~\ref{figmiddleflowfield}, except
  time-averaged and zoomed-in on the BH region that is in inflow
  equilibrium ($3$ inflow times).  The colored (green, black, cyan,
  and blue) thick lines correspond to time-averages of quantities
  $Q=\{\uvec^r,\uvec_t,\beta,b^2/\rho_0\}$, respectively, at values
  $[Q]_t=\{0,-1,2,1\}$.  While near the BH the flow has
  $[b^2/\rho_0]_t\gtrsim 1$ as averaged directly, the dense inflow has
  $b^2/\rho_0\lesssim 1$ at all radii. The inflow occurs in
  geometrically thin streams not accounted for when computing the
  unweighted average.  In summary, the BH is threaded by ordered
  magnetic flux, and the flow exhibits equatorial asymmetry over many
  inflow times.}
\label{figavgflowfield}
\end{figure}

Figure~\ref{figavgflowfield} shows the time-averaged flow-field and
contours for other conditions.  The figure is comparable to the
snapshot shown in Figure~\ref{figmiddleflowfield}. The jet region
contains significant magnetic flux and same-signed polarity field
exists near the BH ready to be accreted.  In the quasi-stationary
state, the BH's magnetic flux oscillates around its saturated
magnitude, whose time-averaged value is determined by some force
balance condition as managed by non-axisymmetric Rayleigh-Taylor
instabilities.

\subsection{Time-Averaged Radial ($r$) Dependence}\label{sec:velvisc}

\begin{figure}
\centering
\includegraphics[width=3.2in,clip]{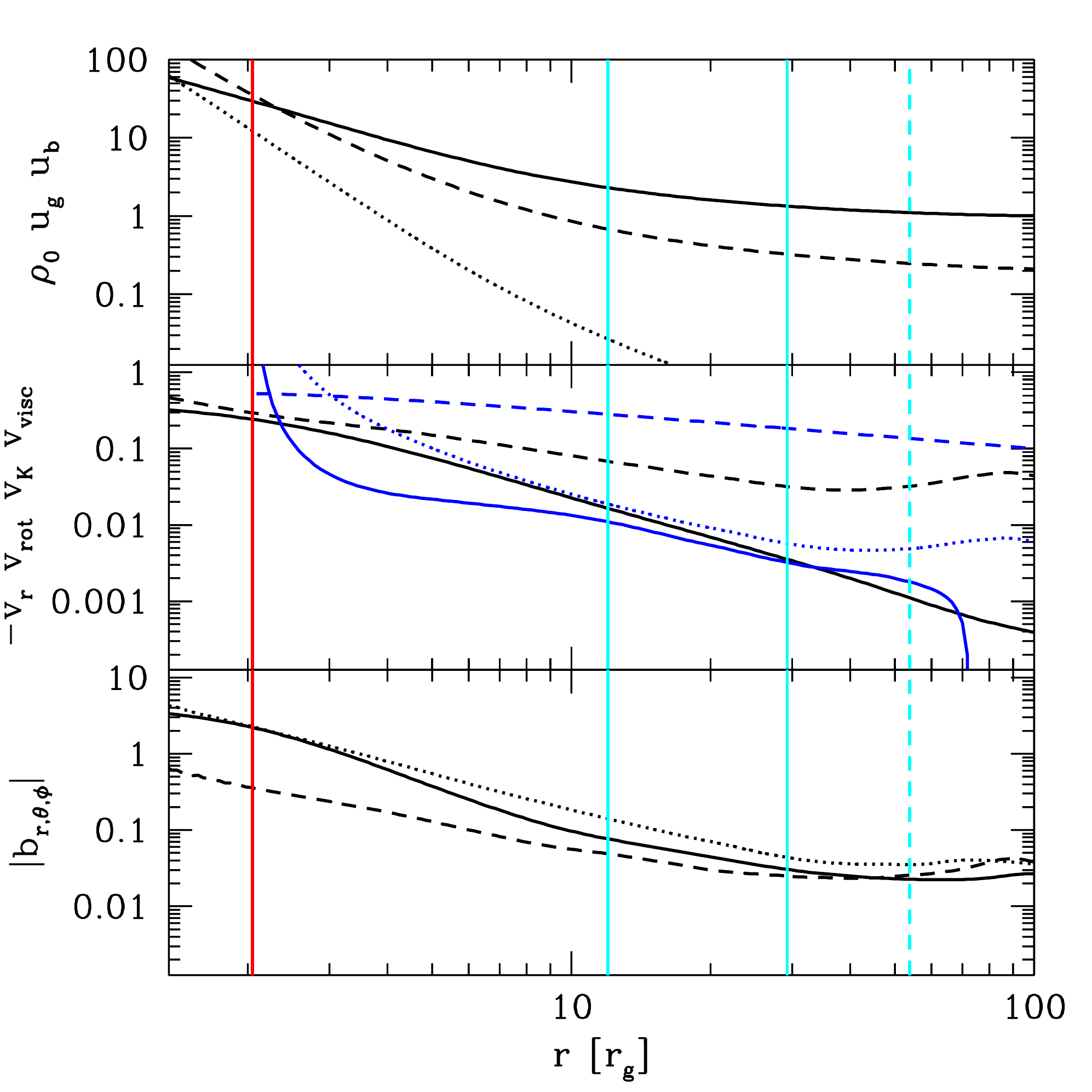}
\caption{The time-angle-averaged densities, 3-velocities, and 4-field
  strengths using a density-weighted average.  Top panel shows
  rest-mass density ($\rho_0$) as black solid line, internal energy
  density ($u_g$) as black short-dashed line, and magnetic energy
  density ($u_b$) as black dotted line.  Middle panel shows negative
  radial velocity ($-v_r$) as black solid line, rotational velocity
  ($v_{\rm rot}$) as black short-dashed line, Keplerian rotational
  velocity ($v_{\rm K}$) as blue short-dashed line, and
  $\alpha$-viscosity theory radial velocity ($v_{\rm visc}$) when
  using $p_{\rm b}$ in denominator for $\alpha=\alpha_b$ (blue solid
  line) and when choosing a fixed $\alpha(\theta^d)^2=0.1$ (blue
  dotted line).  Bottom panel shows comoving 4-field spatial
  components with $r$, $\theta$, and $\phi$ components shows as solid,
  short-dashed, and dotted black lines, respectively.  The vertical
  red line marks the ISCO.  Vertical solid cyan lines show range from
  $r=12r_g$ to $3$ inflow times.  The short-dashed vertical cyan line
  marks a single inflow time.  In summary, dense filaments maintain
  $u_b/\rho_0\lesssim 1$ and $\beta\gg 1$ near the horizon, while a
  similar plot (see bottom-left panel in Figure~\ref{gammie4panel})
  averaged over the entire disk+corona+winds shows $u_b/\rho_0\gg 1$
  and $\beta\ll 1$ near the horizon.  Also, the rotational velocity is
  quite sub-Keplerian.}
\label{rhovelvsr}
\end{figure}

Figure~\ref{rhovelvsr} shows the time-averaged densities,
3-velocities, and comoving 4-fields vs. radius using a
density-weighted average to focus on heavy disk material.  The
solution is in inflow equilibrium ($3$ inflow times; see
section~\ref{sec_infloweq}) only out to $r\sim 30r_g$ and has reached
a single inflow time within $r\sim 55$--$100r_g$ depending upon how
one defines it.  Beyond the BH, the rest-mass and internal energy
densities are quite flat.  The rotational velocity is quite
sub-Keplerian, which is primarily a consequence of thermal pressure
playing an equal role to total gravity, unlike in thinner disks that
are more naturally Keplerian.  This effect is also seen in prior MHD
simulations \citep{ina03,pmw03}.

The GR viscosity estimate for $v_r$ denoted $v_{\rm visc}$ (see above
Eq.~(\ref{alphaeq})) underestimates the simulation $v_r$ when using
the $\alpha$-viscosity with total pressure.  A more accurate match is
found when using magnetic pressure.  Also, choosing $\theta^d\to
\theta^t$ or $\theta^d\to |c_s/v_{\rm rot}|$ still leads to a poor fit
to $v_r$.  Only if we set $\alpha(\theta^d)^2\to 0.1$ at all radii
does $|v_{\rm visc}|\approx |v_r|$ outside the ISCO and inside the
inflow equilibrium region.  The simulation $v_r$ gives $\epsilon\sim
0.1$ for Eq.~(\ref{rmsimple}).

\begin{figure}
\centering
\includegraphics[width=3.2in,clip]{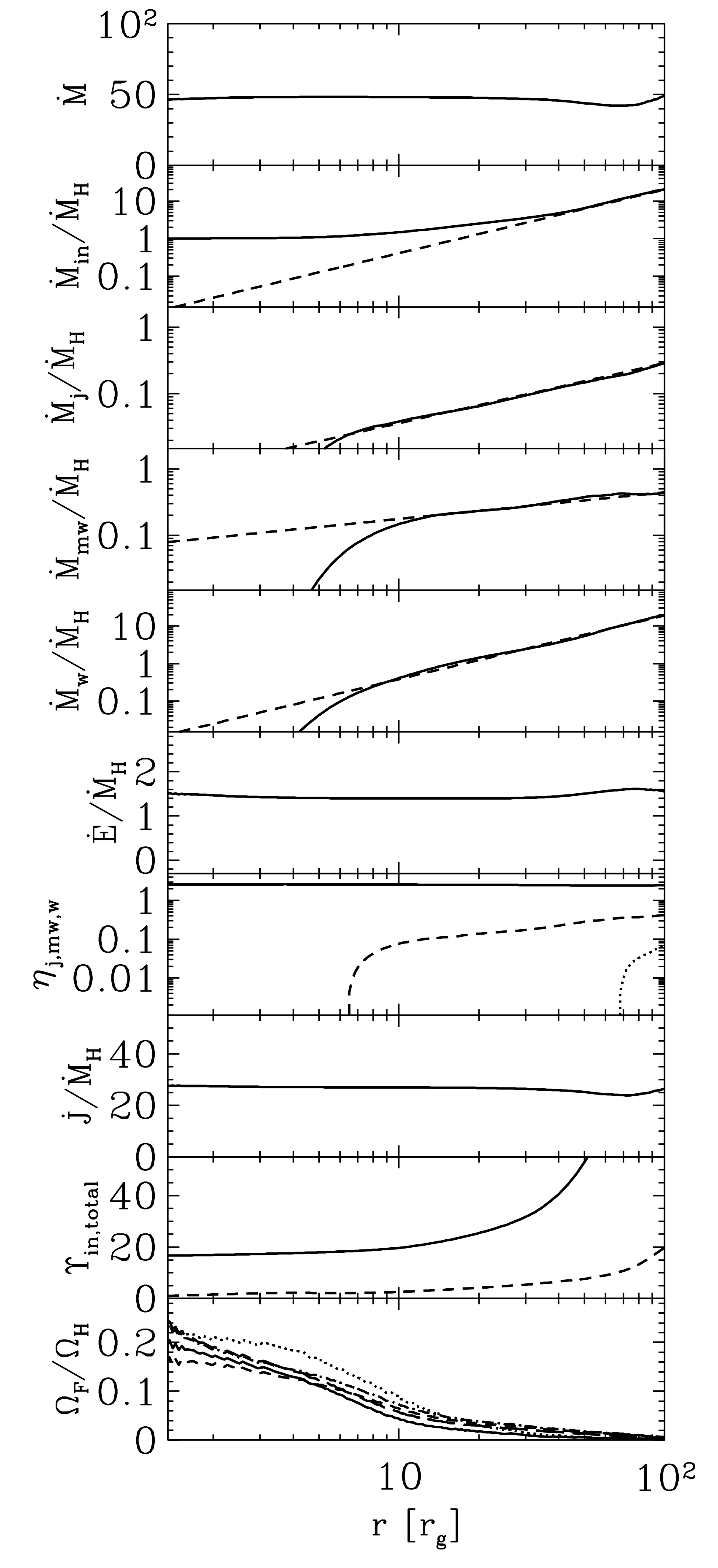}
\caption{The time-averaged angle-integrated fluxes.  From top to
  bottom, panels are: Total mass accretion rate ($\dot{M}$), inflow
  rate ($\dot{M}_{\rm in}$), jet outflow rate ($\dot{M}_{\rm j}$),
  magnetized wind outflow rate ($\dot{M}_{\rm mw}$), entire wind
  outflow rate ($\dot{M}_{\rm w}$), total specific energy accretion
  rate ($\dot{E}/\dot{M}_{\rm H}$), efficiency for the jet (solid
  line) magnetized wind (short-dashed line) and wind (dotted line),
  total specific angular momentum accretion rate
  ($\jmath=\dot{J}/\dot{M}_{\rm H}$), specific magnetic flux
  $\Upsilon$ for the total flow (solid line) and pure inflow with
  $u_r<0$ (short-dashed line), and field line angular rotation
  frequency per unit BH angular frequency ($\Omega_{\rm F}/\Omega_{\rm
    H}$) for time-averaged versions of $\Omega^d_{\rm F}$ (solid
  line), $\Omega^c_{\rm F}$ (short-dashed line), $\Omega^e_{\rm F}$
  (dotted line), $|\Omega^d_{\rm F}|$ (long-dashed line), and
  $|\Omega^c_{\rm F}|$ (dot-short-dashed line).  These $\Omega_{\rm
    F}$ are averaged within the disk+corona part of the flow.
  Power-law fits for mass inflow and outflow rates are shown as
  short-dashed lines.  In summary, inflow equilibrium is achieved over
  a couple decades in radius, and the mass outflows follow a power-law
  behavior.}
\label{fluxvsr}
\end{figure}

Figure~\ref{fluxvsr} shows the fluxes (see section~\ref{fluxes})
vs. radius as well as the field line angular rotation frequency
$\Omega_{\rm F}$ (using various definitions defined in
section~\ref{magneticfluxdiag}).  These quantities are associated with
conserved quantities such that ratios of total fluxes would be
constant along flow-field lines in stationary ideal MHD.  The total
fluxes are constant out to large radii, indicating a single inflow
time is achieved over about $2$ decades in radius.  The true total
fluxes are actually even flatter near the BH if one more accurately
accounts for numerical floor injection as done in \citet{tnm11}, but
this is a small error.  Also shown are the components (inflow, jet,
magnetized wind, and entire wind) of the mass and energy flow.  The
mass inflow and outflow at large radii follow power-laws after
sufficient averaging over turbulent eddies.  The jet efficiency is
order $200\%$ and is constant at large radii.

The winds increase in efficiency with radius, but the jet dominates
the efficiencies of the winds.  Power-law fits over the outer-radial
domain (including the region not actually in inflow equilibrium)
for the mass flow rates are $\dot{M}\propto r^{1.7}$ for the
inflow and entire wind, $\dot{M}\propto r^{0.9}$ for the jet, and
$\dot{M}\propto r^{0.4}$ for the magnetized wind.  At $r\sim 30r_g$,
the entire wind component has an average velocity of order $v_w/c\sim
0.01$ or $v_w\sim 3000$km/s.

The specific magnetic flux $\Upsilon$ measures the total absolute
radial flux. The figure shows that the total radial flux is much
larger than the inflow-only component because the jet harbors most of
the magnetic flux.  The field line angular frequency $\Omega_{\rm
  F}\sim \Omega_{\rm H}/4$ (as in BZ77's paraboloidal model) in the
disk+corona+wind (i.e. ``fdc'' averaging, for full flow except the
highly-magnetized jet).

\begin{figure}
\centering
\includegraphics[width=3.2in,clip]{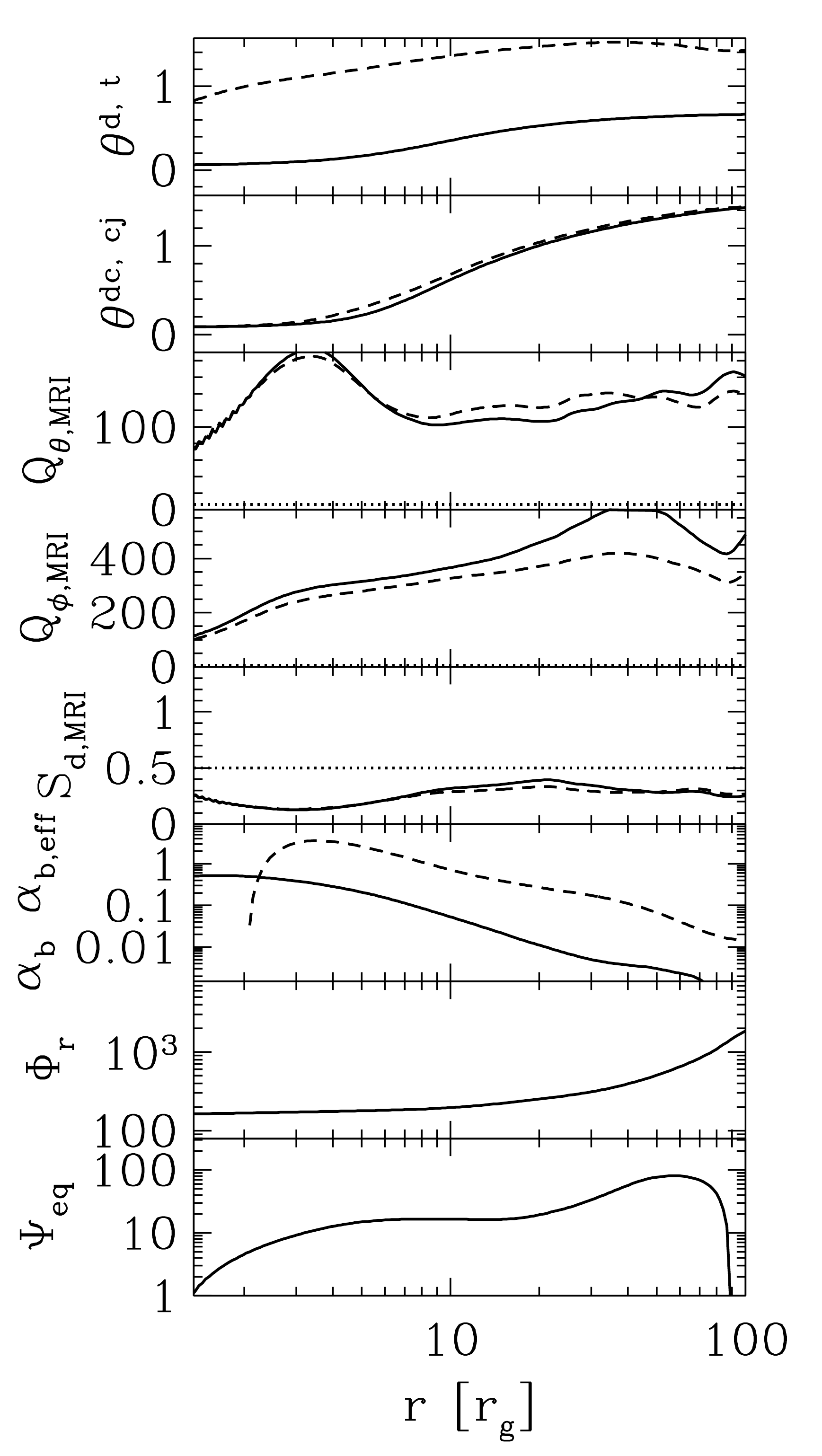}
\caption{Other time-angle-averaged quantities.  From top to bottom,
  panels are: density half-angular thickness ($\theta^d$, solid line)
  and thermal half-angular thickness ($\theta^t$, short-dashed line),
  disk-corona interface angle ($\theta^{dc}$, solid line) and
  corona-jet interface angle ($\theta^{cj}$, short-dashed line),
  number of cells per fastest growing MRI wavelength ($\Qone$, solid
  line ; $\Qoneweak$, short-dashed line) with $\Qone=6$ shown as
  dotted line above where the vertical ($\theta$) MRI is resolved,
  number of cells per fastest growing MRI wavelength ($\Qthree$, solid
  line ; $\Qthreeweak$, short-dashed line) with $\Qthree=6$ shown as
  dotted line above where the azimuthal ($\phi$) MRI is resolved,
  number of fastest growing MRI wavelengths across the full disk
  thickness ($\Qtwo$, solid line ; $\Qtwoweak$, short-dashed line)
  with $\Qtwo=1/2$ shown as dotted line below where the MRI is
  suppressed, viscosity parameter ($\alpha_b$, solid line ;
  $\alpha_{b,\rm eff}$, short-dashed line), radial absolute magnetic
  flux ($\Phi$), and equatorial magnetic flux ($\Psi_{\rm eq}$).  In
  summary, the disk is compressed by the horizon's magnetic flux
  leading to smaller $\theta^d$ as $r$ drops despite little change in
  $\theta^t$.  Also, the linear MRI is suppressed even in the dense
  inflow.}
\label{othersvsr}
\end{figure}

Figure~\ref{othersvsr} shows the time-averages for the disk's
geometric half-angular thickness ($\theta^d$), the thermal
half-thickness ($\theta^t$, using the density-weighted average), flow
interface angular locations, resolution of the MRI wavelength,
approximate $\alpha$ viscosity parameter, and magnetic fluxes
vs. radius.  The magnetic field compresses the disk leading to
decreasing $\theta^d$ with decreasing radius.  While $\theta^d\ll 1$
near the BH, $\theta^t=\arctan{(c_s/v_{\rm rot})}\gtrsim 1$ and
$\arctan{(c_s/v_{\rm K})}\gtrsim 1$.  Hence, the flow is not in
vertical hydrostatic equilibrium due to the strong magnetic field.
Note that using $v_{\rm K}$ instead of $v_{\rm rot}$ in the expression
for $\theta^t$ gives a similarly large thermal half-thickness.  The
disk-corona and corona-jet interfaces trace the path of the
well-collimated jet out to large radii.  The $\Qone\gg 6$ as required
to resolve the MRI \citep{sano04}, while in the inflow equilibrium
region $\Qtwo\lesssim 1/2$, indicating the MRI is suppressed.  Even using
$\Omega\to \Omega_{\rm K}$ in $\Qtwo$ leads to $r_{\Qtwo=1/2}\approx
8r_g$, so the MRI suppression near the BH is not only due to
sub-Keplerian motion but also disk compression by the accumulated
magnetic flux. The horizon's time-averaged radial absolute magnetic
flux is $\Phi_{r,\rm H}\sim 200$ and there is an additional $\Psi_{\rm
  eq}\sim 100$ same polarity magnetic flux available.

\subsubsection{Comparison with Gammie (1999) Model}
\label{sec_gammie}

\begin{figure}
\centering
\includegraphics[width=3.0in,clip]{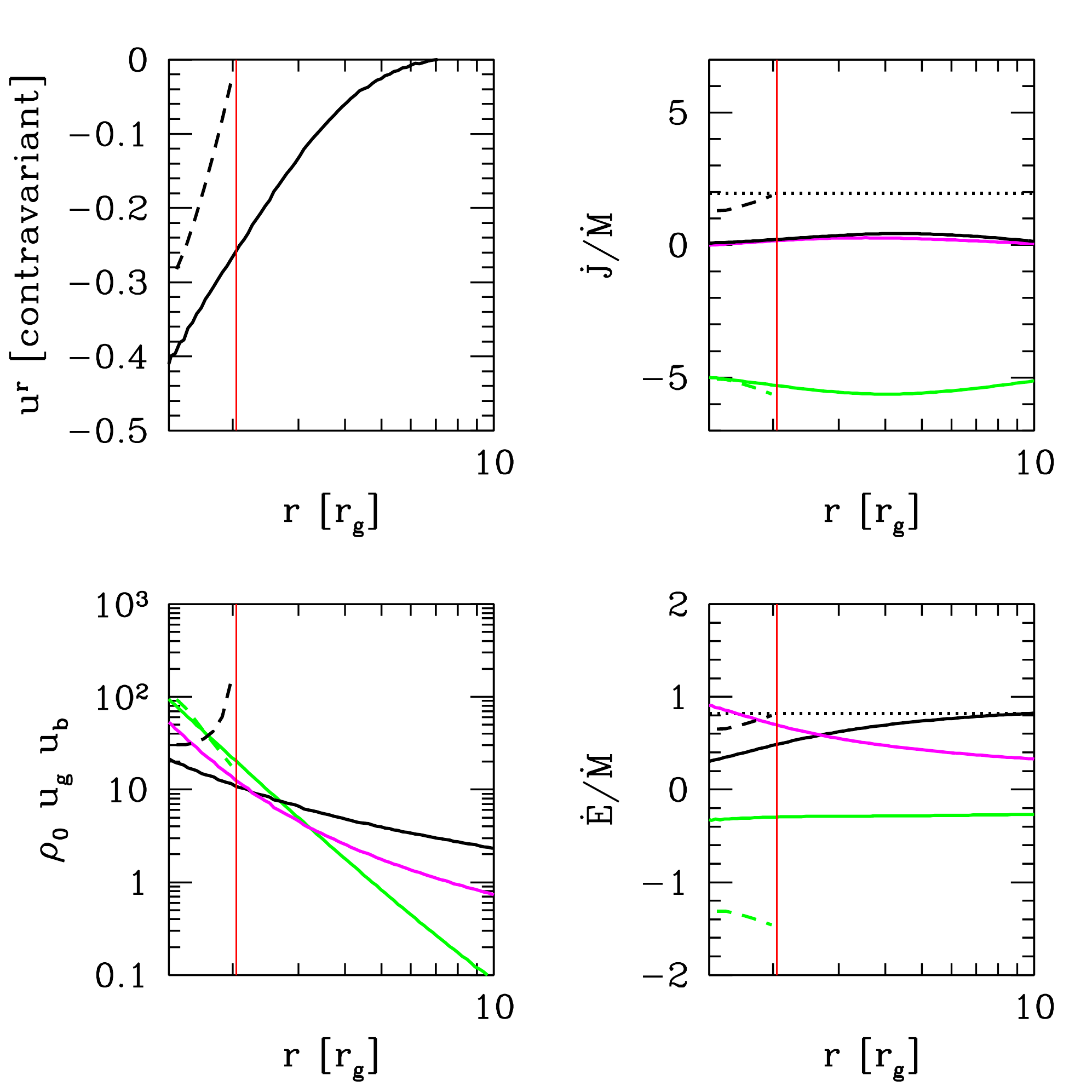}
\caption{Comparison between the accretion flow for the fiducial
  simulation (``fdc'' averaging, focusing on entire flow except the
  highly-magnetized jet, shown by solid lines) and the
  \citet{gammie99} model for an unstratified magnetized inflow within
  the ISCO (shown by dashed lines).  In all panels, red vertical lines
  shows the location of the ISCO.  Top-left panel: Shows the radial
  4-velocity, where the Gammie solution assumes $\uvec^r=0$ at the
  ISCO.  Finite thermal effects lead to non-zero $\uvec^r$ at the ISCO
  for the simulated disk.  Bottom-left panel: Shows the rest-mass
  density ($\rho_0$, black line), the internal energy density ($u_g$,
  magenta line), and magnetic energy density ($u_b$, green line).
  Top-right and bottom-right panels: Show simulation results and
  Gammie solution for the specific particle (PA) flux (black line),
  specific enthalpy (EN) flux (magenta line), and specific
  electromagnetic (EM) flux (green line). Horizontal dotted black
  lines are the Novikov-Thorne thin disk solution
  (i.e. particle/radiation term). This figure is comparable to fig.~10
  for a somewhat thick ($\theta^d\sim 0.2$--$0.25$) disk in
  \citet{mg04} and to fig.~11 for a thin ($\theta^d\sim 0.05$) disk in
  \citet{pmntsm10}.  As for the prior thick disk simulations, the
  model fits $\dot{J}$ and $b^2$ but not $\dot{E}$.}
\label{gammie4panel}
\end{figure}

Figure~\ref{gammie4panel} shows a comparison between the fiducial
model and the \citet{gammie99} model of a magnetized unstratified
accretion flow within the ISCO.  A value of $\Upsilon\approx 8.7$ was
used to match the Gammie model's value of the EM component
of the specific angular momentum at the horizon, while the effective disk thickness was
varied to best fit the magnitude of $u_b$ at the horizon.  All
quantities use ``fdc'' type averaging focusing on all parts of the
flow except the highly-magnetized jet.  Compared to the densities
vs. radius shown in Figure~\ref{rhovelvsr}, here clearly $u_b\gg
\rho_0$ on the horizon due to averaging over the entire
disk+corona+winds.  The densities monotonically increase towards the
BH as the inflow is vertically compressed by magnetic field.

If $\Upsilon$ is only integrated across the time-averaged disk where
$\dot{M}$ is non-zero (i.e. ``fdc'' averaging: full flow except the
highly magnetized part of the jet), then we find $\Upsilon\approx
1.5$.  However, within the heavy dense filaments
(i.e. density-weighted average), we find $\Upsilon\approx 0.1$.  Also,
$\beta\gg 1$ in the dense flow, while $\beta\ll 1$ over the
disk+corona+winds.  This shows that the heavy filamentary parts of the
disk inflow are quite under-magnetized compared to their surroundings,
which is as expected for accretion through efficient magnetic
Rayleigh-Taylor instabilities.

\subsection{Time-Averaged Angular ($\theta$) Dependence}\label{sec:thetadep}

\begin{figure}
\centering
\includegraphics[width=3.2in,clip]{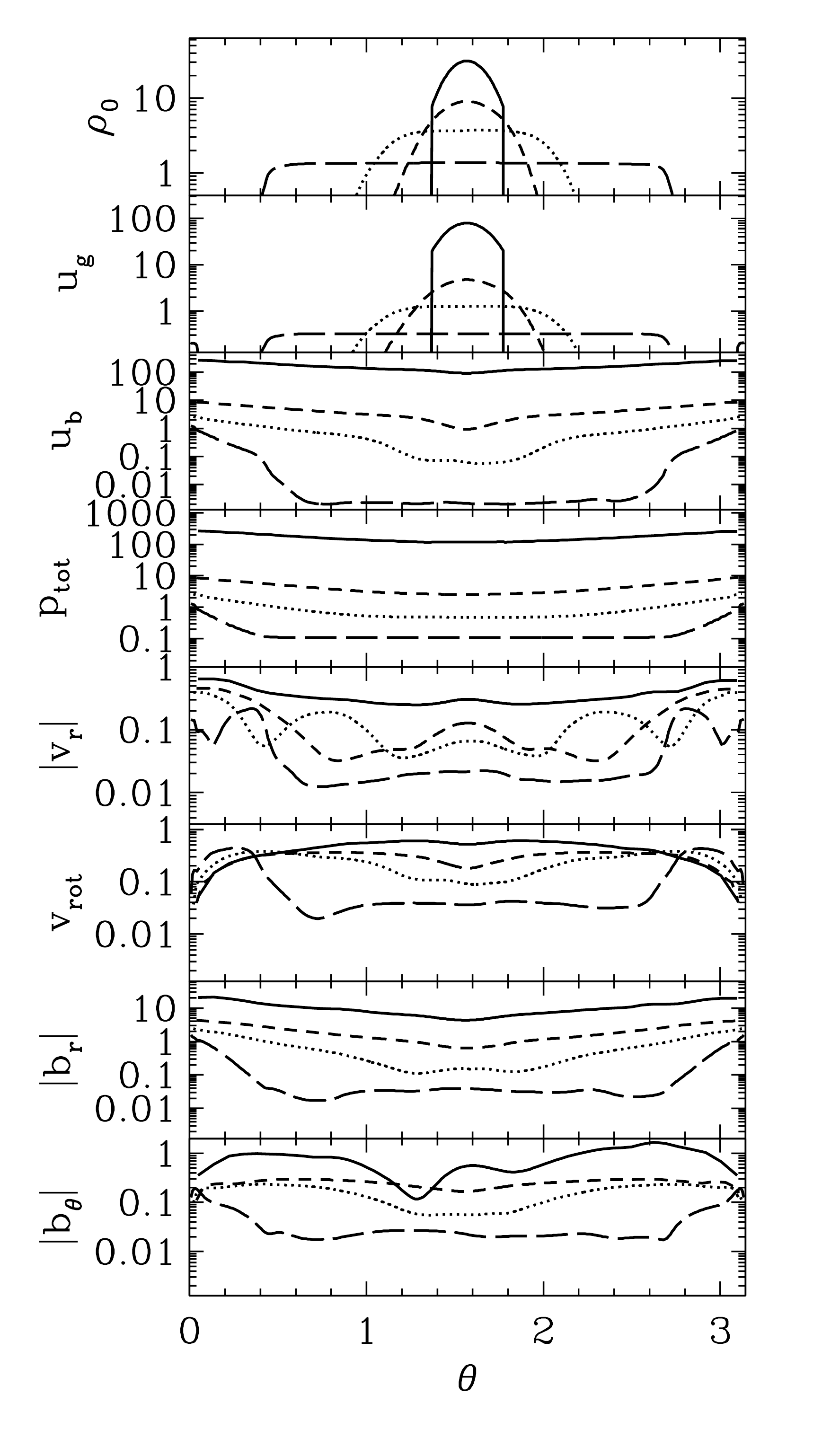}
\caption{Similar quantities as in Figure~\ref{rhovelvsr}, except plotted
  vs. $\theta$ at $r=\{r_{\rm H}/r_g,4,8,30\}r_g$ (respectively:
  solid, short-dashed, dotted, and long dashed lines).  If numerical
  density floors were activated at some space-time point, then
  $\rho_0=u_g=0$ was set there.  This shows up even in these
  time-averaged densities as a sharp drop-off, because the jet
  maintains high magnetizations at the poles during the interval
  time-averaged.  In summary, the disk is compressed by the polar
  magnetic flux as the inflow approaches the horizon, and the total
  pressure is roughly constant with $\theta$.}
\label{rhovelvsh}
\end{figure}

Figure~\ref{rhovelvsh} is similar to Figure~\ref{rhovelvsr} but for
quantities vs. $\theta$ at four different radii.  This also highlights
how the disk flow is compressed as it approaches the horizon.  The
time-averaged density is well-fit by a Gaussian with width
$\theta^d\approx 0.12$ at $r=r_{\rm H}$ even though the time-average
of the instantaneous $\theta^d\sim 0.06$.  This is because the
time-average of density blurs the width of the narrow disk that
oscillates in height about equal to its own height.  Since the
time-averaged density is a Gaussian, the estimate of the thermal
thickness $\theta^t_{\rholab}$ given by Eq.~(\ref{thetateq}) would
have given $\theta^t_{\rholab}\sim \theta^d$ if the disk were in
hydrostatic equilibrium.  That $\theta^d\ll \theta^t$ close to the BH
(as shown in Figure~\ref{othersvsr}) shows that the thermal content
remains relatively unchanged despite significant geometric compression
by the polar magnetic flux.  The figure also shows that the total
pressure is roughly constant with angle as dominated by the magnetic
pressure.  The behavior of $v_\theta,v_\phi$ near the polar axes is
affected by the numerical floor mass injection, although this region
contains little energy or energy flux compared to other angles.

\begin{figure}
\centering
\includegraphics[width=3.2in,clip]{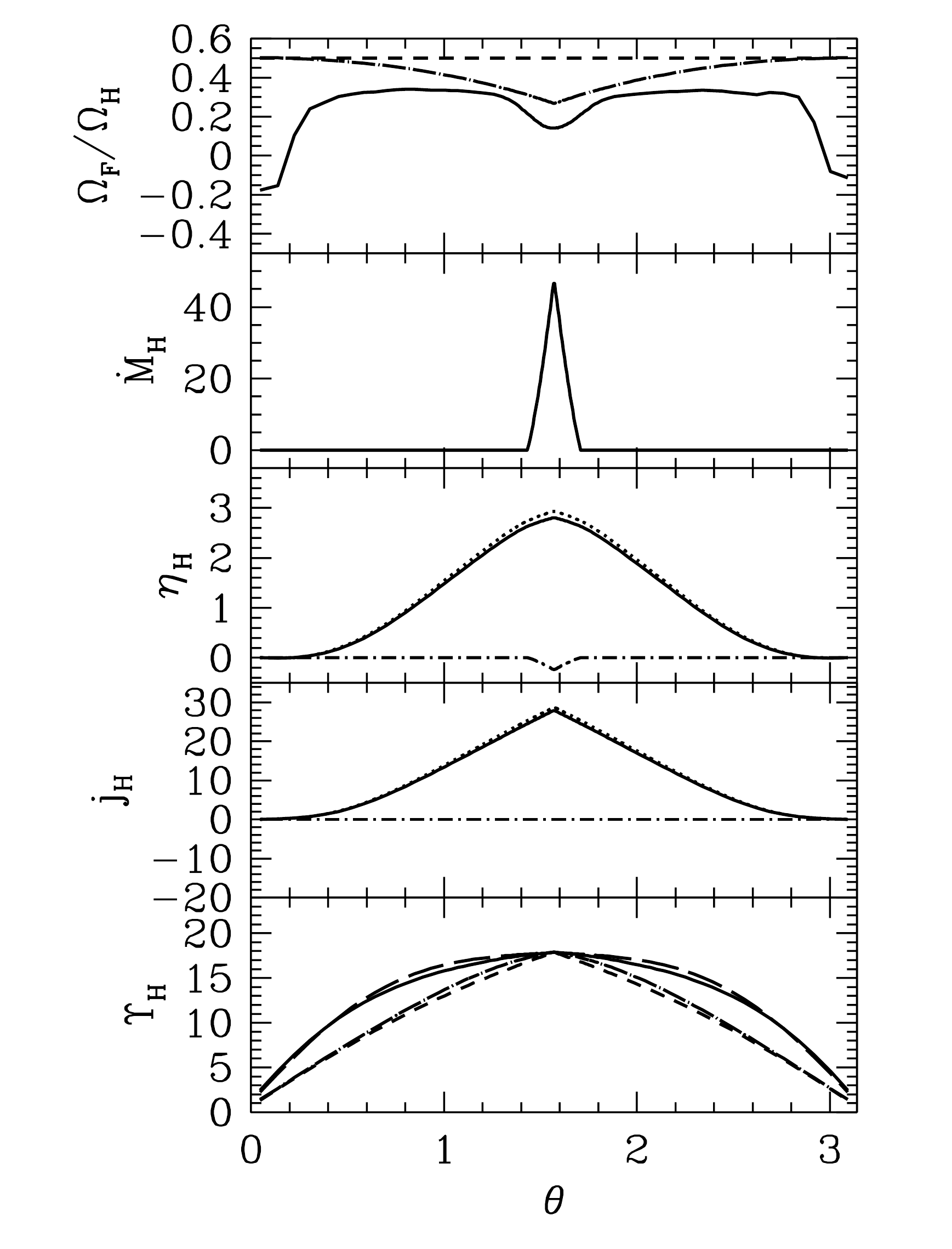}
\caption{Time-$\phi$-averaged quantities and flux integrals on the
  horizon as a function of $\theta$.  From top to bottom: 1) Field
  line rotational angular frequency ($\Omega_{\rm F}/\Omega_{\rm H}$)
  for simulation (solid line), 1st-order-in-spin accurate value for
  monopolar (short-dashed line) and paraboloidal (dot-long-dashed
  line) BZ solutions ; 2) Rest-mass flux ($\dot{M}_{\rm H}$) ; 3)
  Electromagnetic (EM, solid line) and matter (MA, dot-short-dashed
  line) efficiency ($\eta_{\rm H}$), along with the full BZ-type EM
  formula without any renormalization (dotted line) ; 4)
  Electromagnetic (EM, solid line) and matter (MA, dot-short-dashed
  line) specific angular momentum flux (${\jmath}_{\rm H}=\dot{J}_{\rm
    H}/\dot{M}_{\rm H}$), along with the full BZ-type EM formula
  without any renormalization (dotted line); 5) Gammie parameter
  ($\Upsilon_{\rm H}$) for the simulation (solid line), and the BZ
  model for the cases: 0th-order-in-spin accurate monopolar field
  (short-dashed line), 0th-order-in-spin accurate paraboloidal field
  (dot-long-dashed line), and 2nd-order-in-spin accurate monopolar
  field (long-dashed line).  These BZ versions are normalized so total
  magnetic flux is the same as in the simulation.  Notice how the
  2nd-order-in-spin accurate monopolar BZ model fits the simulation
  result quite well.  For the last 3 panels, the divisor is
  (implicitly) $\dot{M}_{\rm H}$ that has been fully angle-integrated
  to a single value.  So, $\eta_{\rm H}$, $j_{\rm H}$, and
  $\Upsilon_{\rm H}$ show the angular dependence of $\dot{E}_{\rm H}$,
  $\dot{J}_{\rm H}$, and $\Psi_{\rm H}$, respectively.  In summary,
  the agreement between the simulation and the BZ picture is
  excellent.}
\label{horizonflux}
\end{figure}

Figure~\ref{horizonflux} shows the horizon's values of quantities
related to the BZ effect \citep{bz77}.  The simulation's fluxes are
computed via Eq.~(\ref{fluxtheta}).  The ``full BZ-type EM formula''
referred to in the figure uses the EM energy flux computed from
equation~33 in \citet{mg04}, which only assumes stationarity and
axisymmetry (rather than also small spin in BZ77) and uses the
simulation's $\Omega_{\rm F}(\theta)$ and $\Bvec^r(\theta)$ on the
horizon.  This figure shows that most of the horizon is highly
magnetized due to accretion occurring through a magnetically
compressed inflow.

The agreement between the simulations and the BZ picture is excellent
for the highly magnetized regions, where roughly $\Omega_{\rm F}\sim
\Omega_{\rm H}/4$ near the disk-jet interface (here, $\Omega_{\rm F}$
is the time-average of Eq.~(\ref{omegaeq})).  While
the simulation is roughly consistent with BZ's paraboloidal solution,
the equatorial $\Omega_{\rm F}$ is somewhat suppressed due to the disk
inflow.  Also, near the polar axes, $\Omega_{\rm F}$ is affected by
ideal MHD effects and numerical floor mass injection.  In the last
$\sim 3$ grid cells, limited resolution affects $\Omega_{\rm
  F}$. However, both the grid and jet collimate, which leads to a
well-resolved $\Omega_{\rm F}$ near the axes at slightly larger radii.
Also, the most energetic part of the jet is near the disk-jet
interface that dominates the dynamics at all radii.  Even if we used a
paraboloidal extension of our simulation data into the region where
the numerical floor injection is important, this only changes the
total efficiency by less than $20\%$.

\subsection{Azimuthal ($\phi$) Time Dependence}\label{sec:timephidep}

\begin{figure*}
\centering

\includegraphics[width=6.6in,clip]{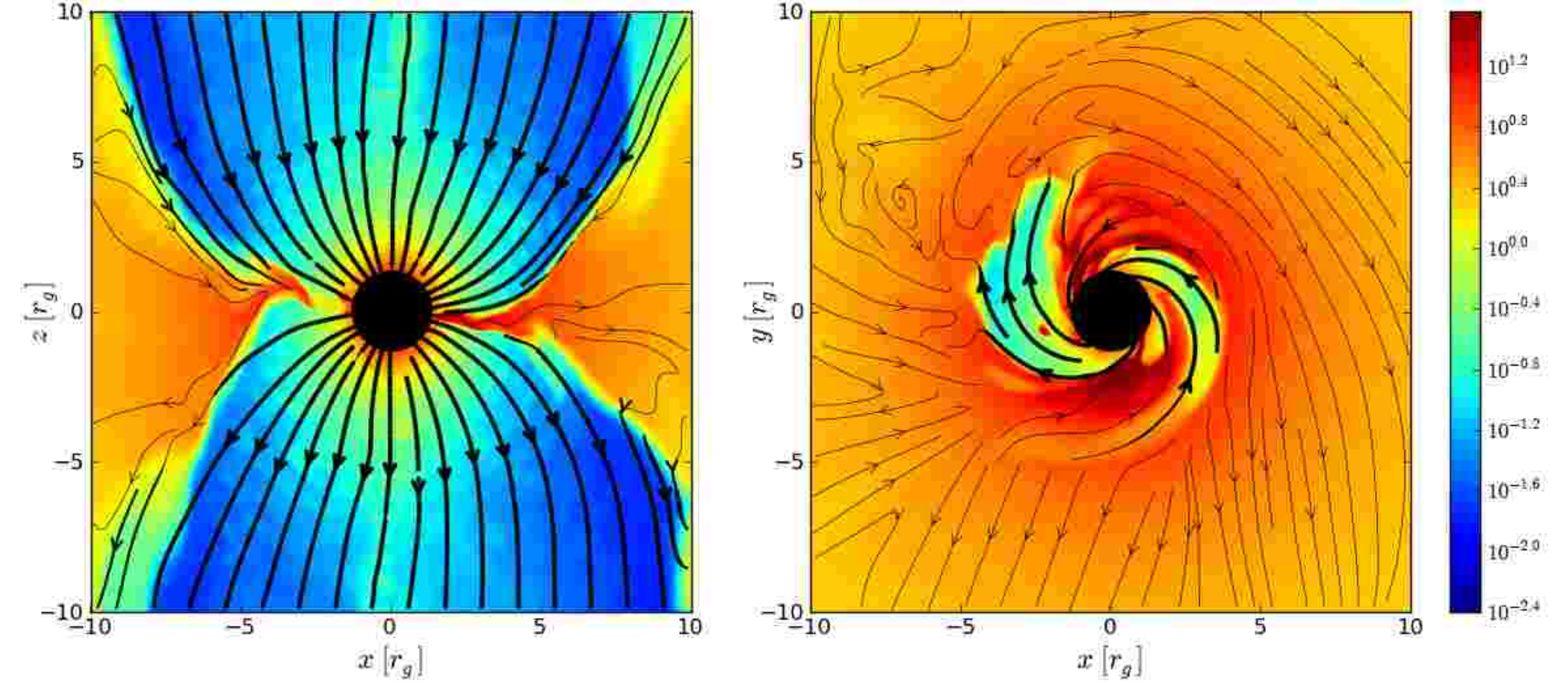}

\caption{Snapshot at $t\approx15172r_g/c$ for the fiducial model
  during one of the larger magnetic Rayleigh-Taylor disruptions, just
  after which the mass accretion rate increases.  Otherwise similar to
  the upper panels in Figure~\ref{evolvedmovie}.  During this
  disruption, $|m|=1,2$ modes appear to dominant the non-axisymmetric
  accretion.}
\label{outburst}
\end{figure*}

Figure~\ref{outburst} shows a large magnetic Rayleigh-Taylor
disruption at $t\approx 15172r_g/c$ once $\Upsilon$ has reached its
saturated value and the flow can no longer accept new magnetic flux on
the BH.  While an atypical looking outburst, other more typical
snapshots, such as at $t\approx11272r_g/c$, show roughly similar
behavior including the same range of density variations.  Such
outbursts occur because magnetic flux is temporarily added to the BH
and exceeds the saturated value, but then that extra flux is ejected
back in magnetic interchange modes.  Here, $|m|=1,2$ modes appear to
dominate.

\begin{figure}
\centering
\includegraphics[width=3.2in,clip]{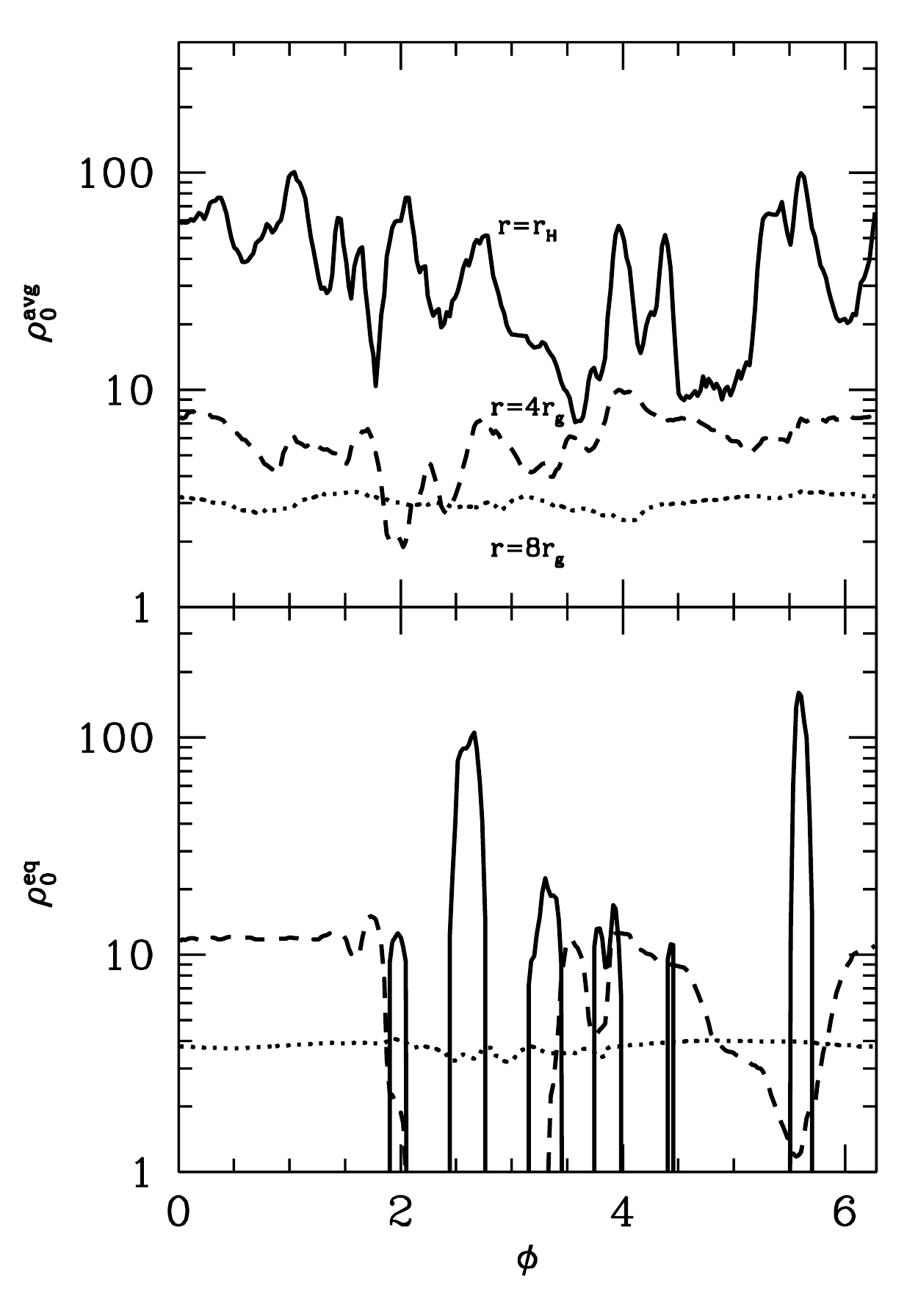}
\caption{Shows rest-mass density at $t\approx15172r_g/c$ (as shown in
  Figure~\ref{outburst}), as $\theta$-averaged over the entire flow
  ($\rho^{\rm avg}_0$) and as cut along the equator ($\rho^{\rm
    eq}_0$) at three radii (in all panels: $r=r_{\rm H}$ shown as
  solid line, $r=4r_g$ shown as short-dashed line, and $r=8r_g$ shown
  as dotted line). The drop-outs in density correspond to where the
  mass is dominated by numerical floor mass injection due to the high
  $u_b/\rho_0$ in such regions, so we have set $\rho_0=0$ there to
  show where the density could be smaller than the floor.  In summary,
  the variations in density become quite large near the BH.}
\label{rhovsphi}
\end{figure}

Figure~\ref{rhovsphi} shows the density vs. $\phi$ at three radii
($r=r_{\rm H},4r_g,8r_g$) for $t\approx15172r_g/c$ (as shown in
Figure~\ref{outburst}).  $\rho_0^{\rm avg}$ shows the average with
weight $w=1$, and $\rho_0^{\rm eq}$ shows the density for a single grid
cell cut exactly at the equator.  Typical density variations averaged
over the flow are up to factors of $10$ near the horizon, while at any
given $\theta-\phi$ angles (here the equator) the density varies
considerably (down to the limits of the numerical density floors).
This shows how the dense part of the flow is forced through the
magnetosphere via magnetic Rayleigh-Taylor modes.

\subsection{Time-Averaged Azimuthal ($\phi$) Dependence}\label{sec:phidep}

\begin{figure}
\centering
\includegraphics[width=3.1in,clip]{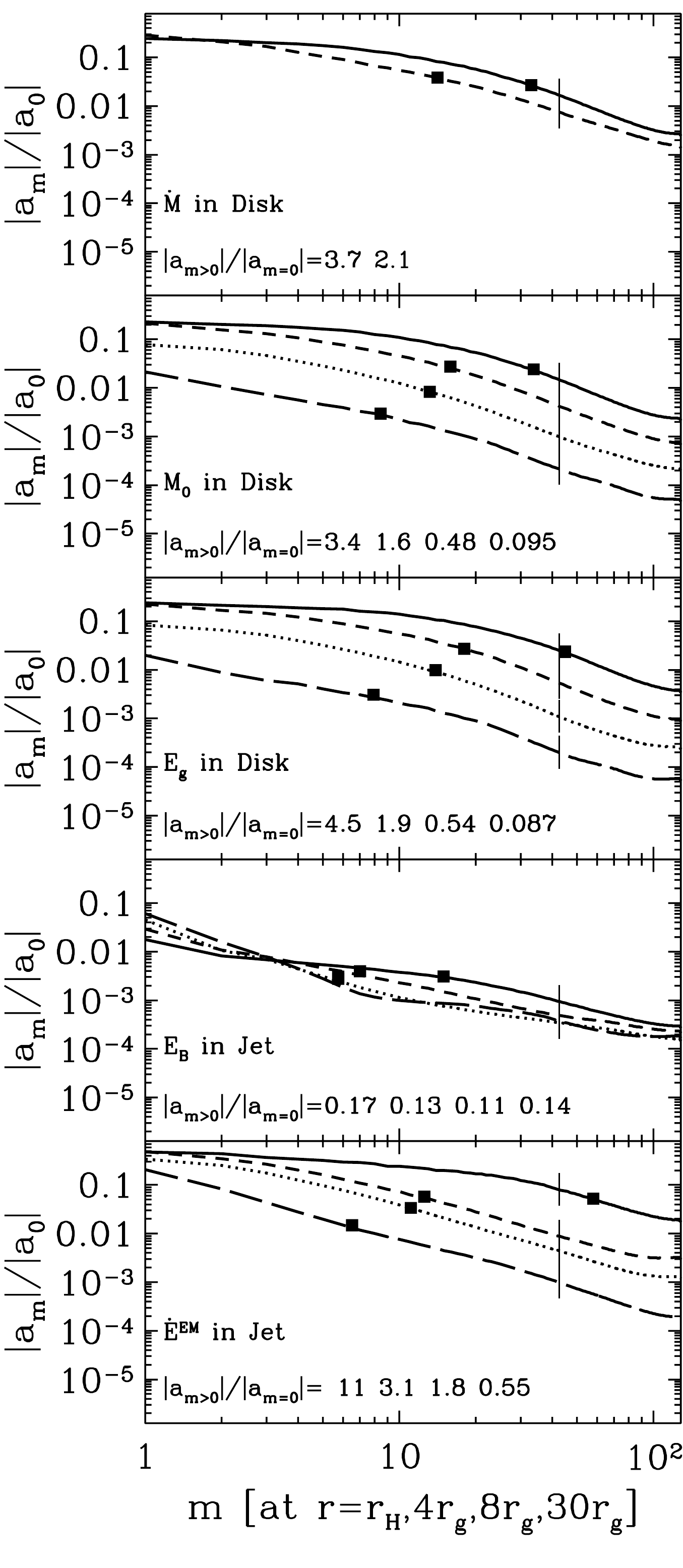}
\caption{Time-averaged Fourier amplitude for $m$ modes.  Each panel
  shows result for $r=r_{\rm H},4r_g,8r_g,30r_g$ (respectively, solid,
  short-dashed, dotted, long-dashed lines), except the top panel that
  only shows results for $r=r_{\rm H},4r_g$ with same line types.  The
  power in all $|m|>0$ modes relative to the $|m|=0$ mode is shown for
  each panel respectively for radii $r=r_{\rm H}, 4r_g, 8r_g, 30r_g$.
  For each line, the vertical bar corresponds to $m_{\rm 6
    cells}=N_\phi/6$, where $m>m_{\rm 6 cells}$ are numerically damped
  on a dynamical time.  For each line, squares mark the correlation
  length's $m=m_{\rm cor}$ mode, so azimuthal structure
  (e.g. turbulence) is resolved if $m_{\rm cor}\lesssim m_{\rm 6
    cells}$.  In summary, azimuthal structures are well-resolved
  across all quantities within the causally-connected region outside
  the horizon.  Changes in $m_{\rm cor}$ with radius in the ``Disk''
  partially track changes in the disk thickness.  Most power is at
  $|m|\sim 1$, but significant power extends up to $|m|\sim 20$ before
  dropping off more rapidly.  Because $|a_{m>0}|\approx |a_{m=0}|$ for
  $\dot{M}$, mass accretion is highly non-axisymmetric (see also
  Figure~\ref{rhovsphi}).}
\label{powervsm}
\end{figure}

Figure~\ref{powervsm} shows the normalized time-average of the Fourier
decomposition in the $\phi$-direction ($[|a_m|]_t/[|a_0|]_t$) using
Eq.~(\ref{am}) for the ``Disk'' and ``Jet'' with quantity $Q$ as
$\rho_0 \uvec^r$, $\rho_0$, $u_g$, $b^2$, and ${T^{\rm EM}}^r_t$ that
gives, respectively, the actual $\dot{M}$, comoving mass $M_0$,
comoving thermal energy $E_g$, comoving electromagnetic energy $E_B$,
and actual electromagnetic energy flux $\dot{E}^{\rm EM}$.  As with
flux ratios, we compute $[|a_m|]_t/[|a_0|]_t$ instead of
$[|a_m/a_0|]_t$ because the latter would exaggerate mode power during
transient moments when $a_0(t)$ becomes small and potentially zero.

We find that there is large $|m|>0$ power in $\dot{M}$ and
$\dot{E}_{\rm EM}$, but there is little power in $|m|>0$ for $b^2$ in
the jet.  Since $|a_{m>0}|\approx |a_{m=0}|$ for $\dot{M}$, this shows
that accretion occurs primarily through non-axisymmetric modes.  Also,
$|a_{m>0}|\approx |a_{m=0}|$ for the jet electromagnetic power even at
$r=8r_g$, so the jet power contains significant non-axisymmetric
structure.  The $|m|=1$ dominates all $|m|>0$, but $|a_m|$ and $|a_1|$
are similar up to $m\approx 20$.

Figure~\ref{powervsm} also shows the azimuthal correlation length
scale (Eq.~(\ref{mcor})), which is always well-resolved beyond the
horizon.  Note that if we compute Eq.~(\ref{am}) as a spectrum of the
averaged flow rather than as an average of the spectrum, then even on
the horizon the correlation lengths are very well-resolved.  The
correlation length scale $\lambda_{\phi,\rm cor}$ is related to the
mode $m_{\rm cor} = 2\pi/\lambda_{\phi,\rm cor}$.  For example, for
$M_0$ (i.e. the density in the disk), we find $m_{\rm cor}\approx
33.8,15.9,13.2,8.5$ at $r=r_{\rm H},4r_g,8r_g,30r_g$, respectively.
With $\theta^d\approx 0.06,0.13,0.29,0.59$, one obtains
$\lambda_{\phi,\rm cor}/\theta^d\approx 3.1,3.0,1.6,1.3$,
respectively.  So as the magnetic flux near the BH vertically
compresses the inflow, the azimuthal correlation length drops, but not
in proportion to the disk height.

\subsection{Quasi-Periodic Oscillations}
\label{sec:qpos}

BH accretion disks are observed to have QPOs roughly classified as
either high-frequency QPOs (HFQPOs) or low-frequency QPOs (LFQPOs).
In BH x-ray binaries, QPOs are seen in only specific states
\citep{fend04a,abr05,rm06}, such as the steep-power law (SPL) state
(related to the very high state and intermediate state, during which
powerful transient jets are observed) \citep{dg03,gn06,omnm09}.  For
BH x-ray binaries, HFQPOs range from 100Hz (GRS1915+105) to 300Hz
(1655-40) corresponding to a period of $\tau\sim 70$--$130r_g/c$.
HFQPO frequencies are sometimes in a 3:2 ratio \citep{psa99}.  QPOs
are not expected to be as easily seen in AGN as for x-ray binaries
\citep{vu05}; but some AGN may have QPOs, such as SgrA*
\citep{2008ApJ...688L..17M,2012arXiv1201.1917D} and RE J1034+396
\citep{gmwd08}.  NS QPOs have similar features \citep{vdk98}.

BH x-ray binary QPOs are not seen (or are very weak) in the high-soft
state that is typically associated with a thin disk dominated by
MHD-MRI turbulence \citep{wij99,r05,rl05,mm08,mbc09}.  The lack of
QPOs in such a state is expected because non-linear turbulence tends
to reduce coherence.  Also, the existence of a bright hard state
defies explanation by standard viscous models that would predict the
bright state should be soft \citep{omnm09}.  This indicates that a
qualitatively new accretion state (like the MCAF state) may be
required.

Analytical models and simulations have demonstrated QPOs through
various mechanisms.  The hope is that QPOs could be used to measure BH
spin and mass \citep{1999PhRvL..82...17S}.  Disk mode oscillations may
cause either HF or LF QPOs \citep{rm09,orm09,orms11}, disk precession
causes LFQPOs in GRMHD simulations of tilted disks \citep{df11}, and
dynamo field oscillations in local MHD simulations show LFQPOs
\citep{dsp10,gg11}. In some cases QPOs may be seen, but their
statistical significance is highly uncertain.  This includes HFQPOs
seen in pseudo-Newtonian MHD simulations \citep{kato04}.  GRMHD
simulations have shown HFQPOs in tilted disks \citep{heni09}.
However, in the latter case higher resolutions were found to eliminate
the QPOs as distinct features because MRI turbulence destroys their
coherence (Henisey, 2010, priv. comm.).  Other GRMHD simulations with radiative transfer show some
HFQPO features \citep{sch06}, but their significance is highly
dependent upon model of ``continuum'' spectrum as a single power-law,
while their power spectrum is identified as a broken power-law.  So
extra apparent power appears near the break.

Our GRMHD simulations also show coherent HFQPOs.  We identify the
coherence of the QPO by its quality factor $Q\approx \nu/(\Delta \nu)$
for frequency $\nu$, where $\Delta\nu$ is the full width at half of
the maximum power (FWHM) (e.g. if the QPO distribution is modelled by
a Lorenzian).  All our thick disk poloidal field simulations at both
resolutions ($136\times 64\times 128$ and $272\times 128\times 256$)
show coherent QPOs during some part of the simulation at various radii
($r=r_{\rm H},4r_g,8r_g,30r_g$), while the QPOs are most coherent for
the high-resolution fiducial model.  This expected resolution
dependence on Q implies that damping at low resolutions can make it
difficult to resolve a coherent QPO.  Radiative transfer
\citep{bm10,2010arXiv1007.4832S,dma11} is necessary to see if these
QPOs are observable.  In this paper, we only consider the MHD
dynamical properties of the HFQPO.

\begin{figure}
\centering
\includegraphics[width=3.2in,clip]{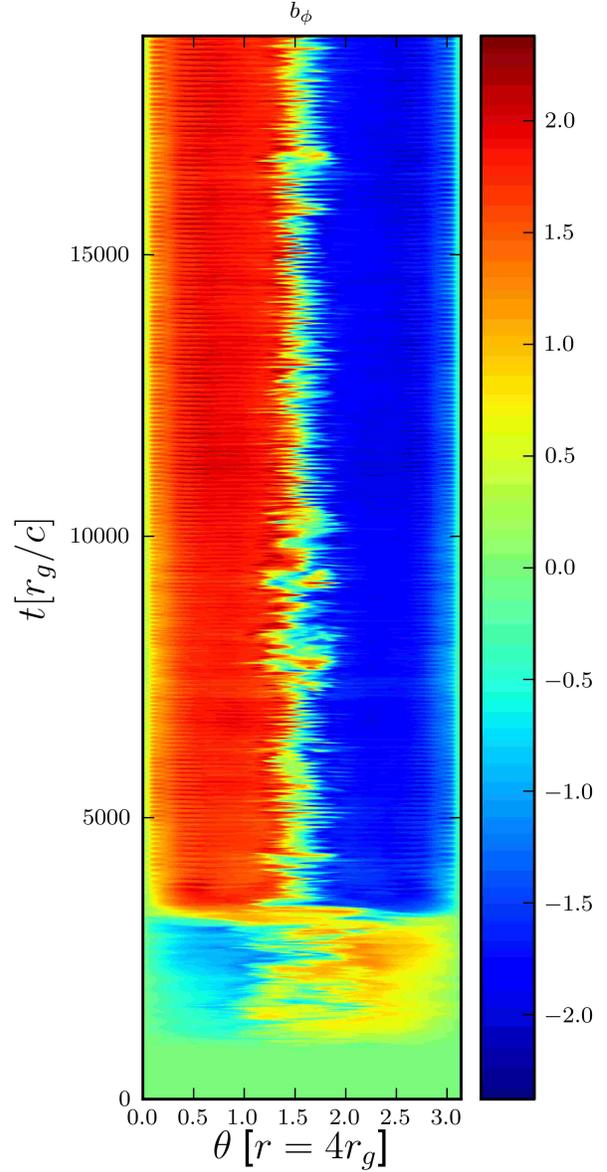}
\caption{$b_\phi$ vs. $t$ and $\theta$ at $r=4r_g$.  The field
  polarity switches at $t\sim 2700r_g/c$, after which magnetic flux
  accumulates and saturates.  After saturation, the jet-disk QPO
  (JD-QPO) becomes coherent, here visible as horizontal stripes at
  late times.}
\label{bphitvsth}
\end{figure}

Figure~\ref{bphitvsth} shows $b_\phi$ averaged over $\phi=0$ to
$\phi=\pi/4$ (with no weighting) vs. $\theta$ and time $t$ at $r=4r_g$
(averaged over $\pm 0.4r_g$ with no weighting) for the fiducial model.
This shows possible QPOs.

\begin{figure}
\centering
\includegraphics[width=3.2in,clip]{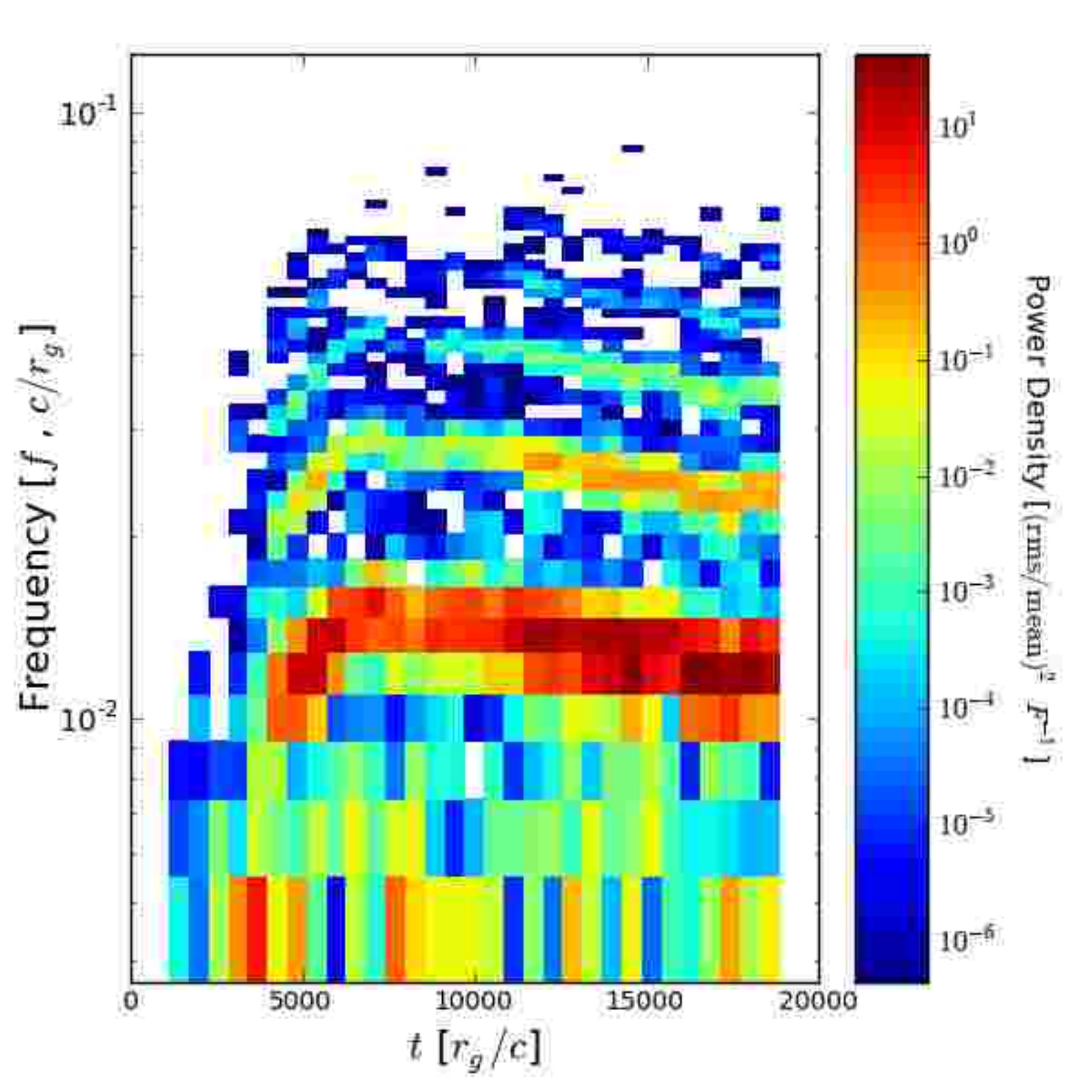}
\caption{Spectrogram for the power density vs. time and frequency for
  $b^2$ (with legend, where white is below lowest power in legend) at
  $r=4r_g$ at $\theta\approx 2.9$ (i.e. deep within the jet).  This
  shows the high $Q$ oscillation starts once flux saturates for
  $t\gtrsim 5000r_g/c$, where $\sim 3$ jet-disk QPO (JD-QPO) modes are
  visible.  The frequency changes little as $\dot{M}$ varies.}
\label{spec}
\end{figure}

Figure~\ref{spec} shows a spectrogram for the power density vs. time
and frequency for $b^2$ at $r=4r_g$ at $\theta\approx 2.9$ (i.e. deep
within the jet).  Power density is shown in commonly-used units: an
integral of power density over a frequency range gives back the square
of the fractional root-mean-squared (rms) amplitude of the variability
in the original time series.

\begin{figure}
\centering
\includegraphics[width=2.5in,clip]{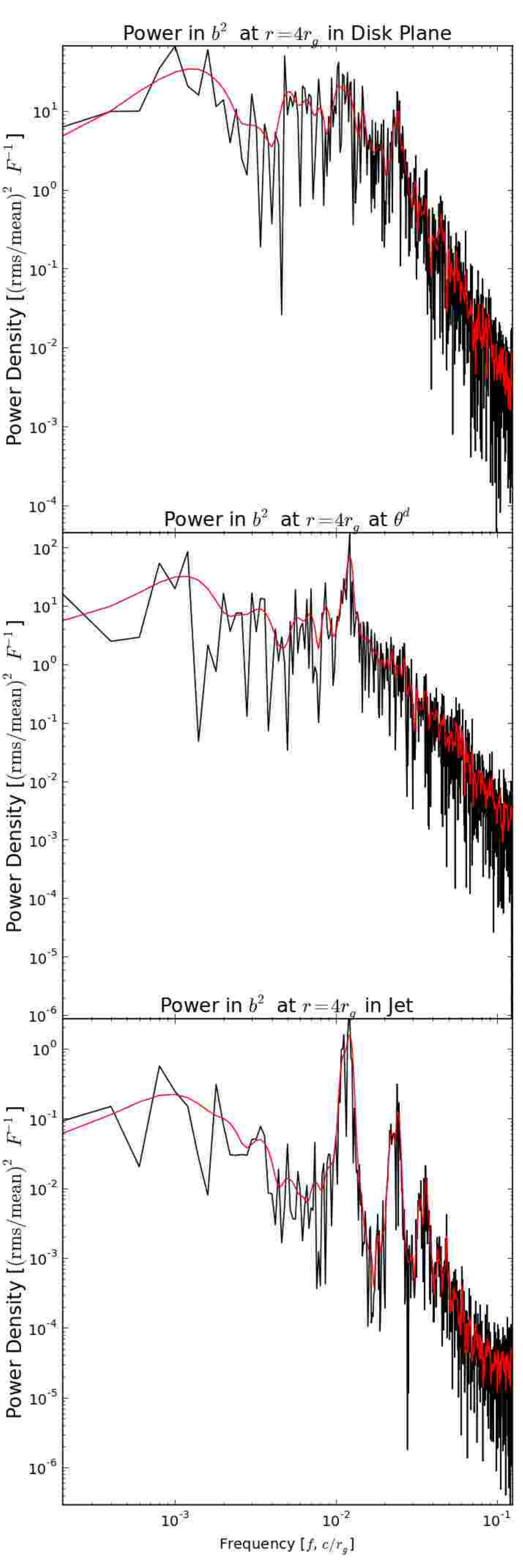}

\caption{Power Density for $b^2$ at $r=4r_g$ in the disk plane (top
  panel), one geometric half-angular thickness ($\theta^d$) above the
  disk plane (middle panel), and deep within the jet at $\theta\approx
  2.9$ (bottom panel) shown as black lines. A smoothed (over 10
  frequencies) version is shown as red lines.  The disk has QPOs with
  quality factor $Q\lesssim 10$, a disk scale-height above the disk
  plane has $Q\sim 10$, while the jet itself shows a few different
  harmonics with up to $Q\sim 100$.  The jet harmonics correspond to
  $|m|=\{1,2,3\}$ modes based upon the field line angular frequency
  (driven by the black hole rotation frequency) at the disk-jet
  magnetospheric interface where the jet-disk QPO (JD-QPO) mechanism
  operates.}
\label{fft}
\end{figure}

Figure~\ref{fft} shows the Fourier transform of $b^2(t)$ at the
equatorial plane (at $\theta=0$), in the disk (at
$\theta=1[\theta^d]_t$, where $[\theta^d]_t$ is the time-averaged
$\theta^d$) and in the jet (at $\theta\approx 2.9$) at $r=4r_g$ as
averaged from $\phi=0$--$\pi/4$ to reduce noise while allowing us to
resolve several $|m|$ modes.  The power density is given in units of
$({\rm rms}/{\rm mean})^2$ multiplied by the total period
($t=12000$--$16000r_g/c$, focusing on the high-Q period) over which
the Fourier transform is performed, as a function of frequency (in
units of $c/r_g$).  We find that $Q\sim 100$ in the jet, $Q\sim 10$
one scale height above the disk plane, and $Q\lesssim 10$ in the disk
plane.  Other radii (e.g. $r=r_{\rm H}$, $8r_g$, and $30r_g$) in the
jet also show similar QPOs as features move out in the jet.

What is the nature of the HFQPOs in our simulations?  Once poloidal
magnetic flux has accumulated and reached its natural saturated limit,
a semi-permeable magnetospheric barrier forms between the heavy disk
inflow and magnetic flux threading the rotating BH and its
jet.  This magnetospheric interface is where magnetic field and
density change somewhat abruptly.  Linear stability analyses of such
magnetospheric interfaces predict them to be unstable to
Rayleigh-Taylor and Kelvin Helmholtz modes, which drive QPOs at
spherical harmonic $m$ modes based upon some rotational frequency
($\Omega$) such that $\Omega_{\rm QPO}\approx m\Omega$
\citep{ln04,2012arXiv1201.5370F}.  These linear stability analyses
predict that low $m$ modes dominate the non-linear dynamics and that
QPOs should appear.  However, these models cannot determine what
controls $\Omega$, the power of the $m$ modes, or the coherence of the
QPO.

Our MCAF simulations validate the prior linear stability analysis.  We
confirm that the magnetospheric interface is indeed unstable and that
low-$|m|$ (primarily $|m|=1$) modes dominate.  We also show that the
unstable jet-disk interface drives coherent HFQPOs.  A movie of
snapshots like Figure~\ref{evolvedmovie} shows that once poloidal
magnetic flux saturates, the jet and disk oscillate together.

Why do such QPOs appear in MCAFs and not MRI-dominated accretion
flows?  QPOs seen in prior MRI-dominated accretion flow simulations
show (e.g.) $m=1$ spiral disk modes extending out in radius
\citep{2012arXiv1201.1917D}. Other QPOs \citep{heni09} were absent at
high resolutions due to decoherence by properly-resolved MRI-driven
MHD turbulence (Henisey, 2010, priv. comm.).  Such MRI-dominated
simulations did not reach saturation of poloidal magnetic flux because
of their limited initial poloidal flux.  Yet, only once the poloidal
flux has saturated does the magnetospheric interface form.  In our
models, the QPOs are driven due to the presence of an unstable {\it
  interface} between the jet and disk where force balance has been
achieved between magnetic forces by the jet and (e.g.) ram forces by
the disk inflow.

What sets the frequency of the QPOs in our simulations?  In prior
magnetospheric QPO models, the QPO is driven as an interface
instability that oscillates at an $m$-mode-based rotational frequency
$\Omega$.  The magnetospheric interface in our simulations is between
the jet and the disk.  The strong field at the disk-jet interface
forces the plasma rotation frequency ($\Omega$) to be similar to the
field line rotational frequency $\Omega_F$, so all $\Omega$ are
similar at the interface.  The QPO frequency is thus set by the BH
spin frequency ($\Omega_{\rm H}=a/(2Mr_{\rm H})$) due to the BH
dragging the field at angular frequency $\Omega_{\rm F}\approx
\Omega_{\rm H}/4$ at the jet-disk interface, as consistent with BZ's
paraboloidal solution (see Figure~\ref{horizonflux}).

The QPO frequencies seen in Figures~\ref{spec},\ref{fft} correspond to
$|m|=1,2,3$ modes at the jet-disk interface.  That is, $\Omega_{\rm
  QPO}\approx m\Omega_{\rm F}\approx m\Omega_{\rm H}/4$.  The $|m|=1$
mode dominates, as shown in Figure~\ref{powervsm}.  So, for example,
the dominant QPO is due to the $|m|=1$ mode that in the fiducial
simulation has period $\tau\sim 70r_g/c$, which agrees with
$\tau\approx 2\pi/(m\Omega_{\rm H}/4)\approx 72.3r_g/c$ for $m=1$ and
$a=0.9375$ from the above analysis.  Other BH spins
(e.g. $|a/M|\lesssim 0.4$) might saturate at $\Omega_{\rm QPO}\sim
m\Omega$ at the (e.g.) ISCO due to the disk dominating the
plasma rotation rate near the disk-jet interface for such models
\citep{mg04,mck05}.

In summary, the period $\tau\sim 70r_g/c$ for $a/M\sim 0.9$
for our $|m|=1$ mode (with longer periods expected for lower BH spins)
is consistent with the range of HFQPOs observed in BH x-ray binaries.
Also, the disk-jet interface (driving the QPO) harbors a large
electromagnetic energy density, so the disk-jet interface can dominate
(non-thermal) synchrotron emission.  Our work shows that coherent HFQPOs may be
initiated by large-scale magnetic flux near the BH, but more work is
required to test their observability.

\section{Dependence Upon Field Geometry and Strength, Spin, and Resolution}
\label{sec:tables}

This section provides results for all our models.  We also take this
opportunity to tabulate results from the brief letters by
\citet{mb09,tnm11}.  Quantities are time-averaged from $T^a_i$ to
$T^a_f$ given in Table~\ref{tbl2}.  The chosen $T^a_i$ is at least
after the dimensionless fluxes through the horizon have reached a
quasi-steady-state.  Typically $T^a_f=T_f$.  Units in tables are
$GM=c=r_g=1$.  Such tables are computed in other works
\citep{mg04,dhkh05,hk06}, and our tables should prove useful for
comparisons with future works.

Summarizing our MB09 models: Results are similar to most prior MHD
simulations of moderately thick disks.  The flows are dominated by the
local MRI, nearly Keplerian, and not efficient.  For MB09D,
$T^a_f=3000$, after which turbulence decays.  The model MB09D focused
on the large-scale jet that is only moderately affected by how the
disk behaves at late time, and the jet's correlation lengths are
well-resolved beyond the horizon.  Unlike our other models, a color
plot (not shown) of $b_\phi$ or $b^2$ vs. $t$ and $\theta$ shows a
butterfly type LFQPO behavior with a period of roughly $15$--$30$
times the orbital period at $r=4r_g$ or $8r_g$, as similar to fig.~10
in \citet{gg11} and figs.~8 and~12 in \citet{dsp10}.

Summarizing our ``thinner disk'' TNM11 models: Results are similar to
the thick disk poloidal field models.  Plotting (not shown) all prior
figures, some similarities and differences are notable.  The TNM11
models are more Keplerian due to $\theta^t<1$.  The efficiencies are
$\eta_{\rm j}\sim 100\%$ for $a/M\ge 0.9$ and $\eta_{\rm mw,w}\sim
30\%$ by $r=100r_g$ for $a/M=0.99$.  On the horizon,
$\Omega_F/\Omega_H$ vs. $\theta$ more closely follows the profile
expected for a paraboloidal field due to the lower numerical density
floors.  $\Qtwo\lesssim 1/2$ inside $r\sim 20r_g$ (dependent upon
initial $\beta$ and simulation duration), indicating that the MRI is
suppressed over smaller radii. The value of $\alpha$ at $r\sim 10r_g$
is larger.  Compared to the thick disk models, the density-weighted
inflow is even more weakly magnetized (i.e. $b^2/\rho_0\ll 1$ and
$\beta\gg 1$).  Also, $\epsilon\sim 0.05$--$0.1$, as used in
Eq.~(\ref{rmsimple}).  No coherent HFQPOs are seen in the thinner disk
models, but the quantities (e.g. $\eta_{\rm H}$) are much more
variable.  The lack of HFQPOs and the higher variability may be due to
the disk being thinner.  However, for some of these models, this could
also be due to lower $N_\theta,N_\phi$ available for capturing
magnetic Rayleigh-Taylor modes, e.g., the azimuthal correlation length
is only marginally resolved with $Q_{m,\rm cor}\le 6$. (However, the
time-averaged quantities are actually well-resolved.)

\subsection{Disk Thickness, Disk-Corona, and Corona-Jet Interfaces}\label{sec:thickness}

\begin{table*}
\caption{Grid Cells across Half-Thickness at Horizon, Half-Thicknesses of Disk, and Location for Interfaces for Disk-Corona and Corona-Jet}
\begin{center}
\begin{tabular}[h]{|l|r|r|r|r|r|r|r|r|r|r|r|r|r|r|r|}
\hline
ModelName         &               $N^d_{\theta,{\rm{}H}}$  &    $\theta^d_{\rm{}H}$  &      $\theta^d_{5}$  &     $\theta^d_{20}$  &     $\theta^d_{100}$  &     $\theta^t_{\rm{}20}$  &     $\theta^{dc}_{\rm{}H}$  &      $\theta^{dc}_{5}$  &     $\theta^{dc}_{20}$  &     $\theta^{dc}_{100}$  &     $\theta^{cj}_{\rm{}H}$  &      $\theta^{cj}_{5}$  &     $\theta^{cj}_{20}$  &     $\theta^{cj}_{100}$  \\
\hline
{\bf              A0.94BfN40}     &                        10   &                    0.061  &               0.17  &                0.53  &                 0.66  &                     1.4   &                       0.089  &                  0.22  &                   1     &                    1.4   &                       0.089  &                  0.3   &                   1     &                    1.4    \\  
A0.94BfN100c1     &               8.4                      &    0.1                  &      0.15            &     0.52             &     0.67              &     1.4                   &     0.067                   &      0.2                &     0.99                &     1.4                  &     0.08                    &      0.27               &     1                   &     1.5                  \\     
A0.94BfN100c2     &               6.5                      &    0.08                 &      0.16            &     0.52             &     0.67              &     1.4                   &     0.057                   &      0.21               &     1                   &     1.4                  &     0.069                   &      0.29               &     1                   &     1.5                  \\     
A0.94BfN100c3     &               8.6                      &    0.11                 &      0.16            &     0.53             &     0.67              &     1.4                   &     0.052                   &      0.2                &     1                   &     1.4                  &     0.078                   &      0.28               &     1.1                 &     1.5                  \\     
A0.94BfN100c4     &               9.5                      &    0.12                 &      0.14            &     0.52             &     0.67              &     1.4                   &     0.043                   &      0.2                &     0.99                &     1.5                  &     0.067                   &      0.26               &     1                   &     1.5                  \\     
A0.94BfN40c5$^*$  &               2.2                      &    0.015                &      0.2             &     0.51             &     0.67              &     1.5                   &     0.026                   &      0.23               &     1.1                 &     1.5                  &     0.03                    &      0.25               &     1.1                 &     1.5                  \\     
\\
{\bf              A-0.94BfN40HR}  &                        10   &                    0.061  &               0.16  &                0.53  &                 0.66  &                     1.4   &                       0.078  &                  0.22  &                   1     &                    1.4   &                       0.08   &                  0.3   &                   1     &                    1.4    \\  
A-0.94BfN30       &               10                       &    0.13                 &      0.23            &     0.53             &     0.67              &     1.4                   &     0.096                   &      0.27               &     1                   &     1.4                  &     0.1                     &      0.43               &     1.1                 &     1.5                  \\     
A-0.5BfN30        &               3                        &    0.043                &      0.28            &     0.51             &     0.66              &     1.3                   &     0.03                    &      0.14               &     0.87                &     1.4                  &     0.039                   &      0.22               &     1.1                 &     1.5                  \\     
A0.0BfN10         &               23                       &    0.56                 &      0.56            &     0.62             &     0.67              &     1.4                   &     0.13                    &      0.21               &     1.1                 &     0                    &     0.26                    &      0.75               &     0                   &     0                    \\     
A0.5BfN30         &               4.3                      &    0.059                &      0.31            &     0.53             &     0.67              &     1.3                   &     0.036                   &      0.13               &     0.9                 &     1.4                  &     0.047                   &      0.23               &     1.1                 &     1.5                  \\     
A0.94BfN30        &               8.9                      &    0.11                 &      0.14            &     0.51             &     0.66              &     1.4                   &     0.067                   &      0.19               &     0.97                &     1.4                  &     0.076                   &      0.26               &     1                   &     1.4                  \\     
A0.94BfN30r       &               8.4                      &    0.1                  &      0.14            &     0.51             &     0.66              &     1.4                   &     0.065                   &      0.19               &     0.96                &     1.4                  &     0.072                   &      0.26               &     1                   &     1.5                  \\     
\\
A0.94BpN100       &               9.3                      &    0.12                 &      0.16            &     0.51             &     0.65              &     1.4                   &     0.09                    &      0.21               &     0.96                &     1.4                  &     0.1                     &      0.29               &     1                   &     1.4                  \\     
\\
A-0.94BtN10       &               24                       &    0.58                 &      0.63            &     0.66             &     0.66              &     1.1                   &     0.47                    &      0.23               &     0.00065             &     0                    &     0.19                    &      0.00049            &     0                   &     0                    \\     
A-0.5BtN10        &               23                       &    0.57                 &      0.61            &     0.64             &     0.65              &     0.99                  &     0.57                    &      0.4                &     0.03                &     0                    &     0.21                    &      0.00041            &     0                   &     0                    \\     
A0.0BtN10         &               23                       &    0.59                 &      0.63            &     0.66             &     0.66              &     1.1                   &     0.58                    &      0.45               &     0.02                &     0.00076              &     0.38                    &      0.016              &     0                   &     0                    \\     
A0.5BtN10         &               23                       &    0.6                  &      0.64            &     0.66             &     0.67              &     1.2                   &     0.59                    &      0.56               &     0.15                &     0.00038              &     0.54                    &      0.22               &     0.019               &     0                    \\     
A0.94BtN10        &               23                       &    0.53                 &      0.63            &     0.66             &     0.65              &     1.1                   &     0.5                     &      0.38               &     0.16                &     0.026                &     0.4                     &      0.13               &     0.055               &     0                    \\     
{\bf              A0.94BtN10HR}   &                        47   &                    0.57   &               0.64  &                0.67  &                 0.66  &                     1.2   &                       0.6    &                  0.63  &                   0.44  &                    0.62  &                       0.59   &                  0.41  &                   0.29  &                    0.034  \\  
\\
MB09D             &               5.5                      &    0.1                  &      0.21            &     0.25             &     0.55              &     0.3                   &     0.11                    &      0.55               &     0.68                &     0.63                 &     0.31                    &      0.76               &     1.1                 &     1.3                  \\     
MB09Q             &               12                       &    0.21                 &      0.3             &     0.36             &     0.37              &     0.4                   &     0.28                    &      0.54               &     0.58                &     0.62                 &     0.78                    &      1.2                &     1.3                 &     1.1                  \\     
\\
A-0.9N100         &               13                       &    0.17                 &      0.17            &     0.3              &     0.46              &     0.42                  &     0.17                    &      0.16               &     0.41                &     0.51                 &     0.2                     &      0.34               &     0.92                &     1.3                  \\     
A-0.5N100         &               6.1                      &    0.08                 &      0.13            &     0.28             &     0.47              &     0.46                  &     0.094                   &      0.12               &     0.36                &     0.53                 &     0.11                    &      0.25               &     0.86                &     1.3                  \\     
A-0.2N100         &               3.5                      &    0.048                &      0.1             &     0.25             &     0.48              &     0.43                  &     0.055                   &      0.08               &     0.32                &     0.63                 &     0.066                   &      0.18               &     0.85                &     1.3                  \\     
A0.0N100          &               4.4                      &    0.06                 &      0.13            &     0.29             &     0.48              &     0.47                  &     0.06                    &      0.11               &     0.32                &     0.6                  &     0.076                   &      0.23               &     0.91                &     1.4                  \\     
A0.1N100          &               4.9                      &    0.066                &      0.13            &     0.3              &     0.51              &     0.47                  &     0.053                   &      0.097              &     0.31                &     0.45                 &     0.072                   &      0.22               &     0.88                &     1.3                  \\     
A0.2N100          &               4.2                      &    0.057                &      0.12            &     0.29             &     0.49              &     0.51                  &     0.053                   &      0.1                &     0.3                 &     0.25                 &     0.068                   &      0.21               &     0.82                &     1.3                  \\     
A0.5N100          &               3                        &    0.042                &      0.11            &     0.26             &     0.51              &     0.46                  &     0.041                   &      0.1                &     0.32                &     0.36                 &     0.055                   &      0.21               &     0.79                &     1.3                  \\     
A0.9N25           &               5.2                      &    0.068                &      0.15            &     0.33             &     0.46              &     0.74                  &     0.068                   &      0.15               &     0.36                &     0.53                 &     0.081                   &      0.27               &     0.83                &     1.3                  \\     
A0.9N50           &               5.9                      &    0.077                &      0.17            &     0.3              &     0.47              &     0.59                  &     0.075                   &      0.16               &     0.33                &     0.62                 &     0.09                    &      0.3                &     0.85                &     1.3                  \\     
A0.9N100          &               5.3                      &    0.07                 &      0.16            &     0.29             &     0.47              &     0.51                  &     0.068                   &      0.16               &     0.31                &     0.44                 &     0.083                   &      0.29               &     0.86                &     1.3                  \\     
A0.9N200          &               5.2                      &    0.068                &      0.15            &     0.3              &     0.44              &     0.49                  &     0.066                   &      0.15               &     0.39                &     0.55                 &     0.08                    &      0.29               &     0.87                &     1.3                  \\     
{\bf              A0.99N100}      &                        8.2  &                    0.11   &               0.19  &                0.33  &                 0.45  &                     0.59  &                       0.1    &                  0.2   &                   0.38  &                    0.56  &                       0.11   &                  0.34  &                   0.88  &                    1.3    \\  
\hline
\hline
\end{tabular}
\end{center}
\label{tbl3}
\end{table*}

Table~\ref{tbl3} shows the number of grid cells ($N^d_{\theta, \rm
  H}$) at the horizon that cover the angular span of the disk's
geometric half-angular thickness ($\theta^d_{\rm H}$) ($\theta^d$ is
computed by Eq.~(\ref{thicknesseq})), the disk thickness at the
horizon ($\theta^d_{\rm H}$) and other radii $r$ ($\theta^d_{r/r_g}$),
the thermal half-angular thickness ($\theta^t_{20}$ is
$\theta^t_{\rholab}$ at $r=20r_g$ as computed by
Eq.~(\ref{thetateq})), the disk-corona interface angular locations at
the same radii ($\theta^{dc}_{r/r_g}$, defined by where $\beta=1$),
and the corona-jet interface angular locations at the same radii
($\theta^{cj}_{r/r_g}$, defined by where $b^2/\rho_0=1$).  See
section~\ref{integrations} for details.  In summary, jets in the
thicker disk models collimate better.

The rest-mass density geometric half-angular thickness ($\theta^d$) is
quite different than the thermal half-angular thickness
($\theta^t\equiv \arctan{(c_s/v_{\rm rot})}$), the latter being an
estimate of $\theta^d$ if gas pressure balances vertical gravity.
Close to the BH, $\theta^d\ll \theta^t$ is controlled by magnetic
forces, and $\theta^t\sim \theta^d$ only at large radii.  About $6$
cells should span the full disk to resolve it at a basic level, and
most 3D models satisfy this even on the horizon ($\sim 20$ cells for
the fiducial model).  The ``magnetospheric radius'' ($r_m$) is not a
sharp boundary, and instead magnetic forces gradually compress the
disk.

\subsection{Mass Accretion and Ejection Rates}\label{sec:mdot}

\begin{table*}
\caption{Rest-Mass Accretion and Ejection Rates}
\begin{center}
\begin{tabular}[h]{|l|r|r|r|r|r|r|r|r|r|}
\hline
ModelName         &               $\dot{M}_{\rm{}H}$  &    $\dot{M}_{\rm{}in,i}-\dot{M}_{\rm{}H}$  &      $\dot{M}_{\rm{}in,o}-\dot{M}_{\rm{}H}$  &    $\dot{M}_{\rm{}j}$  &       $\dot{M}_{\rm{}mw,i}$  &    $\dot{M}_{\rm{}mw,o}$  &     $\dot{M}_{\rm{}w,i}$  &    $\dot{M}_{\rm{}w,o}$  \\
\hline
{\bf              A0.94BfN40}     &                   48   &                                       20     &                                       250  &                   6.9     &                      6.8  &                      17    &                     19   &                     250  \\  
A0.94BfN100c1     &               37                  &    16                                      &      170                                     &    6.2                 &       4.6                    &    14                     &     15                    &    170                   \\   
A0.94BfN100c2     &               38                  &    15                                      &      210                                     &    6.6                 &       5.1                    &    20                     &     13                    &    210                   \\   
A0.94BfN100c3     &               39                  &    17                                      &      200                                     &    6.5                 &       6.3                    &    20                     &     15                    &    190                   \\   
A0.94BfN100c4     &               38                  &    18                                      &      180                                     &    5.4                 &       6.4                    &    16                     &     16                    &    190                   \\   
A0.94BfN40c5$^*$  &               4                   &    15                                      &      280                                     &    1.7                 &       4.3                    &    11                     &     12                    &    280                   \\   
\\
{\bf              A-0.94BfN40HR}  &                   49   &                                       21     &                                       250  &                   6.6     &                      6.5  &                      18    &                     20   &                     240  \\  
A-0.94BfN30       &               98                  &    20                                      &      250                                     &    6.8                 &       8.6                    &    26                     &     17                    &    250                   \\   
A-0.5BfN30        &               110                 &    6.8                                     &      220                                     &    2e-6                &       0.53                   &    7.2                    &     3.8                   &    220                   \\   
A0.0BfN10         &               160                 &    3.6                                     &      62                                      &    0                   &       0.0049                 &    0.033                  &     5.1                   &    100                   \\   
A0.5BfN30         &               110                 &    7.6                                     &      140                                     &    0                   &       0.2                    &    3.6                    &     4.3                   &    140                   \\   
A0.94BfN30        &               41                  &    18                                      &      160                                     &    7.6                 &       6.4                    &    21                     &     15                    &    150                   \\   
A0.94BfN30r       &               41                  &    17                                      &      170                                     &    5.7                 &       6.4                    &    19                     &     15                    &    160                   \\   
\\
A0.94BpN100       &               46                  &    21                                      &      210                                     &    9.6                 &       7.2                    &    24                     &     19                    &    200                   \\   
\\
A-0.94BtN10       &               280                 &    1.9                                     &      230                                     &    0                   &       0.0044                 &    1.7e-6                 &     1                     &    230                   \\   
A-0.5BtN10        &               93                  &    2.2                                     &      84                                      &    0                   &       0.017                  &    0.0007                 &     2                     &    86                    \\   
A0.0BtN10         &               180                 &    5                                       &      220                                     &    0                   &       0.089                  &    0.00083                &     4.1                   &    220                   \\   
A0.5BtN10         &               200                 &    5.5                                     &      180                                     &    0                   &       0.29                   &    0.043                  &     4.6                   &    180                   \\   
A0.94BtN10        &               190                 &    16                                      &      420                                     &    0.00095             &       0.46                   &    0.16                   &     14                    &    410                   \\   
{\bf              A0.94BtN10HR}   &                   330  &                                       -0.83  &                                       410  &                   0.0085  &                      1.6  &                      0.59  &                     9.6  &                     440  \\  
\\
MB09D             &               0.082               &    0.62                                    &      -0.016                                  &    0.0002              &       0.0011                 &    0.015                  &     0.77                  &    0.35                  \\   
MB09Q             &               0.94                &    0.65                                    &      0.13                                    &    3.1e-5              &       0.02                   &    0.061                  &     0.74                  &    1.9                   \\   
\\
A-0.9N100         &               18                  &    1.6                                     &      18                                      &    0.32                &       0.76                   &    7.2                    &     1.4                   &    20                    \\   
A-0.5N100         &               15                  &    1.4                                     &      21                                      &    0.15                &       0.54                   &    6.8                    &     1.3                   &    25                    \\   
A-0.2N100         &               15                  &    1.3                                     &      19                                      &    0.041               &       0.31                   &    5.5                    &     1.1                   &    23                    \\   
A0.0N100          &               17                  &    1.3                                     &      14                                      &    0.015               &       0.4                    &    6.7                    &     1.2                   &    19                    \\   
A0.1N100          &               14                  &    1.3                                     &      14                                      &    0.028               &       0.36                   &    5.7                    &     1.2                   &    19                    \\   
A0.2N100          &               15                  &    1.3                                     &      14                                      &    0.063               &       0.35                   &    5.8                    &     1.2                   &    19                    \\   
A0.5N100          &               13                  &    1.8                                     &      16                                      &    0.27                &       0.72                   &    8                      &     1.7                   &    21                    \\   
A0.9N25           &               13                  &    4.3                                     &      20                                      &    0.8                 &       2.2                    &    15                     &     3.7                   &    22                    \\   
A0.9N50           &               16                  &    5.6                                     &      23                                      &    0.88                &       3                      &    16                     &     4.9                   &    25                    \\   
A0.9N100          &               11                  &    3.9                                     &      21                                      &    0.61                &       2                      &    11                     &     3.5                   &    24                    \\   
A0.9N200          &               7.5                 &    2.8                                     &      16                                      &    0.5                 &       1.5                    &    8.6                    &     2.4                   &    17                    \\   
{\bf              A0.99N100}      &                   9.8  &                                       3.7    &                                       14   &                   0.4     &                      2.2  &                      11    &                     3.3  &                     16   \\  
\hline
\hline
\end{tabular}
\end{center}
\label{tbl5}
\end{table*}

Table~\ref{tbl5} shows $\dot{M}$ computed via the first row of
Eq.~(\ref{Dotsmej}) through the BH horizon ($\dot{M}_{\rm
  H}$), inflow-only mass accretion rate ($\dot{M}_{\rm in}$) at an
inner radius of $r_i=10r_g$ giving $\dot{M}_{\rm in,i}$ and an outer
radius of $r_o=50r_g$ (except model MB09Q that uses $r_0=30r_g$ due to
its limited radial domain) giving $\dot{M}_{\rm in,o}$.  Also shown
are $\dot{M}$ through the jet ($\dot{M}_j$), the magnetized wind at
$r_i$ ($\dot{M}_{\rm mw,i}$) and $r_o$ ($\dot{M}_{\rm mw,o}$), and the
entire wind at the same radii ($\dot{M}_{\rm w,i}$ and $\dot{M}_{\rm
  w,o}$).

Most models show massive winds, and the poloidal models show some mass
carried in the magnetized wind and jet.  Our models are in
inflow-outflow equilibrium with $\dot{M}_{\rm in,o}-\dot{M}_{\rm
  H}\approx \dot{M}_{\rm w,o}+\dot{M}_{\rm j}$ (i.e. all outflowing material, that is
not in the jet, is part of the ``entire wind'').  Little mass is
accreted in the 2D poloidal models because magnetic flux accumulates
and fully (except through periodic reconnection events) suspends the
inflow.

\subsection{Energy Efficiency of Hole, Jet, and Winds}\label{sec:edot1}

\begin{table*}
\caption{Percent Energy Efficiency: BH, Jet, Winds, and NT}
\begin{center}
\begin{tabular}[h]{|l|r|r|r|r|r|r|r|r|r|r|r|r|}
\hline
ModelName         &               $\eta_{\rm{}H}$  &      $\eta^{\rm{}EM}_{\rm{}H}$  &       $\eta^{\rm{}MAKE}_{\rm{}H}$  &      $\eta^{\rm{}PAKE}_{\rm{}H}$  &     $\eta^{\rm{}EN}_{\rm{}H}$  &      $\eta_{\rm{}j}$  &      $\eta^{\rm{}EM}_j$  &      $\eta^{\rm{}MAKE}_{\rm{}j}$  &      $\eta_{\rm{}mw,o}$  &      $\eta_{\rm{}w,o}$  &      $\eta_{\rm{}NT}$  \\
\hline
{\bf              A0.94BfN40}     &                235    &                          264     &                            -29.2  &                            63.4  &                          -92.5  &                242    &                   228    &                            13.5   &                   27.9   &                  -6.48  &                 17.9  \\  
A0.94BfN100c1     &               323              &      379                        &       -56.1                        &      83.1                         &     -139                       &      333              &      322                 &      11                           &      26.4                &      -7.21              &      17.9              \\    
A0.94BfN100c2     &               298              &      340                        &       -42.3                        &      74.2                         &     -116                       &      304              &      293                 &      11.3                         &      34.1                &      -4.45              &      17.9              \\    
A0.94BfN100c3     &               295              &      337                        &       -41.7                        &      74.3                         &     -116                       &      299              &      288                 &      10.7                         &      35.5                &      -2.95              &      17.9              \\    
A0.94BfN100c4     &               323              &      368                        &       -44.1                        &      73.2                         &     -117                       &      323              &      313                 &      9.44                         &      31.6                &      4.21               &      17.9              \\    
A0.94BfN40c5$^*$  &               1170             &      1240                       &       -65.5                        &      86.4                         &     -152                       &      945              &      925                 &      19.9                         &      109                 &      205                &      17.9              \\    
\\
{\bf              A-0.94BfN40HR}  &                244    &                          274     &                            -30.4  &                            64.3  &                          -94.7  &                251    &                   238    &                            12.1   &                   27.5   &                  -7.01  &                 3.85  \\  
A-0.94BfN30       &               87.8             &      122                        &       -34.3                        &      73.7                         &     -108                       &      101              &      95.3                &      5.38                         &      15.6                &      -12.6              &      3.85              \\    
A-0.5BfN30        &               -13.2            &      39.4                       &       -52.7                        &      100                          &     -153                       &      -1.51            &      3.27                &      -4.78                        &      1.15                &      -13.7              &      4.51              \\    
A0.0BfN10         &               -17              &      -0.114                     &       -16.9                        &      24.1                         &     -41                        &      0                &      0                   &      0                            &      -5.94               &      -12.2              &      5.72              \\    
A0.5BfN30         &               -17.1            &      38.8                       &       -55.9                        &      85.8                         &     -142                       &      -0.796           &      5.59                &      -6.38                        &      -6.9                &      -18                &      8.21              \\    
A0.94BfN30        &               353              &      420                        &       -67.1                        &      88.6                         &     -156                       &      361              &      348                 &      12.8                         &      32.6                &      -11.4              &      17.9              \\    
A0.94BfN30r       &               323              &      386                        &       -62.7                        &      86.3                         &     -149                       &      334              &      326                 &      8.37                         &      26.1                &      -14.4              &      17.9              \\    
\\
A0.94BpN100       &               238              &      291                        &       -52.8                        &      83.8                         &     -137                       &      241              &      220                 &      20.5                         &      40.6                &      1.94               &      17.9              \\    
\\
A-0.94BtN10       &               -17.8            &      -0.703                     &       -17.1                        &      24.7                         &     -41.8                      &      0                &      0                   &      0                            &      -9e-7               &      -17.6              &      3.85              \\    
A-0.5BtN10        &               -13.6            &      -0.287                     &       -13.3                        &      19.8                         &     -33.1                      &      0                &      0                   &      0                            &      -7e-5               &      -13.2              &      4.51              \\    
A0.0BtN10         &               -18.7            &      -0.077                     &       -18.7                        &      21.9                         &     -40.6                      &      0                &      0                   &      0                            &      -0.006              &      -18.7              &      5.72              \\    
A0.5BtN10         &               -19.2            &      -0.246                     &       -19                          &      26.8                         &     -45.8                      &      0                &      0                   &      0                            &      -0.018              &      -19                &      8.21              \\    
A0.94BtN10        &               -22.8            &      -0.162                     &       -22.7                        &      33.2                         &     -55.9                      &      0.0008           &      0.0004              &      0.0004                       &      0.033               &      -23.4              &      17.9              \\    
{\bf              A0.94BtN10HR}   &                -28.2  &                          -0.569  &                            -27.6  &                            8.71  &                          -36.3  &                0.007  &                   0.006  &                            0.002  &                   0.032  &                  -19.5  &                 17.9  \\  
\\
MB09D             &               14.4             &      3.08                       &       11.4                         &      24.9                         &     -13.5                      &      2.96             &      2.7                 &      0.269                        &      0.485               &      -0.739             &      16.7              \\    
MB09Q             &               5.19             &      0.199                      &       4.99                         &      22.9                         &     -17.9                      &      0.008            &      0.004               &      0.003                        &      0.431               &      0.576              &      17.9              \\    
\\
A-0.9N100         &               32.3             &      39.8                       &       -7.51                        &      74.6                         &     -82.2                      &      23.4             &      20.8                &      2.61                         &      8.49                &      9.59               &      3.9               \\    
A-0.5N100         &               17.2             &      11.4                       &       5.77                         &      41.7                         &     -36                        &      9.4              &      8.69                &      0.706                        &      7.18                &      8.03               &      4.51              \\    
A-0.2N100         &               6.64             &      0.906                      &       5.73                         &      34.5                         &     -28.8                      &      1.12             &      1.04                &      0.079                        &      4.78                &      5.38               &      5.15              \\    
A0.0N100          &               4.51             &      -0.884                     &       5.4                          &      33.7                         &     -28.3                      &      0.052            &      0.035               &      0.017                        &      3.78                &      4.23               &      5.72              \\    
A0.1N100          &               5.04             &      -0.242                     &       5.29                         &      32.4                         &     -27.1                      &      0.342            &      0.292               &      0.051                        &      4.16                &      4.5                &      6.06              \\    
A0.2N100          &               7.34             &      1.89                       &       5.46                         &      34.9                         &     -29.4                      &      1.73             &      1.64                &      0.088                        &      5.04                &      5.38               &      6.46              \\    
A0.5N100          &               30.8             &      24.3                       &       6.48                         &      38.1                         &     -31.6                      &      19.6             &      18.3                &      1.25                         &      10.3                &      10.8               &      8.21              \\    
A0.9N25           &               113              &      112                        &       1.46                         &      53.3                         &     -51.8                      &      88.8             &      80.3                &      8.55                         &      24.1                &      25.7               &      15.6              \\    
A0.9N50           &               102              &      99.3                       &       2.91                         &      51.3                         &     -48.3                      &      77.7             &      70.3                &      7.47                         &      24.4                &      24.8               &      15.6              \\    
A0.9N100          &               101              &      97.2                       &       4.19                         &      50                           &     -45.8                      &      77.6             &      70.1                &      7.44                         &      24.1                &      24.7               &      15.6              \\    
A0.9N200          &               121              &      117                        &       4.28                         &      52                           &     -47.7                      &      96.1             &      87.5                &      8.67                         &      25.1                &      26                 &      15.6              \\    
{\bf              A0.99N100}      &                130    &                          127     &                            2.89   &                            64.4  &                          -61.5  &                103    &                   95.5   &                            7.41   &                   26.4   &                  27     &                 26.4  \\  
\hline
\hline
\end{tabular}
\end{center}
\label{tbl6}
\end{table*}

Table~\ref{tbl6} shows the flow efficiency computed via
Eq.~(\ref{eff}) (see section~\ref{integrations}), where
$\eta=\eta^{\rm EM}+\eta^{\rm MAKE}$ and $\eta^{\rm MAKE}=\eta^{\rm
  PAKE}+\eta^{\rm EN}$.  These efficiencies are computed at the
horizon ($\eta_{\rm H}$), for the jet ($\eta_{\rm j}$), for the
magnetized wind ($\eta_{\rm mw}$), and for the entire wind ($\eta_{\rm
  w}$).  The winds are measured at $r_i$ and $r_o$ given above, while
the jet is measured at $r_o$.  The radial asymptotic efficiency is
$\eta_\infty\sim \eta_{\rm j} + \eta_{\rm mw,o}$, because these are
unbound outflows with roughly constant $\eta$ by the measured radius.
BH and jet efficiencies are shown decomposed into EM, MAKE, PAKE, and EN terms.
The radiative efficiency ($\eta_{\rm NT}$) for the Novikov-Thorne (NT)
model is shown for comparison.

Many models with $|a/M|\gtrsim 0.9$ show greater than 100\% efficiency
(and up to about $300\%$ for the 3D models) for the BH energy
extraction and jet at larger radii.  This is much higher than the
efficiencies of order the NT efficiency seen in other 3D GRMHD
simulations that start off with limited poloidal magnetic flux
(e.g. \citealt{hk06}). In the thick disk models, the MAKE efficiency
is negative even though the particle term ($\eta^{\rm PAKE}_{\rm
  H}\sim -\rho_0 \uvec_t$) is positive.  This negative $\eta_{\rm
  H}^{\rm MAKE}$ is due to a high specific enthalpy (i.e. $\eta^{\rm
  EN}_{\rm H}\propto(u_g + p_g)/\rho_0\gtrsim 1$).  The negative
$\eta_{\rm H}^{\rm MAKE}$ means the BH is accreting thermo-kinetically
unbound material as can occur in a Bondi flows or does occur in ADAFs
\citep{nar94}.  For the thick disk models, part of the negative
$\eta_{\rm H}^{\rm MAKE}$ is due the initial torus being marginally
unbound, which contributes $\sim -26\%$ to $\eta_{\rm H}^{\rm MAKE}$.

The dependence upon initial $\beta$ is relatively weak for the thick
disk models.  $\beta$ varies by factors of $2$--$3$, while the
efficiency $\eta\propto B_r^2$ varies by much less than factors of
$2$--$3$ expected if the initial $\beta\propto 1/B_r^2$ completely
controlled the flux on the BH.  $\Upsilon\propto B_r$ (discussed
later) shows even less dispersion vs. initial $(\beta)^{-1/2}$.
Similar insensitivity to initial $\beta$ is seen in \citet{ina03}.

Also, notice there are very similar results between the flipping and
non-flipping poloidal field, which shows that the flipping models
have plenty of constant polarity flux unlike the MB09D model.

The 2D axisymmetric simulations (for all thick disk poloidal cases,
but only showing A0.94BfN40c5* in tables) show relatively higher
efficiencies than otherwise similar 3D models.  Mass is not accreted
except during infrequent penetrations of the magnetic barrier via
magnetic reconnection.

Despite the presence of well-ordered poloidal field near the BH, we
see no evidence for significant energy extraction by any ergospheric
type disk threaded by magnetic flux that would convert spin energy to
MA energy in the disk and then to EM energy out in the magnetic field
\citep{pc90} unlike suggested to be present sometimes in other
simulations \citep{pih09}.  Such an effect would lead to $\eta^{\rm
  MAKE}_{\rm H}\gg 0$ (i.e. outgoing MAKE energy on the horizon).  For
this effect to be dynamically important, one should find $\eta^{\rm
  MAKE}_{\rm H}\gtrsim \eta^{\rm EM}_{\rm H}$, while in all cases we
find $\eta^{\rm MAKE}_{\rm H}\ll \eta^{\rm EM}_{\rm H}$ except MB09
models give $\eta^{\rm MAKE}_{\rm H}\gtrsim \eta^{\rm EM}_{\rm H}$.
However, in the MB09 models all of that horizon MAKE energy is
dissipated in the disk. So in all cases, the jet power is dominated by
the EM term, i.e. $\eta^{\rm EM}_{\rm H}\sim \eta_{\rm j}$.  In
summary, we find that the energy reaching large radii is dominated by
the EM power produced via the magnetic flux penetrating the horizon as
in the BZ mechanism.

\subsection{Energy Efficiency of Magnetized and Entire Winds}\label{sec:edot2}

\begin{table*}
\caption{Percent Energy Efficiency: Magnetized Wind and Entire Wind}
\begin{center}
\begin{tabular}[h]{|l|r|r|r|r|r|r|r|r|r|r|r|r|r|}
\hline
ModelName         &               $\eta_{\rm{}mw,i}$  &      $\eta^{\rm{}EM}_{\rm{}mw,i}$  &      $\eta^{\rm{}MAKE}_{\rm{}mw,i}$  &      $\eta_{\rm{}mw,o}$  &      $\eta^{\rm{}EM}_{\rm{}mw,o}$  &     $\eta^{\rm{}MAKE}_{\rm{}mw,o}$  &       $\eta_{\rm{}w,i}$  &      $\eta^{\rm{}EM}_{\rm{}w,i}$  &      $\eta^{\rm{}MAKE}_{\rm{}w,i}$  &      $\eta_{\rm{}w,o}$  &      $\eta^{\rm{}EM}_{\rm{}w,o}$  &      $\eta^{\rm{}MAKE}_{\rm{}w,o}$  \\
\hline
{\bf              A0.94BfN40}     &                   7.49   &                             5.9    &                               1.59   &                   27.9   &                             13    &                               14.9    &                  -20.1  &                            6.66   &                              -26.8  &                  -6.48  &                            13.5   &                              -20    \\  
A0.94BfN100c1     &               8.89                &      6.5                           &      2.39                            &      26.4                &      12.3                          &     14.2                            &       -19.9              &      7.24                         &      -27.1                          &      -7.21              &      8.16                         &      -15.4                          \\     
A0.94BfN100c2     &               9.39                &      7.36                          &      2.03                            &      34.1                &      14.6                          &     19.5                            &       -18.3              &      8.14                         &      -26.5                          &      -4.45              &      9.99                         &      -14.4                          \\     
A0.94BfN100c3     &               11.6                &      8.07                          &      3.57                            &      35.5                &      15.4                          &     20.1                            &       -17.8              &      8.82                         &      -26.6                          &      -2.95              &      11.6                         &      -14.5                          \\     
A0.94BfN100c4     &               11.9                &      6.98                          &      4.92                            &      31.6                &      14.8                          &     16.8                            &       -20                &      7.77                         &      -27.8                          &      4.21               &      12.3                         &      -8.08                          \\     
A0.94BfN40c5$^*$  &               -4.03               &      15.4                          &      -19.5                           &      109                 &      43.6                          &     64.9                            &       -33.6              &      15.9                         &      -49.5                          &      205                &      42.7                         &      163                            \\     
\\
{\bf              A-0.94BfN40HR}  &                   8.47   &                             5.57   &                               2.9    &                   27.5   &                             12.6  &                               14.9    &                  -20.4  &                            6.44   &                              -26.9  &                  -7.01  &                            13.1   &                              -20.1  \\  
A-0.94BfN30       &               3.36                &      3.71                          &      -0.346                          &      15.6                &      7.55                          &     8.01                            &       -19                &      4.04                         &      -23                            &      -12.6              &      7.62                         &      -20.2                          \\     
A-0.5BfN30        &               -1.19               &      2.1                           &      -3.29                           &      1.15                &      6.48                          &     -5.33                           &       -13.2              &      2.19                         &      -15.4                          &      -13.7              &      6.06                         &      -19.7                          \\     
A0.0BfN10         &               -13.3               &      0.068                         &      -13.4                           &      -5.94               &      0.087                         &     -6.03                           &       -16.8              &      0.087                        &      -16.9                          &      -12.2              &      0.26                         &      -12.4                          \\     
A0.5BfN30         &               -3.82               &      1.54                          &      -5.37                           &      -6.9                &      3.92                          &     -10.8                           &       -14.8              &      1.42                         &      -16.2                          &      -18                &      4.09                         &      -22.1                          \\     
A0.94BfN30        &               11.2                &      7.26                          &      3.9                             &      32.6                &      15.4                          &     17.2                            &       -19.7              &      7.89                         &      -27.6                          &      -11.4              &      16.7                         &      -28.1                          \\     
A0.94BfN30r       &               9.03                &      6.62                          &      2.41                            &      26.1                &      11.9                          &     14.2                            &       -21                &      7.14                         &      -28.1                          &      -14.4              &      13.3                         &      -27.7                          \\     
\\
A0.94BpN100       &               10.8                &      7.24                          &      3.52                            &      40.6                &      18.4                          &     22.2                            &       -16.9              &      8.23                         &      -25.1                          &      1.94               &      20.6                         &      -18.7                          \\     
\\
A-0.94BtN10       &               -0.164              &      -0.017                        &      -0.147                          &      -9e-7               &      -2e-7                         &     -7e-7                           &       -17.6              &      -0.148                       &      -17.4                          &      -17.6              &      0.019                        &      -17.7                          \\     
A-0.5BtN10        &               -0.392              &      -0.014                        &      -0.378                          &      -7e-5               &      -3e-5                         &     -4e-5                           &       -13.5              &      -0.081                       &      -13.4                          &      -13.2              &      0.04                         &      -13.2                          \\     
A0.0BtN10         &               -0.457              &      0.026                         &      -0.483                          &      -0.006              &      -9e-5                         &     -0.006                          &       -18.7              &      0.054                        &      -18.8                          &      -18.7              &      0.057                        &      -18.7                          \\     
A0.5BtN10         &               -1.36               &      0.077                         &      -1.44                           &      -0.018              &      0.003                         &     -0.022                          &       -19                &      0.098                        &      -19.1                          &      -19                &      0.072                        &      -19.1                          \\     
A0.94BtN10        &               -0.413              &      0.09                          &      -0.503                          &      0.033               &      0.012                         &     0.021                           &       -22.9              &      0.109                        &      -23                            &      -23.4              &      0.113                        &      -23.5                          \\     
{\bf              A0.94BtN10HR}   &                   -1.66  &                             0.162  &                               -1.83  &                   0.032  &                             0.04  &                               -0.008  &                  -23.1  &                            0.187  &                              -23.3  &                  -19.5  &                            0.177  &                              -19.7  \\  
\\
MB09D             &               3.15                &      3.42                          &      -0.271                          &      0.485               &      1.28                          &     -0.792                          &       5.99               &      6.18                         &      -0.184                         &      -0.739             &      1.36                         &      -2.09                          \\     
MB09Q             &               2.49                &      1.65                          &      0.839                           &      0.431               &      0.985                         &     -0.555                          &       4.84               &      2.17                         &      2.67                           &      0.576              &      1.14                         &      -0.567                         \\     
\\
A-0.9N100         &               3.97                &      2.26                          &      1.71                            &      8.49                &      3.72                          &     4.77                            &       5.41               &      2.28                         &      3.12                           &      9.59               &      3.85                         &      5.75                           \\     
A-0.5N100         &               3.57                &      2.05                          &      1.52                            &      7.18                &      3.39                          &     3.78                            &       5.3                &      2.07                         &      3.23                           &      8.03               &      3.51                         &      4.52                           \\     
A-0.2N100         &               2.85                &      1.54                          &      1.32                            &      4.78                &      2.52                          &     2.26                            &       4.82               &      1.56                         &      3.26                           &      5.38               &      2.61                         &      2.77                           \\     
A0.0N100          &               2.88                &      1.67                          &      1.21                            &      3.78                &      2.37                          &     1.41                            &       4.5                &      1.7                          &      2.79                           &      4.23               &      2.45                         &      1.78                           \\     
A0.1N100          &               2.97                &      1.73                          &      1.24                            &      4.16                &      2.78                          &     1.38                            &       4.56               &      1.75                         &      2.81                           &      4.5                &      2.82                         &      1.69                           \\     
A0.2N100          &               3.42                &      2.09                          &      1.32                            &      5.04                &      3.42                          &     1.63                            &       4.94               &      2.1                          &      2.83                           &      5.38               &      3.49                         &      1.89                           \\     
A0.5N100          &               5.77                &      4.3                           &      1.47                            &      10.3                &      6.85                          &     3.46                            &       7.45               &      4.33                         &      3.12                           &      10.8               &      6.95                         &      3.83                           \\     
A0.9N25           &               9.8                 &      7.01                          &      2.79                            &      24.1                &      12.2                          &     11.9                            &       11.7               &      7.09                         &      4.66                           &      25.7               &      12.3                         &      13.4                           \\     
A0.9N50           &               10.6                &      7.86                          &      2.78                            &      24.4                &      13.1                          &     11.3                            &       13                 &      7.94                         &      5.03                           &      24.8               &      13.1                         &      11.7                           \\     
A0.9N100          &               11                  &      8.21                          &      2.84                            &      24.1                &      13.3                          &     10.8                            &       13.4               &      8.27                         &      5.13                           &      24.7               &      13.4                         &      11.3                           \\     
A0.9N200          &               11.4                &      9.09                          &      2.26                            &      25.1                &      14.3                          &     10.8                            &       13                 &      9.22                         &      3.73                           &      26                 &      14.5                         &      11.5                           \\     
{\bf              A0.99N100}      &                   11.8   &                             8.69   &                               3.08   &                   26.4   &                             13.8  &                               12.5    &                  13.6   &                            8.79   &                              4.85   &                  27     &                            14     &                              13     \\  
\hline
\hline
\end{tabular}
\end{center}
\label{tbl7}
\end{table*}

Table~\ref{tbl7} is similar to Table~\ref{tbl6}, but the magnetized
wind (``mw'') and entire wind (``w'') efficiencies (both EM and MAKE
decompositions) are shown as evaluated at $r_i$ and $r_o$ given
earlier.  The BH and jet are dominated by EM power, the magnetized
wind has EM power similar to MAKE power, and the wind is dominated by
MAKE power (especially at larger radii).  The magnetized wind, which
can reach large radii, is fairly efficient in the poloidal models.
However, the efficiency is much less than the BH and jet efficiency.
So the BH dominates the disk's magnetized wind, unlike in some models
\citep{ga97,lop99}.

\subsection{Angular Momentum Flux for Hole, Jet,  and Winds}\label{sec:ldot12}

A table (not shown) similar to Table~\ref{tbl6}, except for the
specific angular momentum flux (as computed via the third row of
Eq.~(\ref{Dotsmej})), shows that the poloidal field models have net
extraction of angular momentum from the BH.  The jet carries
most of the angular momentum and most of that is in EM form.  The
toroidal field models have small angular momentum flux because no
steady magnetized winds or jets emerge.  A table (not shown) similar
to Table~\ref{tbl7}, except for the specific angular momentum flux,
shows that the wind's angular momentum flux follows our prior
discussions for the efficiency of the winds, except that the EM term
tends to dominate the MA term for the magnetized wind.

\subsection{Spin-Up Parameter}\label{sec:spar}

\begin{table*}
\caption{Spin-Up Parameter: BH, Jet, Winds, and NT}
\begin{center}
\begin{tabular}[h]{|l|r|r|r|r|r|r|r|r|r|r|r|r|}
\hline
ModelName         &               $s_{\rm{}H}$  &      $s^{\rm{}EM}_{\rm{}H}$  &       $s^{\rm{}MA}_{\rm{}H}$  &      $s^{\rm{}PA}_{\rm{}H}$  &       $s^{\rm{}EN}_{\rm{}H}$  &       $s_{\rm{}j}$  &      $s^{\rm{}EM}_j$  &       $s^{\rm{}MA}_{\rm{}j}$  &      $s_{\rm{}mw,o}$  &      $s_{\rm{}w,o}$  &      $s_{\rm{}NT}$  \\
\hline
{\bf              A0.94BfN40}     &             -23.5  &                       -21.1   &                       -2.37  &                       -0.609  &                       -1.76   &             -19.5  &                -16.8   &                       -2.65  &                -3.1   &               -4.03  &              0.411  \\  
A0.94BfN100c1     &               -28           &      -25                     &       -3.01                   &      -0.27                   &       -2.74                   &       -22.6         &      -20.3            &       -2.31                   &      -3.47            &      -5.67           &      0.411          \\     
A0.94BfN100c2     &               -26.6         &      -23.6                   &       -2.95                   &      -0.478                  &       -2.47                   &       -21.1         &      -18.6            &       -2.44                   &      -3.45            &      -2.92           &      0.411          \\     
A0.94BfN100c3     &               -25.7         &      -22.7                   &       -2.97                   &      -0.484                  &       -2.48                   &       -20.2         &      -17.8            &       -2.43                   &      -4.23            &      -4.44           &      0.411          \\     
A0.94BfN100c4     &               -26.8         &      -24.1                   &       -2.75                   &      -0.468                  &       -2.29                   &       -20.8         &      -18.4            &       -2.34                   &      -3.11            &      -6              &      0.411          \\     
A0.94BfN40c5$^*$  &               -111          &      -105                    &       -6.41                   &      -0.747                  &       -5.66                   &       -107          &      -103             &       -4.75                   &      -11.6            &      -67.1           &      0.411          \\     
\\
{\bf              A-0.94BfN40HR}  &             23.5   &                       21.1    &                       2.31   &                       0.573   &                       1.74    &             19.4   &                16.9    &                       2.49   &                3.19   &               7.76   &              6      \\  
A-0.94BfN30       &               11.4          &      9.54                    &       1.89                    &      0.264                   &       1.63                    &       8.81          &      6.67             &       2.14                    &      2.55             &      5.01            &      6              \\     
A-0.5BfN30        &               12.7          &      11.5                    &       1.23                    &      -0.144                  &       1.37                    &       5.62          &      4.91             &       0.714                   &      6.97             &      8.5             &      4.84           \\     
A0.0BfN10         &               0.013         &      0.032                   &       -0.019                  &      -0.009                  &       -0.009                  &       -0            &      -0               &       -0                      &      0.061            &      0.404           &      3.46           \\     
A0.5BfN30         &               -11.5         &      -10.2                   &       -1.3                    &      -0.005                  &       -1.3                    &       -5.15         &      -4.55            &       -0.599                  &      -6.54            &      -8.05           &      1.98           \\     
A0.94BfN30        &               -30.6         &      -27.5                   &       -3.08                   &      -0.162                  &       -2.92                   &       -25           &      -22.3            &       -2.62                   &      -4.06            &      -10.3           &      0.411          \\     
A0.94BfN30r       &               -28.9         &      -25.9                   &       -3                      &      -0.205                  &       -2.79                   &       -23.1         &      -20.7            &       -2.45                   &      -3.63            &      -8.54           &      0.411          \\     
\\
A0.94BpN100       &               -25           &      -22.3                   &       -2.72                   &      -0.214                  &       -2.5                    &       -21.3         &      -18.7            &       -2.56                   &      -5.07            &      -5.97           &      0.411          \\     
\\
A-0.94BtN10       &               3.29          &      0.205                   &       3.09                    &      1.98                    &       1.11                    &       1.88          &      0                &       1.88                    &      1.88             &      3.2             &      6              \\     
A-0.5BtN10        &               2.39          &      0.103                   &       2.29                    &      1.61                    &       0.678                   &       1             &      0                &       1                       &      1                &      2.18            &      4.84           \\     
A0.0BtN10         &               1.15          &      0.004                   &       1.15                    &      0.751                   &       0.396                   &       -0            &      -0               &       -0                      &      0.0004           &      1.06            &      3.46           \\     
A0.5BtN10         &               -0.639        &      -0.297                  &       -0.342                  &      -0.208                  &       -0.134                  &       -1            &      -0               &       -1                      &      -1               &      -0.793          &      1.98           \\     
A0.94BtN10        &               -1.53         &      -0.232                  &       -1.3                    &      -0.716                  &       -0.584                  &       -1.88         &      -0.0003          &       -1.87                   &      -1.88            &      -1.63           &      0.411          \\     
{\bf              A0.94BtN10HR}   &             -2.04  &                       -0.316  &                       -1.72  &                       -1.38   &                       -0.336  &             -1.88  &                -0.002  &                       -1.87  &                -1.88  &               -2.33  &              0.411  \\  
\\
MB09D             &               -0.204        &      -0.069                  &       -0.135                  &      -0.039                  &       -0.096                  &       -1.99         &      -0.137           &       -1.85                   &      -8.63            &      -15.3           &      0.492          \\     
MB09Q             &               0.103         &      -0.046                  &       0.149                   &      0.132                   &       0.017                   &       -1.88         &      -0.0005          &       -1.87                   &      -5.39            &      -7.63           &      0.411          \\     
\\
A-0.9N100         &               5.92          &      3.17                    &       2.75                    &      0.896                   &       1.86                    &       3.03          &      1.15             &       1.88                    &      -0.151           &      3.52            &      5.9            \\     
A-0.5N100         &               5.28          &      3.63                    &       1.65                    &      1.03                    &       0.625                   &       2.61          &      1.54             &       1.06                    &      -0.819           &      1.52            &      4.84           \\     
A-0.2N100         &               3.4           &      2.19                    &       1.21                    &      0.816                   &       0.399                   &       1.17          &      0.765            &       0.404                   &      -0.761           &      0.879           &      4.02           \\     
A0.0N100          &               1.03          &      0.004                   &       1.03                    &      0.726                   &       0.304                   &       -0.045        &      -0.045           &       -0.0004                 &      -3.19            &      -0.6            &      3.46           \\     
A0.1N100          &               -0.546        &      -1.03                   &       0.481                   &      0.339                   &       0.141                   &       -0.649        &      -0.448           &       -0.201                  &      -3.91            &      -2.35           &      3.18           \\     
A0.2N100          &               -1.53         &      -1.98                   &       0.451                   &      0.323                   &       0.128                   &       -1.37         &      -0.954           &       -0.418                  &      -4.28            &      -2.41           &      2.89           \\     
A0.5N100          &               -5.05         &      -5.15                   &       0.099                   &      0.096                   &       0.002                   &       -4.43         &      -3.27            &       -1.16                   &      -6.39            &      -4              &      1.98           \\     
A0.9N25           &               -8.5          &      -7.31                   &       -1.18                   &      -0.511                  &       -0.674                  &       -6.5          &      -4.33            &       -2.17                   &      -7.77            &      -4.49           &      0.58           \\     
A0.9N50           &               -7.46         &      -6.44                   &       -1.01                   &      -0.45                   &       -0.565                  &       -6.05         &      -3.9             &       -2.15                   &      -8.34            &      -4.72           &      0.58           \\     
A0.9N100          &               -7.28         &      -6.35                   &       -0.925                  &      -0.429                  &       -0.496                  &       -6.05         &      -3.92            &       -2.13                   &      -8.74            &      -5.41           &      0.58           \\     
A0.9N200          &               -8.64         &      -7.57                   &       -1.07                   &      -0.477                  &       -0.591                  &       -7.24         &      -4.98            &       -2.26                   &      -10.1            &      -4.42           &      0.58           \\     
{\bf              A0.99N100}      &             -6.8   &                       -5.57   &                       -1.22  &                       -0.394  &                       -0.828  &             -5.51  &                -3.34   &                       -2.16  &                -9.7   &               -4.57  &              0.111  \\  
\hline
\hline
\end{tabular}
\end{center}
\label{tbl15}
\end{table*}

Table~\ref{tbl15} is similar to Table~\ref{tbl6} except the BH spin-up
parameter is computed via Eq.~(\ref{spinevolve}), where $s=s^{\rm
  EM}+s^{\rm MA}$ and $s^{\rm MA}=s^{\rm PA}+s^{\rm EN}$.  One can
recover the specific angular momentum flux from this spin-up parameter
and efficiency given in Table~\ref{tbl6}.  The thick disk poloidal
models have a spin-up parameter that is extremely negative relative to $a/M$, unlike the
NT thin disk and MB09 models.  The BH is always rapidly spinning down
in absolute spin magnitude for all models except the MB09D and the
A0.5BtN10 toroidal field model.  In all cases, the $a/M=0$ models are
spinning up, although quite weakly with the model A0.0BfN10 as
compared to models A0.0BtN10 and A0.0N100.  In summary, the thermal
pressure and magnetic field are capable of greatly decreasing the BH's
spin.

\subsection{$\alpha$ Viscosity and Suppression of the MRI}\label{sec:alphamri}

\begin{table*}
\caption{Viscosities, Grid Cells per Correlation lengths and MRI Wavelengths, MRI Wavelengths per full Disk Height, and Radii for MRI Suppression}
\begin{center}
\begin{tabular}[h]{|l|r|r|r|r|r|r|r|r|r|r|r|r|r|r|r|}
\hline
ModelName         &               $\alpha_{b,\rm{}eff}$  &      $\alpha_b$  &       $\alpha_{b,\rm{}M2}$  &       $\alpha_{b,\rm{}mag}$  &       $\bfrac{Q_{n,\rm{}cor,}}{{}_{\{\rho_0,b^2\}}}$  &    $\bfrac{Q_{l,\rm{}cor,}}{{}_{\{\rho_0,b^2\}}}$  &    $\bfrac{Q_{m,\rm{}cor,}}{{}_{\{\rho_0,b^2\}}}$  &   $Q_{\theta,\rm{}MRI,\{i,  o\}}$  &   $Q_{\phi,\rm{}MRI,\{i,  o\}}$  &    $S_{\rm{}d,\rm{}MRI,\{i,  o\}}$  &    $\bfrac{r_{\{S_{\rm{}d},S_{\rm{}d,\rm{}weak}\}}}{{\  }_{\rm{}MRI=1/2}}$  \\
\hline
{\bf              A0.94BfN40}     &                      0.39   &           0.022   &                     0.02    &                      0.26    &                                               22,  21                                              &    6,                                              12  &                         19,    18  &                       110,   120  &                         380,   540  &                                                    0.33,               0.34  &        $>55$,  $>55$  \\  
A0.94BfN100c1     &               0.35                   &      0.029       &       0.029                 &       0.36                   &       13,                                             13   &                                               4,   7                                               &   12,                       11     &   56,                     53     &    210,                      220    &    0.29,                                                0.38                &     64,      44      \\     
A0.94BfN100c2     &               0.3                    &      0.029       &       0.028                 &       0.3                    &       13,                                             13   &                                               4,   7                                               &   6,                        6      &   48,                     61     &    98,                       100    &    0.38,                                                0.34                &     $>44$,   $>44$   \\     
A0.94BfN100c3     &               0.32                   &      0.03        &       0.026                 &       0.31                   &       14,                                             13   &                                               4,   7                                               &   3,                        3      &   54,                     62     &    47,                       59     &    0.31,                                                0.33                &     47,      42      \\     
A0.94BfN100c4     &               0.29                   &      0.022       &       0.018                 &       0.24                   &       13,                                             13   &                                               3,   8                                               &   2,                        2      &   36,                     48     &    24,                       31     &    0.47,                                                0.43                &     14,      72      \\     
A0.94BfN40c5$^*$  &               0.089                  &      0.0053      &       0.0032                &       0.051                  &       22,                                             19   &                                               3,   22                                              &   1,                        1      &   1300,                   660    &    1,                        1      &    0.025,                                               0.058               &     55,      60      \\     
\\
{\bf              A-0.94BfN40HR}  &                      0.088  &           -0.017  &                     -0.017  &                      -0.24   &                                               22,  21                                              &    6,                                              12  &                         20,    18  &                       110,   140  &                         380,   560  &                                                    0.36,               0.29  &        $>54$,  $>54$  \\  
A-0.94BfN30       &               0.12                   &      -0.04       &       -0.031                &       -0.37                  &       15,                                             13   &                                               4,   7                                               &   10,                       9      &   42,                     45     &    180,                      260    &    0.44,                                                0.46                &     210,     190     \\     
A-0.5BfN30        &               0.2                    &      -0.11       &       -0.11                 &       -0.65                  &       13,                                             13   &                                               5,   6                                               &   15,                       16     &   35,                     23     &    200,                      290    &    0.5,                                                 0.91                &     -,       -       \\     
A0.0BfN10         &               0.77                   &      -0.012      &       $\sim                 0$      &                      -0.011  &                                               15,  15                                              &    12,                                             8   &                         13,    13  &                       110,   120  &                         110,   170  &                                                    0.22,               0.19  &        $>88$,  $>88$  \\  
A0.5BfN30         &               0.39                   &      0.12        &       0.12                  &       0.63                   &       14,                                             13   &                                               3,   7                                               &   16,                       19     &   47,                     18     &    240,                      270    &    0.39,                                                1.2                 &     18,      18      \\     
A0.94BfN30        &               0.36                   &      0.024       &       0.022                 &       0.25                   &       13,                                             13   &                                               3,   7                                               &   12,                       12     &   35,                     28     &    210,                      370    &    0.46,                                                0.75                &     13,      23      \\     
A0.94BfN30r       &               0.28                   &      0.024       &       0.023                 &       0.19                   &       13,                                             13   &                                               3,   6                                               &   12,                       12     &   31,                     34     &    190,                      330    &    0.52,                                                0.6                 &     11,      13      \\     
\\
A0.94BpN100       &               0.45                   &      0.028       &       0.024                 &       0.3                    &       13,                                             13   &                                               4,   8                                               &   12,                       11     &   51,                     100    &    180,                      270    &    0.32,                                                0.19                &     130,     100     \\     
\\
A-0.94BtN10       &               0.14                   &      0.053       &       $\sim                 0$      &                      0.4     &                                               15,  13                                              &    17,                                             10  &                         20,    6   &                       19,    10   &                         40,    32   &                                                    1.4,                2.7   &        -,      -      \\  
A-0.5BtN10        &               0.17                   &      0.096       &       0.01                  &       0.38                   &       15,                                             13   &                                               18,  10                                              &   20,                       7      &   23,                     12     &    48,                       40     &    1.2,                                                 2.1                 &     -,       -       \\     
A0.0BtN10         &               0.24                   &      0.06        &       0.016                 &       0.47                   &       15,                                             13   &                                               17,  9                                               &   20,                       8      &   19,                     15     &    61,                       64     &    1.5,                                                 1.7                 &     -,       -       \\     
A0.5BtN10         &               0.48                   &      0.067       &       0.022                 &       0.44                   &       15,                                             15   &                                               15,  12                                              &   15,                       9      &   31,                     23     &    87,                       83     &    0.9,                                                 1.1                 &     -,       -       \\     
A0.94BtN10        &               0.4                    &      0.062       &       0.0072                &       0.34                   &       14,                                             14   &                                               16,  11                                              &   19,                       10     &   19,                     16     &    58,                       63     &    1.5,                                                 1.7                 &     -,       -       \\     
{\bf              A0.94BtN10HR}   &                      0.75   &           0.066   &                     0.02    &                      0.38    &                                               28,  25                                              &    29,                                             24  &                         25,    18  &                       71,    50   &                         170,   170  &                                                    0.8,                1     &        -,      -      \\  
\\
MB09D             &               0.014                  &      0.01        &       0.0079                &       0.23                   &       18,                                             14   &                                               4,   4                                               &   4,                        3      &   3,                      3      &    4,                        4      &    7.5,                                                 8.1                 &     -,       -       \\     
MB09Q             &               0.099                  &      0.076       &       0.054                 &       0.35                   &       22,                                             19   &                                               12,  9                                               &   3,                        3      &   11,                     11     &    9,                        9      &    3,                                                   2.9                 &     -,       -       \\     
\\
A-0.9N100         &               0.25                   &      0.034       &       0.18                  &       0.49                   &       23,                                             22   &                                               9,   11                                              &   5,                        5      &   110,                    35     &    22,                       21     &    0.29,                                                0.84                &     22,      20      \\     
A-0.5N100         &               0.51                   &      0.067       &       0.19                  &       0.48                   &       32,                                             31   &                                               9,   9                                               &   4,                        3      &   170,                    33     &    25,                       22     &    0.18,                                                0.84                &     25,      22      \\     
A-0.2N100         &               0.68                   &      0.12        &       0.17                  &       0.46                   &       32,                                             29   &                                               8,   7                                               &   3,                        3      &   140,                    22     &    23,                       21     &    0.21,                                                1.4                 &     20,      18      \\     
A0.0N100          &               0.62                   &      0.21        &       0.18                  &       0.52                   &       31,                                             29   &                                               9,   10                                              &   4,                        3      &   110,                    32     &    23,                       21     &    0.28,                                                1.1                 &     21,      18      \\     
A0.1N100          &               0.67                   &      0.2         &       0.21                  &       0.49                   &       31,                                             31   &                                               9,   10                                              &   3,                        3      &   140,                    35     &    28,                       23     &    0.23,                                                1                   &     21,      19      \\     
A0.2N100          &               0.92                   &      0.35        &       0.23                  &       0.57                   &       32,                                             31   &                                               8,   8                                               &   3,                        3      &   150,                    34     &    29,                       22     &    0.21,                                                0.94                &     24,      23      \\     
A0.5N100          &               0.74                   &      0.24        &       0.18                  &       0.53                   &       31,                                             28   &                                               3,   7                                               &   3,                        3      &   90,                     30     &    25,                       20     &    0.32,                                                1.1                 &     16,      15      \\     
A0.9N25           &               1.2                    &      0.13        &       0.26                  &       0.59                   &       29,                                             27   &                                               8,   10                                              &   4,                        4      &   160,                    280    &    48,                       45     &    0.21,                                                0.16                &     $>120$,  $>120$  \\     
A0.9N50           &               0.96                   &      0.16        &       0.21                  &       0.55                   &       28,                                             27   &                                               8,   10                                              &   3,                        4      &   130,                    88     &    30,                       23     &    0.25,                                                0.4                 &     34,      32      \\     
A0.9N100          &               0.92                   &      0.17        &       0.21                  &       0.54                   &       29,                                             28   &                                               8,   10                                              &   4,                        4      &   120,                    42     &    30,                       22     &    0.27,                                                0.79                &     23,      21      \\     
A0.9N200          &               0.53                   &      0.18        &       0.15                  &       0.46                   &       27,                                             26   &                                               8,   9                                               &   3,                        4      &   75,                     24     &    26,                       24     &    0.42,                                                1.5                 &     14,      12      \\     
{\bf              A0.99N100}      &                      0.83   &           0.2     &                     0.2     &                      0.53    &                                               22,  20                                              &    6,                                              8   &                         7,     7   &                       110,   67   &                         66,    54   &                                                    0.33,               0.56  &        26,     20     \\  
\hline
\hline
\end{tabular}
\end{center}
\label{tbl4}
\end{table*}

Table~\ref{tbl4} shows the $\alpha$ viscosity parameters computed via
Eq.~(\ref{alphaeq}) using various averaging as described below
Eq.~(\ref{alphaeq}) in section~\ref{sec_infloweq}, where $\alpha_a$
includes the flow with $b^2/\rho_0<1$ and $\alpha_b$ focuses on the
highest density parts of the disk flow by using averaging weight
$w=\rholab$.  $\alpha$ is averaged over $r=12$--$20r_g$ for all models
except MB09 models that are averaged from $r=10$--$12r_g$.  If a
component's contribution to $\alpha$ is less than $10\%$, then we set
that component to $\sim 0$ in the table.  Also provided is $Q_{p,\rm
  cor}$ ($p=n,l,m$) computed via Eq.~(\ref{Qmcor}) for quantities
$\rho_0$ and $b^2$ at $r=8r_g$ for the ``Disk'' component.  We also
show $\Qone,\Qthree$, and $\Qtwo$ at $r^{\rm dcden}_{\rm i}$ and
$r^{\rm dcden}_{\rm o}$ given in Table~\ref{tbl14} as computed via
Eq.~(\ref{q1mri}) and Eq.~(\ref{q2mri}).  The convergence quality
measures ($\alpha_{\rm mag}$, $Q_{nlm,\rm cor}$, $\Qone$, and $\Qthree$) are
discussed in section~\ref{sec:resolution}.  The radius
($r_{\Qtwo=1/2}$ -- first occurrence of $\Qtwo=1/2$ outside $4r_g$ and
up to a one inflow radius) within which the MRI is suppressed is
provided, where ``-'' indicates that $\Qtwo>1/2$ implying no strong
MRI suppression.  $\Qtwo$ and $\Qtwoweak$ (focusing more on heavy
disk) give similar results.

The radially-averaged $\alpha$ depends upon the accumulated
magnetic flux, disk thickness, and BH spin.  Within the high-density
disk, the thick disk poloidal field models have $|\alpha_b|\sim 0.02$,
thick disk toroidal field models have $\alpha_b\sim 0.05$--$0.07$, and
thinner disk poloidal field models have $\alpha_b\sim 0.05$--$0.2$.
Our thick toroidal field models have higher $\alpha$ than found in
other works for which the MRI is not suppressed, such as $\alpha\sim
0.025$ found by \citet{bas11}.  Also, (not shown) $\alpha_a>\alpha_b$,
so the less dense and higher magnetized corona has larger $\alpha$ as
expected. MB09D, that has decaying turbulence, has $\alpha_b\sim
0.01$.

$\alpha$ and $\alpha_{\rm eff}$ are often quite different for our new
thick/thinner poloidal field models and thick toroidal field models,
which shows that local viscous $\alpha$-disk theory is inaccurate.
For the poloidal field models, $\alpha$-disk theory fails for
$r\lesssim 10r_g$ due to disk compression (leading to smaller
$\theta^d$ than a hydrostatic solution would have) by magnetic flux.
Even at larger radii within the inflow equilibrium region where
$\theta^d$ is not compressed much by magnetic flux, $\alpha$-disk
theory still does poorly.  Using $\alpha_b\propto p_{\rm b}$ leads to
somewhat better agreement for some of the poloidal field models. The
overall lack of agreement shows that large-scale magnetic torques
through magnetic {\it confinement} (rather than direct magnetic
torques within the heavy inflow) play an important role.  Convection
may also play an important role.  For the toroidal field models, using
$\alpha_b\propto p_{\rm tot}$ gives a radial scaling for $v_{\rm
  visc}$ similar to $v_r$ but too small by a factor of $\sim 10$,
while $\alpha_b\propto p_b$ gives far too large $v_{\rm visc}$
compared to $v_r$.  Summarizing, in all our new models, using $p_{\rm
  tot}$ leads to too small $v_r=v_{\rm visc}$ despite often having a reasonable
radial scaling.  Our poloidal and toroidal field simulations have
$\alpha_{\rm eff}\sim 0.2$--$1$ (which indicates the actual effective
viscous timescale) that is large enough to be consistent with
observations \citep{2007MNRAS.376.1740K}.

By contrast, the same table diagnostics for the multi-field loop thin
($\theta^d\sim 0.07$) disk model A0HR07 in \citet{pmntsm10} give
$\alpha_b\sim \alpha_{b,\rm eff}\sim 0.04$ for $r=7$--$9r_g$, so
$\alpha$-disk theory works quite well for those multi-loop field thin
disk models.  Also, for $\theta^d\sim 0.07$, \citet{bas11} found
$\alpha\sim 0.025$ and $\alpha_{\rm eff}\sim 0.1$ (from their fig.~8
around $r\sim 10r_g$ where $G\approx 2$), so $\alpha$-disk theory is
holding (at best) marginally well.  The tables show that our older MB09 models
show reasonable agreement with $\alpha$-disk theory.

$\alpha_{b,\rm M2}$ dominates the local stress contribution to
$\alpha_b$ in many cases.  Note that averaging $|\alpha_{\rm mag}|$
gives values of $\sim 0.4$--$0.6$ for all our simulations (including
2D models), but that neglects the direction of angular momentum
transport.  The PA (the Reynolds stress) term can be computed as
$\alpha_{\rm PA}\approx \alpha - \alpha_{\rm M2}$ from the table since
other terms are small.  Interestingly, $\alpha_b$, $\alpha_{b,\rm
  M2}$, and $\alpha_{b,\rm mag}$ are negative for $a/M\le 0$ for the
thick disk poloidal field models, and such models even have $v_{\rm
  rot}$ the same sign as $a/M$ (i.e. flow reversal) in the heavy disk
inflow even out to $r\sim 40r_g$ (e.g. in model A-0.94BfN30).  This
flow reversal behavior was validated by restarting the A0.94BfN40
model at $t=8000r_g/c$ with $a/M=-0.9375$ that produced model
A-0.94BfN40HR. (Compared to A-0.94BfN30, A-0.94BfN40HR shows much less
variance in time-dependent quantities -- probably due to
lower-resolution models allowing some of the opposite polarity
magnetic flux to reach the BH causing, e.g., the time-averaged
$\eta_{\rm H}$ to be smaller.)  Recall that $\alpha_{\rm mag}\propto
-b_r b_\phi$, which is just a term related to the Poynting flux.
Evidently, BH angular momentum is being dumped even into the heavy
disk.  Yet, in all cases, mass continues to accrete due to large-scale
stresses by magnetic flux and also possibly due to convection.
Interestingly, the thick disk toroidal field models are dominated by
positive $\alpha_{b,\rm PA}$ (Reynolds stress).  The thick disk
$a/M=0$ poloidal field model A0.0BfN10 has non-negligible negative
$\alpha_{b,\rm PA}$. Also, some of the thinner disk poloidal field
models have (negative) Reynolds stress that is up to $50\%$ of
$|\alpha_{b,\rm M2}|$.  $\alpha_{b,\rm EN}$ contributes little to
$\alpha_b$, except for the thick disk toroidal field models for which
$\alpha_{b,\rm EN}\lesssim \alpha_{b,\rm M2}$, both at about $20\%$
contribution to the total $\alpha$.

The MRI is not suppressed ($\Qtwo\gtrsim 1$) for thick disks at low
spin, flows with initially toroidally-dominated field, or MB09 type
limited poloidal flux or quadrupolar field models.  However, for
poloidal field models where magnetic flux saturates, the MRI is
suppressed within $r_{\Qtwo=1/2}\sim 20$--$200r_g$ for many thick disk
models and within $r_{\Qtwo=1/2}\sim 20r_g$ for the thinner disk
models (the radius depends upon the initial $\beta$ and duration of
simulation).  The MRI is suppressed in the A0.0BfN10 model as due to
the small initial $\beta$ and much of the magnetic field staying in
the dense flow (instead of a jet) where $\Qtwo$ is measured.

\begin{figure}
\centering
\includegraphics[width=3.2in,clip]{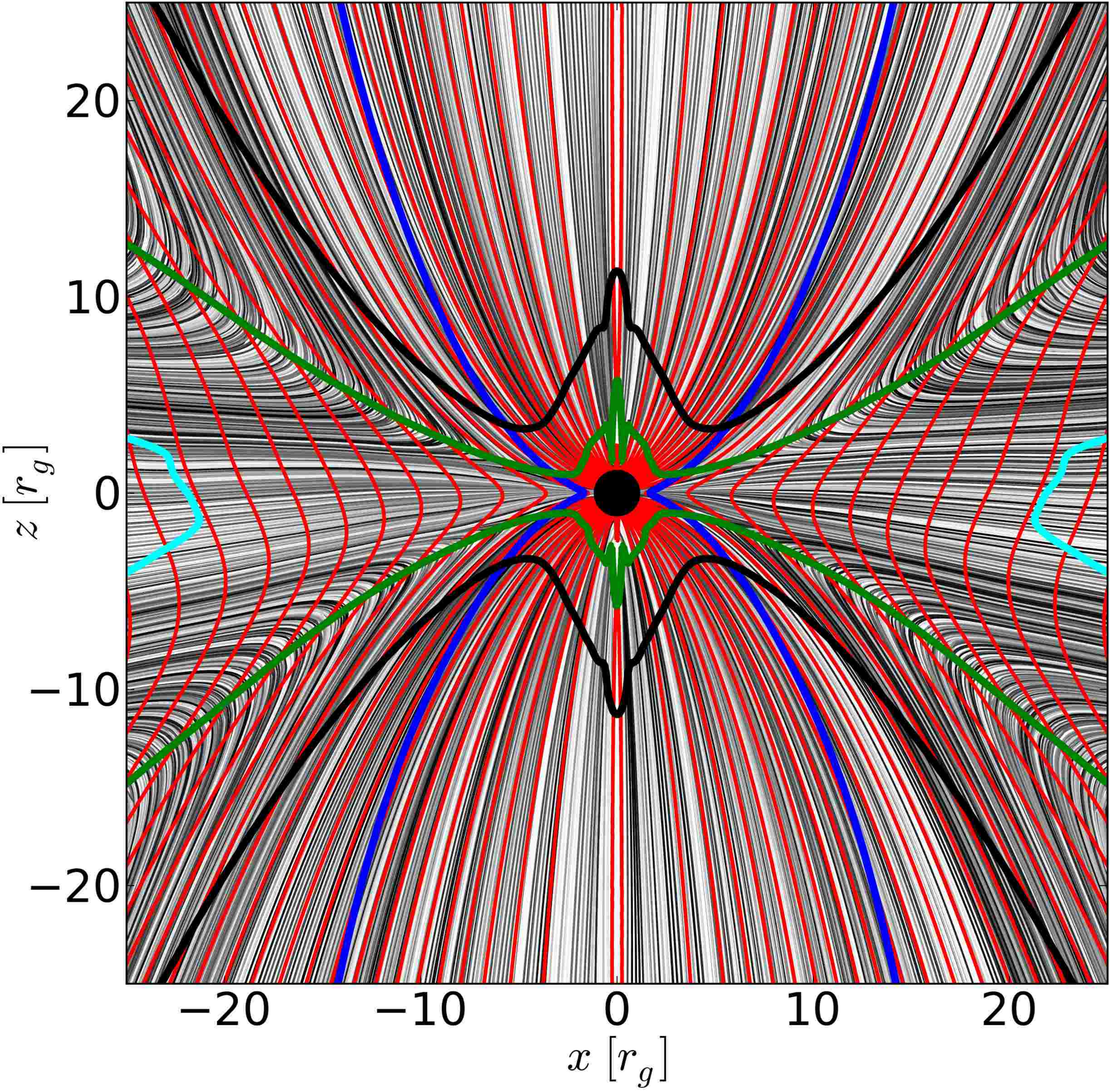}
\caption{Flow-field (as in Figure~\ref{figavgflowfield} for our
  fiducial thick disk model) for model A0.99N100 with $a/M=0.99$ and a
  thinner disk.  The thinner disks show less magnetic flux threading
  the BH.  The dense inflow accretes onto the BH through the ordered
  magnetic flux via efficient non-axisymmetric magnetic
  Rayleigh-Taylor instabilities.}
\label{figavgflowfieldsasha99}
\end{figure}

The suppression of the MRI is due to both the field strength and the
angular velocity.  The thinner disk models have accumulated much
magnetic flux near the BH as most evident from
Figure~\ref{figavgflowfieldsasha99} (similar to
Figure~\ref{figavgflowfield}) showing the time-averaged flow-field for
A0.99N100.  This suggests that thinner disks accrete (or hold onto)
magnetic flux as easily as thicker disks.  However, the thick flows
are more sub-Keplerian, as evident when computing $\Qtwo$ using
$\Omega_{\rm rot}\to\Omega_{\rm K}$.  Then, the thinner models would
still have $r_{\Qtwo=1/2}\sim 20r_g$ because they are somewhat
Keplerian, while the thick models would have $r_{\Qtwo=1/2}\sim
8r_g$. The MB09Q and toroidal field models have $\Qtwo\sim$ few
because poloidal flux build-up is competing against its destruction
since there is no net flux except in small patches.  Our high
resolution toroidal field model shows smaller $\Qtwo$, so even higher
resolutions might lead to $\Qtwo\sim 0.3$ as in the poloidal field
models.  MB09D probably has $\Qtwo\sim 8$ simply because turbulence is
under-resolved.

\subsection{Magnetic Flux}\label{sec:magneticflux}

\begin{table*}
\caption{Absolute Magnetic Flux per unit: Rest-Mass Fluxes, Initial Magnetic Fluxes, Available Magnetic Fluxes, and BH Magnetic Flux}
\begin{center}
\begin{tabular}[h]{|l|r|r|r|r|r|r|r|r|r|r|r|r|r|r|r|}
\hline
ModelName         &               $\Upsilon_{\rm{}H}$  &    $\Upsilon_{\rm{}in,i}$  &    $\Upsilon_{\rm{}in,o}$  &    $\Upsilon_{\rm{}j}$  &    $\Upsilon_{\rm{}mw,i}$  &    $\Upsilon_{\rm{}mw,o}$  &    $\Upsilon_{\rm{}w,i}$  &    $\Upsilon_{\rm{}w,o}$  &    $\left|\frac{\Psi_{\rm{}H}}{\Psi_1(t=0)}\right|$  &     $\left|\frac{\Psi_{\rm{}H}}{\Psi_2(t=0)}\right|$  &    $\left|\frac{\Psi_{\rm{}H}}{\Psi_3(t=0)}\right|$  &     $\left|\frac{\Psi_{\rm{}H}}{\Psi_a}\right|$  &     $\left|\frac{\Psi_{\rm{}H}}{\Psi_s}\right|$  &      $\left|\frac{\Phi_{\rm{}H}}{\Psi_{\rm{}fH}}\right|$  \\
\hline
{\bf              A0.94BfN40}     &                    17   &                       2.6  &                       7.9  &                    16   &                       2.7  &                       1.1  &                      6.7  &                      16   &                                                 21    &                                                 1    &                                                 0.16  &                                            0.46  &                                            0.047  &                                                    1    \\  
A0.94BfN100c1     &               18                   &    2.6                     &    14                      &    17                   &    2.9                     &    2.7                     &    6.8                    &    28                     &    31                                                &     1.5                                               &    0.24                                              &     0.65                                         &     0.065                                        &      1                                                    \\   
A0.94BfN100c2     &               18                   &    2.7                     &    12                      &    16                   &    3.1                     &    3                       &    7.3                    &    28                     &    30                                                &     1.5                                               &    0.23                                              &     0.63                                         &     0.043                                        &      1                                                    \\   
A0.94BfN100c3     &               18                   &    2.6                     &    14                      &    16                   &    3                       &    2.3                     &    6.8                    &    29                     &    31                                                &     1.5                                               &    0.25                                              &     0.7                                          &     0.063                                        &      1                                                    \\   
A0.94BfN100c4     &               18                   &    2.2                     &    13                      &    16                   &    1.6                     &    1.4                     &    5.7                    &    29                     &    32                                                &     1.6                                               &    0.25                                              &     0.72                                         &     0.047                                        &      1                                                    \\   
A0.94BfN40c5$^*$  &               37                   &    11                      &    13                      &    40                   &    15                      &    4.9                     &    18                     &    20                     &    13                                                &     0.67                                              &    0.1                                               &     0.17                                         &     0.6                                          &      1                                                    \\   
\\
{\bf              A-0.94BfN40HR}  &                    17   &                       2.5  &                       8.1  &                    16   &                       2.2  &                       1.2  &                      6.4  &                      17   &                                                 27    &                                                 1.1  &                                                 0.18  &                                            0.56  &                                            0.048  &                                                    1    \\  
A-0.94BfN30       &               11                   &    2.8                     &    8.4                     &    9.4                  &    3.5                     &    1.3                     &    8.8                    &    16                     &    19                                                &     0.78                                              &    0.12                                              &     0.41                                         &     0.024                                        &      1                                                    \\   
A-0.5BfN30        &               16                   &    14                      &    14                      &    8.4                  &    23                      &    25                      &    16                     &    17                     &    29                                                &     1.3                                               &    0.2                                               &     0.61                                         &     0.037                                        &      1                                                    \\   
A0.0BfN10         &               8.9                  &    9.7                     &    15                      &    0                    &    1600                    &    350                     &    56                     &    31                     &    9.4                                               &     0.44                                              &    0.068                                             &     0.11                                         &     0.19                                         &      1.1                                                  \\   
A0.5BfN30         &               15                   &    14                      &    14                      &    8.3                  &    45                      &    35                      &    17                     &    16                     &    24                                                &     1.1                                               &    0.18                                              &     0.5                                          &     1.6e-5                                       &      1                                                    \\   
A0.94BfN30        &               19                   &    2                       &    7.8                     &    18                   &    2                       &    0.98                    &    5.6                    &    15                     &    17                                                &     0.84                                              &    0.13                                              &     0.36                                         &     1.3e-5                                       &      1                                                    \\   
A0.94BfN30r       &               19                   &    2.2                     &    8.3                     &    17                   &    2.2                     &    0.77                    &    5.9                    &    15                     &    17                                                &     0.82                                              &    0.13                                              &     0.33                                         &     1.2e-5                                       &      1                                                    \\   
\\
A0.94BpN100       &               17                   &    2.4                     &    12                      &    16                   &    2.7                     &    2.9                     &    6                      &    24                     &    6.7e-5                                            &     0                                                 &    0                                                 &     1.3e-5                                       &     1.3e-5                                       &      1                                                    \\   
\\
A-0.94BtN10       &               0.74                 &    0.93                    &    1.2                     &    0                    &    21                      &    0.11                    &    16                     &    2.6                    &    6e3                                               &     3e2                                               &    3e2                                               &     0.46                                         &     0.27                                         &      4.3                                                  \\   
A-0.5BtN10        &               1                    &    1.3                     &    1.6                     &    0                    &    19                      &    0.85                    &    9.1                    &    3.1                    &    8e3                                               &     1e2                                               &    11                                                &     0.55                                         &     0.4                                          &      4.5                                                  \\   
A0.0BtN10         &               1.3                  &    1.6                     &    2.3                     &    0                    &    15                      &    18                      &    11                     &    4.5                    &    2e4                                               &     2e3                                               &    1.9                                               &     3.7                                          &     1.6                                          &      3.5                                                  \\   
A0.5BtN10         &               2.1                  &    2.7                     &    3.1                     &    0                    &    30                      &    14                      &    18                     &    6.7                    &    4e3                                               &     2e2                                               &    1.9                                               &     13                                           &     0.57                                         &      3.5                                                  \\   
A0.94BtN10        &               1.2                  &    1.5                     &    2                       &    0.02                 &    11                      &    8.4                     &    5.8                    &    4.1                    &    7e3                                               &     2e3                                               &    5e3                                               &     0.36                                         &     0.27                                         &      3.1                                                  \\   
{\bf              A0.94BtN10HR}   &                    2.1  &                       2.7  &                       3.1  &                    0.1  &                       14   &                       8    &                      15   &                      6.7  &                                                 4e4   &                                                 1e4  &                                                 4e3   &                                            4.4   &                                            3.2    &                                                    3.8  \\  
\\
MB09D             &               2.1                  &    0.38                    &    1                       &    1.5                  &    8                       &    5.2                     &    0.7                    &    2.1                    &    0.61                                              &     0                                                 &    0                                                 &     0.97                                         &     0.97                                         &      1                                                    \\   
MB09Q             &               0.58                 &    1.4                     &    2.4                     &    0.05                 &    8.9                     &    11                      &    2.6                    &    3.9                    &    2e6                                               &     0                                                 &    0                                                 &     0.67                                         &     1.2                                          &      2.6                                                  \\   
\\
A-0.9N100         &               6.6                  &    2.7                     &    4.6                     &    4.9                  &    16                      &    23                      &    13                     &    16                     &    0.16                                              &     0                                                 &    0                                                 &     0.16                                         &     0.27                                         &      1.1                                                  \\   
A-0.5N100         &               9.9                  &    3.9                     &    3.5                     &    7.1                  &    21                      &    21                      &    14                     &    13                     &    0.24                                              &     0                                                 &    0                                                 &     0.24                                         &     0.4                                          &      1.1                                                  \\   
A-0.2N100         &               12                   &    6.4                     &    3.7                     &    6.8                  &    26                      &    23                      &    15                     &    13                     &    0.29                                              &     0                                                 &    0                                                 &     0.29                                         &     0.49                                         &      1                                                    \\   
A0.0N100          &               11                   &    8.1                     &    8                       &    5.1                  &    24                      &    24                      &    15                     &    17                     &    0.29                                              &     0                                                 &    0                                                 &     0.29                                         &     0.47                                         &      1                                                    \\   
A0.1N100          &               12                   &    7.6                     &    5.9                     &    6.5                  &    26                      &    23                      &    15                     &    14                     &    0.4                                               &     0                                                 &    0                                                 &     0.4                                          &     0.53                                         &      1                                                    \\   
A0.2N100          &               12                   &    5.9                     &    4.1                     &    8                    &    23                      &    22                      &    14                     &    14                     &    0.3                                               &     0                                                 &    0                                                 &     0.3                                          &     0.51                                         &      1                                                    \\   
A0.5N100          &               13                   &    2.7                     &    3.8                     &    10                   &    13                      &    15                      &    9.9                    &    11                     &    0.4                                               &     0                                                 &    0                                                 &     0.4                                          &     0.55                                         &      1                                                    \\   
A0.9N25           &               11                   &    2.8                     &    9.1                     &    9.3                  &    9.9                     &    25                      &    8.6                    &    24                     &    0.15                                              &     0                                                 &    0                                                 &     0.15                                         &     0.21                                         &      1                                                    \\   
A0.9N50           &               10                   &    2.4                     &    5.1                     &    8.7                  &    9.1                     &    20                      &    8                      &    17                     &    0.22                                              &     0                                                 &    0                                                 &     0.22                                         &     0.34                                         &      1                                                    \\   
A0.9N100          &               10                   &    2.3                     &    4.4                     &    8.7                  &    8.9                     &    16                      &    7.6                    &    13                     &    0.26                                              &     0                                                 &    0                                                 &     0.26                                         &     0.41                                         &      1                                                    \\   
A0.9N200          &               11                   &    2.3                     &    4.2                     &    9.7                  &    8.6                     &    11                      &    7.7                    &    11                     &    0.33                                              &     0                                                 &    0                                                 &     0.33                                         &     0.59                                         &      1                                                    \\   
{\bf              A0.99N100}      &                    9.1  &                       2.4  &                       5.5  &                    7.7  &                       7.9  &                       16   &                      7.5  &                      16   &                                                 0.23  &                                                 0    &                                                 0     &                                            0.23  &                                            0.36   &                                                    1    \\  
\hline
\hline
\end{tabular}
\end{center}
\label{tbl11}
\end{table*}

Table~\ref{tbl11} shows the value of $\Upsilon$ (computed via the
Eq.~(\ref{equpsilon})) on the horizon ($\Upsilon_{\rm H}$), in the
inflow-only ($u_r<0$) regions at $r_i$ and $r_o$ defined already
($\Upsilon_{\rm in,i}$ and $\Upsilon_{\rm in,o}$ respectively), in the
jet ($\Upsilon_j$), in the magnetized wind at $r_i$ and $r_o$
($\Upsilon_{\rm mw,i}$ and $\Upsilon_{\rm mw,o}$ respectively), and in
the entire wind at $r_i$ and $r_o$ ($\Upsilon_{\rm mw,i}$ and
$\Upsilon_{\rm mw,o}$ respectively).  Only $\Upsilon_j$ uses the
non-local normalization of $\dot{M}_{\rm H}$.

In the poloidal field models, $\Upsilon$ is dominated by the magnetic
flux threading the BH polar regions as shown by comparing
$\Upsilon_{\rm H}$ and $\Upsilon_j$.  Also evident is that $\Upsilon$
decreases from $r_o$ to $r_i$ (shown by $\Upsilon_{\rm in,o}$ and
$\Upsilon_{\rm in,i}$, respectively), but then rises near the horizon.
The flux at large radii acts a reservoir. At smaller radii some flux
gets accreted, and on the horizon much of the mass gets drained.  This
leaves the ordered part of the magnetic flux giving high
$\Upsilon_{\rm H}$.

The magnetized wind and entire wind values of $\Upsilon$ are
normalized by the local mass accretion rate for the inflow component
at those radii (i.e.  $\dot{M}_{\rm in,i}$ and $\dot{M}_{\rm in,o}$).
This shows that the local mass-loading can be large in the wind and
small in the magnetized wind.  The entire wind is just part of the
overall circularizing flow and eventually feeds the pure inflow and
feeds the inflow value of $\Upsilon$.

Also, notice that $\Upsilon$ is quite similar between the flipping and
non-flipping poloidal field, which shows that the flipping models have
plenty of constant polarity flux unlike the MB09D model.  This also
shows that even with different initial conditions, the flux threading
the BH saturates in some type of force balance between (e.g.) the
inner magnetospheric pressure against the exterior disk gas+ram
pressure.  This unstable balance results in time-dependent magnetic
Rayleigh-Taylor events that regulate the magnetic flux on the horizon
by carrying magnetic flux back into the disk and wind.

Table~\ref{tbl11} also shows the magnetic flux (computed as in
section~\ref{magneticfluxdiag}) on the horizon normalized in various
ways.  The values $\Psi_{\rm H}/\Psi_1(t=0)$, $\Psi_{\rm
  H}/\Psi_2(t=0)$, and $\Psi_{\rm H}/\Psi_3(t=0)$ are for
normalizations by the extrema (label of $1$ for the extrema at
smallest radii, and so forth) in the magnetic flux in the initial
disk.  For the flipping poloidal field geometry, we show only the
first $3$ extrema since only those are relevant over the entire
evolution.  A value of zero indicates no relevant value, as for the
non-flipping poloidal field geometries.  For the toroidal field
models, this just shows the growth of the initially small magnetic
flux threading the equator.

The value $\Psi_{\rm H}/\Psi_a$ (as other quantities, this is
$[\Psi_{\rm H}(t)/[\Psi_a]_t]_t$ in the table) is [the magnetic flux
on the horizon] per unit [flux available on the BH plus beyond the BH
of the same polarity of magnetic flux].  This ratio shows how much
more flux is available to the BH, and a value of $\Psi_{\rm
  H}/\Psi_a\sim 1$ would mean the magnetic flux beyond the BH is only
of opposite polarity -- so that the BH has as much flux as it can get
of the same field polarity.  The value $\Psi_{\rm H}/\Psi_s$ is
similar to $\Psi_{\rm H}/\Psi_a$, except it shows how much magnetic
flux is available to the BH of any polarity within the stagnation
radius.  The stagnation radius ($r_s$) is defined by where
$0=\int\rho_0 \uvec^r dA_{\theta\phi}$ over the full flow within
$b^2/\rho_0<1$ (so includes the disk inflow and entire wind outflows),
because the whole flow has not reached inflow equilibrium there.  This
is roughly where $u_r=0$ as weighted for the whole massive flow.  If
the density-weighted radial lab-frame 3-velocity $v_r=0$ inside this
radius, then that radius is used instead for $r_s$.  Note that
$\Psi_a$ and $\Psi_s$ have been time-averaged before forming a ratio
in the table.  Lastly, the value $|\Phi_{\rm H}/\Psi_{\rm tH}|$ shows
the value of the absolute magnetic flux per unit absolute of the
extremum of the signed magnetic flux, where the ratio itself is
time-averaged since the measurements are exactly co-spatial.  As
discussed in section~\ref{magneticfluxdiag}, this roughly measures the
vector spherical harmonic $l$ mode (e.g. $l=1$ is dipolar).

For the polarity-flipping poloidal field models, the values of
$\Psi_{\rm H}/\Psi_i\gtrsim 1$ if the magnetic flux on the BH has
already exceeded the $i$-th loop's initial magnetic flux.  So, in our
polarity flip simulations, the table shows that magnetic flux has
accumulated and been destroyed already twice in the simulation with
the flipping model settling on the third extremum (i.e. 3rd field
loop).  Longer times lead to the next polarity being accreted.  This
shows that ordered flux is easily transported through the accretion
disk, and large-scale field that penetrates the corona (see, e.g.,
\citealt{2008ApJ...677.1221R}) is not required for flux transport.
Movies of the field lines (contours of $\Avpotvec_\phi$ or streamlines
of $B^i$) do show that magnetic field lines at higher latitudes
initially transport more readily than those at the equator as seen in
more idealized models \citep{bhk09trans}.  However, eventually the
magnetic flux threading the equator does accrete as well, and in
steady state we cannot identify anything special about the corona
vs. the disk in accretion of ordered magnetic flux.
\citet{bhk09trans} suggest their simulations may have reached magnetic
flux saturation near the BH (i.e. saturated $\Upsilon_{\rm H}$
and $\Psi_{\rm H}$), but they also found that most of the magnetic
flux moves out to large radii rather than accreting.  We suggest that
this is due to their small initial torus leading to significant
outflow of magnetic flux (as in their fig.~2, fig.~3, and animation).

The values of $\Psi_{\rm H}/\Psi_a$ and $\Psi_{\rm H}/\Psi_s$ show
that there is plenty of magnetic flux available (including of the same
polarity) for the BH in most models.  These are time-averages, while
in general each polarity loop that is accreted starts off at quite low
values of $\Psi_{\rm H}/\Psi_a$ until nearing the next polarity when
$\Psi_{\rm H}/\Psi_a\sim 1$ and then the field polarity inversion
occurs on the horizon.

The MB09D model, typical of most torus-based MHD simulations in the
literature, was time-averaged over an early period of accretion.
However, already by $t\sim 2000r_g/c$ the $\Psi_{\rm H}/\Psi_a\to 1$
and $\Psi_{\rm H}/\Psi_s\to 1$, indicating there is no more available
magnetic flux with the same polarity (either at all or within the part
of the ingoing flow, respectively).  Only the accretion of another
$\sim 5\times$ the same polarity magnetic flux would have led to a
MCAF state and much higher $\eta$.  Also, some prior MHD simulations
may have been run for too short of a duration.  Depending upon the
initial conditions, the magnetic flux can accumulate over longer times
than other quantities, so looking at only energy and angular momentum
fluxes can be misleading.  In short, the choice of initial conditions
and short duration can limit $\Upsilon_{\rm H}$ and so the
efficiencies.

The value of $|\Phi_{\rm H}/\Psi_{\rm tH}|$ shows the average
approximate vector spherical harmonic $|l|$-mode.  All poloidal field
models have unity as expected, while the MB09Q large-scale quadrupolar
field model has $l\sim 3$ as expected given the large-scale field is
quadrupolar ($l=2$) and there is an equatorial MHD turbulent disk.

\begin{figure}
\centering
\includegraphics[width=3.2in,clip]{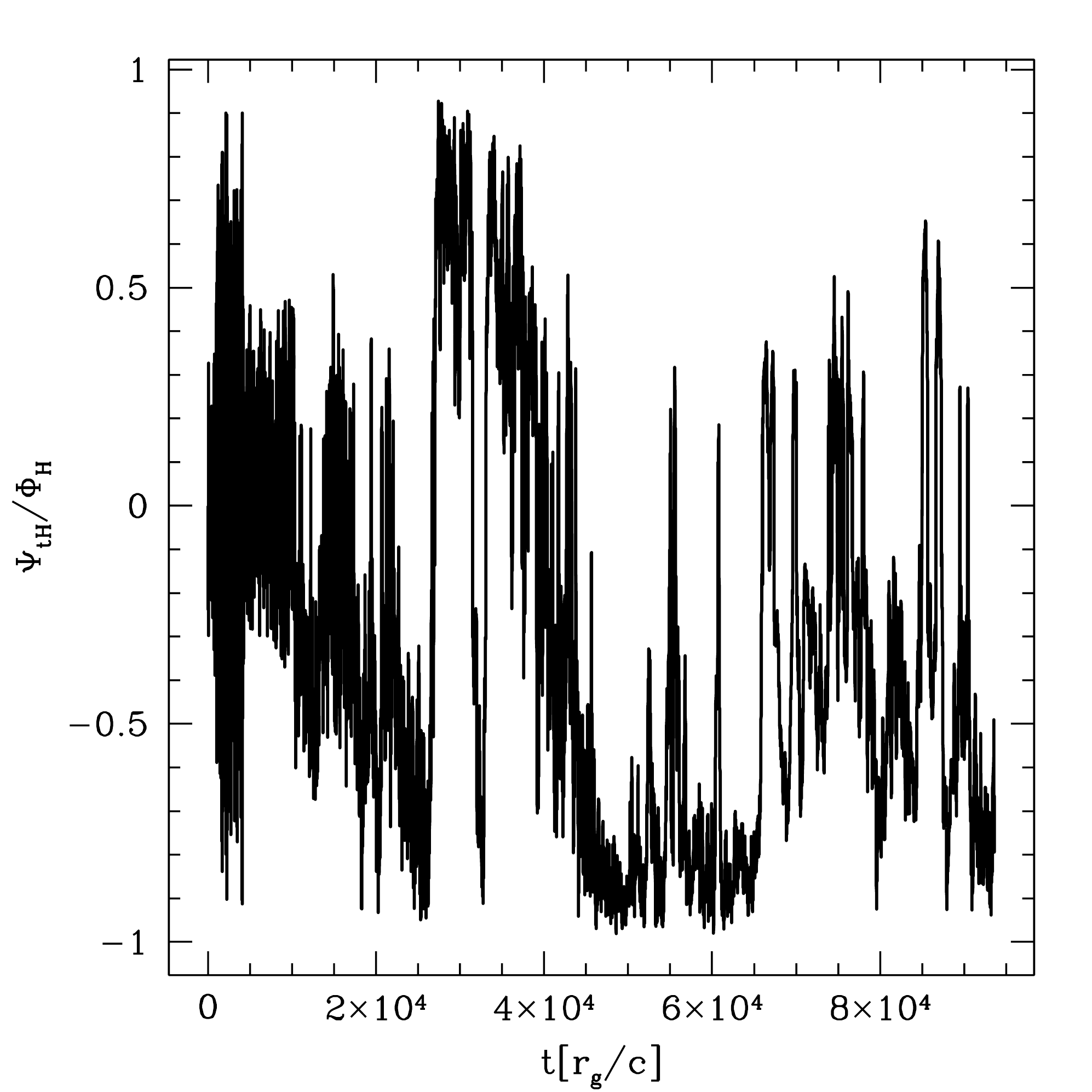}
\caption{Value of $\Psi_{\rm tH}/\Phi_{\rm H}$ corresponding to the
  approximate value of $1/l$ for the vector spherical harmonic
  expansion $l$-mode for model A0.94BtN10 (similar results for
  A0.94BtN10HR).  This shows that well-ordered dipolar ($|\Psi_{\rm
    tH}/\Phi_{\rm H}|\approx 1$, where $|\Psi_{\rm tH}/\Phi_{\rm
    H}|\approx 1/2$ corresponds to quadrupolar) field on the BH
  appears for somewhat sustained durations.  In summary, only weak
  relativistic jets are produced in our simulations that start with
  mostly toroidal field, but much higher resolutions might allow this
  emergent large-scale dipolar field to launch persistent
  lightly-loaded relativistic jets.}
\label{psiophitoroidal}
\end{figure}

\begin{figure}
\centering
\includegraphics[width=3.2in,clip]{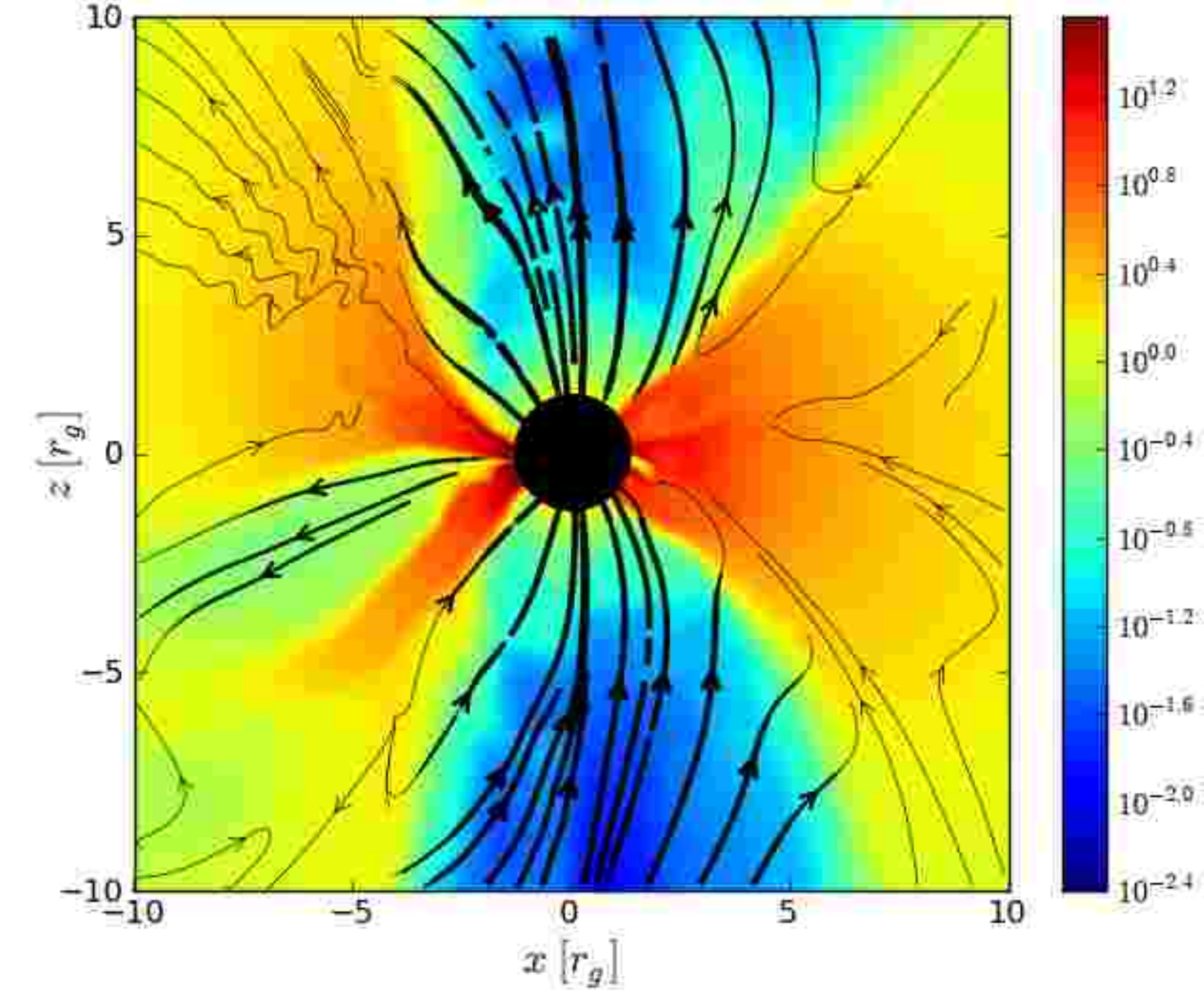}
\caption{Like upper-left panel in Figure~\ref{evolvedmovie} and
  Figure~\ref{outburst}, but for initially toroidally-dominated
  magnetic field model A0.94BtN10.  This shows ``spontaneous''
  formation of large-scale and dipolar magnetic flux near the BH at
  $t\approx 62828r_g/c$ -- as also occurs at other times when
  $|\Psi_{\rm tH}/\Psi_{\rm H}|\sim 1$ (see
  Figure~\ref{psiophitoroidal}).  The magnetic field lines are shown
  as thick (thin) black lines when they are lightly (heavily)
  mass-loaded.  Only lightly mass-loaded field lines would permit
  ultrarelativistic jets.  The magnetic field and low-density (blue in
  plot) wind/jet regions extend out to (not shown) $r\sim 50r_g$.}
\label{largescalefield}
\end{figure}

Interestingly, the toroidal models show episodes with well-ordered
{\it and} large-scale dipolar $|l|\approx 1$ field on the horizon.
For model A0.94BtN10 (A0.94BtN10HR is similar),
Figure~\ref{psiophitoroidal} shows $\Psi_{\rm tH}/\Phi_{\rm H}$
(approximate $1/l$ for the vector spherical harmonic $l$ mode).  The
large-scale dipolar field is evident in Figure~\ref{largescalefield},
where the large-scale dipolar flux extends out to $r\sim 50r_g$.

\begin{figure}
\centering
\includegraphics[width=3.2in,clip]{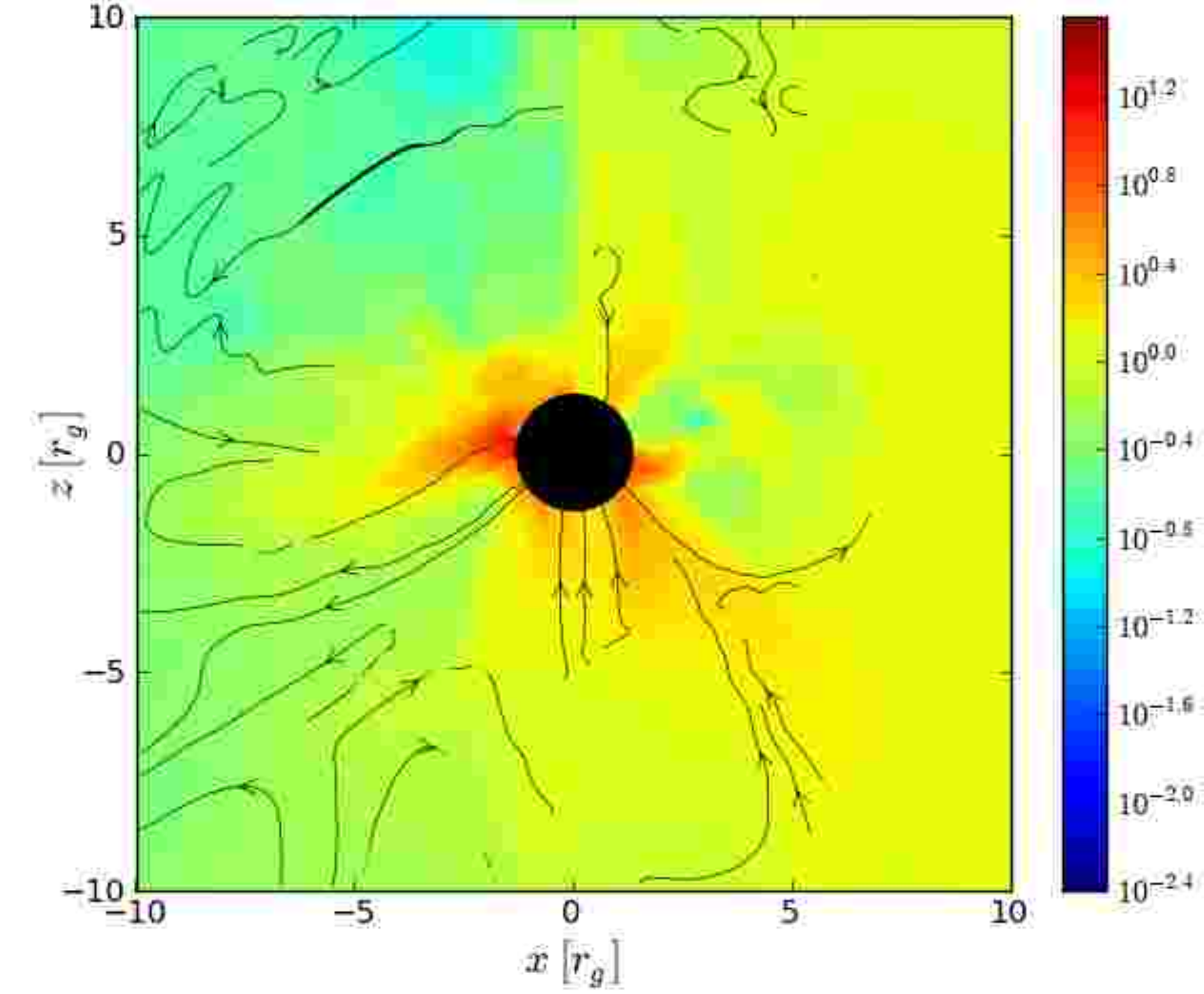}
\caption{Like Figure~\ref{largescalefield}, but typical snapshot at
  $t\approx 40164r_g/c$ when $|\Psi_{\rm tH}/\Psi_{\rm H}|\ll 1$ while
  the field is not dipolar or large-scale.  Magnetic field lines are
  strongly mass-loaded (i.e. field lines shown as thin lines).  Field
  lines appear/disappear because (in this slice) they bend in/out of
  the plotted plane.  In such thick flows, mass inflow occurs from
  quite random directions with no strong preference for the plane of
  the disk/BH rotation.}
\label{typicalnolargescalefield}
\end{figure}

By contrast, Figure~\ref{typicalnolargescalefield} shows the typical
flow at times when $|\Psi_{\rm tH}/\Phi_{\rm H}|\ll 1$. The magnetic
field is strongly mass-loaded and not particularly aligned with any
rotational axis.  The high-density material inflows from quite random
directions.

During the episodes of large-scale dipolar flux formation in the
toroidal field models, only weakly powered highly relativistic jets
emerge due to too low $\Upsilon\sim 3$. Higher resolutions, higher
$\Upsilon$, or somewhat thinner disks might promote more persistent
jet formation, which is required by some reconnection-based jet models
\citep{um10,mu12}.

\subsection{Power-Law Fits for Radial Dependence}\label{sec:raddep}

The power-law scaling of quantities in the flow is important for
testing accretion flow models. RIAFs such as ADAFs have $\rho\propto
r^{-3/2}$ and $p_g\propto r^{-5/2}$, while CDAFs have $\rho\propto
r^{-1/2}$, and ADIOSs vary over this range.  ADAFs have $c_s/v_{\rm
  K}\sim 0.5$--$0.7$ and $c_s/v_\phi\gtrsim 1$.

Viscous simulations and RIAF models show similar radial scalings for
density, velocity, and mass accretion rate
\citep{sto99,igu00,igu02,mck02}.  For example, flows with
$\theta^d\sim 0.2$ show an inflow $\dot{M}\propto r$ \citep{hb02}.
For viscous simulations, \citet{ia00} found $\rho\propto r^{-1/2}$ to
$r^{-0.7}$ for $\Gamma=5/3$--$4/3$, respectively.  For relatively thin
disk injection 3D MHD simulations giving $c_s/v_{\rm rot}\sim 0.4$,
\citet{ina03} found $\rho\propto r^{-1}$ and $v_{\rm rot}\propto r^{-1/2}$.
In 2D MHD simulations with $c_s/v_{\rm rot}\sim 0.6$, \citet{sto01}
(i.e. Run F) found $\rho\propto$ constant, $p_g\propto r^{-1.5}$ and
$v_r\propto r^{-1}$.  However, these MHD simulations involved
relatively thin disks with $\theta^t_{\rholab}\lesssim 0.6$.  For MHD
simulations of truly thick ($\theta^t\sim 1$ or $c_s/v_{\rm rot}\gg 1$)
flows, \citet{pmw03} found $\rho\propto r^{-0.72}$, $p_g\propto
r^{-1.5}$, and $v_{\rm rot}\sim 0.1v_{\rm K}$ -- a sub-Keplerian flow.

\begin{table*}
\caption{Inner and Outer Radii for Least-Square Fits, Disk+Corona Stagnation Radius, and Fitted Power-Law Indices for Disk and Wind Flows}
\begin{center}
\begin{tabular}[h]{|l|r|r|r|r|r|r|r|r|r|r|r|r|r|}
\hline
ModelName      &               $r^{\rm{}dcden}_{\rm{}i}$  &   $r^{\rm{}dcden}_{\rm{}o}$  &   $r^{\rm{}dcden}_{\rm{}f}$  &    $r^{\rm{}dcden}_{\rm{}s}$  &    $\rho_0$               &                      $p_g$                  &                      $|b|$                  &                      $r^{\rm{}w}_{\rm{}i}$  &   $r^{\rm{}w}_{\rm{}o}$  &    $\dot{M}_{\rm{}in}-\dot{M}_{\rm{}H}$  &                     $\dot{M}_{\rm{}mw}$    &                     $\dot{M}_{\rm{}w}$    \\
\hline
{\bf           A0.94BfN40}     &                          12  &                          29  &                          55   &                          310  &                      ${-0.61}_{\pm{}0.05}$  &                      ${-0.83}_{\pm{}0.07}$  &                      ${-1.1}_{\pm{}0.02}$   &                      20  &                      42   &                                     ${1.3}_{\pm{}0.04}$   &                      ${0.5}_{\pm{}0.08}$   &                     ${1.3}_{\pm{}0.04}$   \\  
A0.94BfN100c1  &               12                         &   27                         &   280                        &    280                        &    ${-0.63}_{\pm{}0.09}$  &                      ${-0.9}_{\pm{}0.1}$    &                      ${-0.85}_{\pm{}0.1}$   &                      20                     &   40                     &    ${1.3}_{\pm{}0.03}$                   &                     ${0.6}_{\pm{}0.1}$     &                     ${1.3}_{\pm{}0.02}$   \\                    
A0.94BfN100c2  &               12                         &   27                         &   44                         &    340                        &    ${-0.62}_{\pm{}0.08}$  &                      ${-0.9}_{\pm{}0.1}$    &                      ${-0.8}_{\pm{}0.1}$    &                      20                     &   40                     &    ${1.4}_{\pm{}0.04}$                   &                     ${0.8}_{\pm{}0.1}$     &                     ${1.5}_{\pm{}0.03}$   \\                    
A0.94BfN100c3  &               12                         &   27                         &   590                        &    280                        &    ${-0.6}_{\pm{}0.08}$   &                      ${-0.87}_{\pm{}0.1}$   &                      ${-0.8}_{\pm{}0.2}$    &                      20                     &   40                     &    ${1.4}_{\pm{}0.1}$                    &                     ${0.6}_{\pm{}0.2}$     &                     ${1.4}_{\pm{}0.1}$    \\                    
A0.94BfN100c4  &               12                         &   29                         &   280                        &    310                        &    ${-0.63}_{\pm{}0.08}$  &                      ${-0.9}_{\pm{}0.1}$    &                      ${-0.8}_{\pm{}0.1}$    &                      20                     &   40                     &    ${1.5}_{\pm{}0.1}$                    &                     ${0.4}_{\pm{}0.3}$     &                     ${1.6}_{\pm{}0.1}$    \\                    
\\
{\bf           A-0.94BfN40HR}  &                          12  &                          30  &                          54   &                          320  &                      ${-0.62}_{\pm{}0.06}$  &                      ${-0.85}_{\pm{}0.08}$  &                      ${-0.98}_{\pm{}0.06}$  &                      20  &                      46   &                                     ${1.4}_{\pm{}0.04}$   &                      ${0.58}_{\pm{}0.05}$  &                     ${1.4}_{\pm{}0.04}$   \\  
A-0.94BfN30    &               12                         &   30                         &   310                        &    310                        &    ${-0.68}_{\pm{}0.07}$  &                      ${-0.9}_{\pm{}0.1}$    &                      ${-1}_{\pm{}0.2}$      &                      20                     &   52                     &    ${1.5}_{\pm{}0.08}$                   &                     ${0.58}_{\pm{}0.04}$   &                     ${1.5}_{\pm{}0.08}$   \\                    
A-0.5BfN30     &               12                         &   30                         &   320                        &    410                        &    ${-0.62}_{\pm{}0.08}$  &                      ${-0.9}_{\pm{}0.1}$    &                      ${-0.94}_{\pm{}0.09}$  &                      20                     &   58                     &    ${1.4}_{\pm{}0.2}$                    &                     ${3}_{\pm{}2}$         &                     ${2.2}_{\pm{}0.1}$    \\                    
A0.0BfN10      &               12                         &   30                         &   88                         &    83                         &    ${-0.68}_{\pm{}0.07}$  &                      ${-0.93}_{\pm{}0.1}$   &                      ${-1.1}_{\pm{}0.1}$    &                      20                     &   75                     &    ${1.3}_{\pm{}0.1}$                    &                     -                      &                     ${1.7}_{\pm{}0.02}$   \\                    
A0.5BfN30      &               12                         &   30                         &   450                        &    14000                      &    ${-0.55}_{\pm{}0.08}$  &                      ${-0.84}_{\pm{}0.09}$  &                      ${-0.8}_{\pm{}0.1}$    &                      20                     &   58                     &    ${0.8}_{\pm{}0.3}$                    &                     ${4}_{\pm{}1}$         &                     ${1.4}_{\pm{}0.4}$    \\                    
A0.94BfN30     &               12                         &   27                         &   310                        &    14000                      &    ${-0.66}_{\pm{}0.09}$  &                      ${-0.9}_{\pm{}0.1}$    &                      ${-0.88}_{\pm{}0.05}$  &                      20                     &   40                     &    ${1.4}_{\pm{}0.06}$                   &                     ${0.57}_{\pm{}0.03}$   &                     ${1.4}_{\pm{}0.05}$   \\                    
A0.94BfN30r    &               12                         &   27                         &   430                        &    14000                      &    ${-0.61}_{\pm{}0.08}$  &                      ${-0.8}_{\pm{}0.1}$    &                      ${-1}_{\pm{}0.02}$     &                      20                     &   40                     &    ${1.3}_{\pm{}0.09}$                   &                     ${0.5}_{\pm{}0.2}$     &                     ${1.2}_{\pm{}0.09}$   \\                    
\\
A0.94BpN100    &               12                         &   30                         &   290                        &    14000                      &    ${-0.65}_{\pm{}0.08}$  &                      ${-0.9}_{\pm{}0.1}$    &                      ${-0.8}_{\pm{}0.1}$    &                      20                     &   52                     &    ${1.3}_{\pm{}0.04}$                   &                     ${0.56}_{\pm{}0.04}$   &                     ${1.2}_{\pm{}0.03}$   \\                    
\\
A-0.94BtN10    &               12                         &   30                         &   240                        &    240                        &    ${-0.64}_{\pm{}0.07}$  &                      ${-0.88}_{\pm{}0.09}$  &                      ${-1.3}_{\pm{}0.1}$    &                      20                     &   100                    &    ${2.1}_{\pm{}0.2}$                    &                     -                      &                     ${2.2}_{\pm{}0.3}$    \\                    
A-0.5BtN10     &               12                         &   30                         &   160                        &    160                        &    ${-0.64}_{\pm{}0.08}$  &                      ${-0.9}_{\pm{}0.1}$    &                      ${-1.2}_{\pm{}0.07}$   &                      20                     &   100                    &    ${1.3}_{\pm{}0.2}$                    &                     -                      &                     ${1.4}_{\pm{}0.2}$    \\                    
A0.0BtN10      &               12                         &   30                         &   190                        &    190                        &    ${-0.61}_{\pm{}0.07}$  &                      ${-0.83}_{\pm{}0.09}$  &                      ${-1.1}_{\pm{}0.08}$   &                      20                     &   100                    &    ${1.4}_{\pm{}0.2}$                    &                     -                      &                     ${1.5}_{\pm{}0.2}$    \\                    
A0.5BtN10      &               12                         &   30                         &   200                        &    200                        &    ${-0.67}_{\pm{}0.08}$  &                      ${-0.9}_{\pm{}0.1}$    &                      ${-1.2}_{\pm{}0.08}$   &                      20                     &   100                    &    ${1.7}_{\pm{}0.2}$                    &                     ${-3}_{\pm{}2}$        &                     ${1.8}_{\pm{}0.2}$    \\                    
A0.94BtN10     &               12                         &   30                         &   230                        &    230                        &    ${-0.6}_{\pm{}0.08}$   &                      ${-0.8}_{\pm{}0.1}$    &                      ${-1}_{\pm{}0.08}$     &                      20                     &   100                    &    ${1.4}_{\pm{}0.2}$                    &                     ${-1.6}_{\pm{}0.4}$    &                     ${1.4}_{\pm{}0.2}$    \\                    
{\bf           A0.94BtN10HR}   &                          12  &                          30  &                          180  &                          170  &                      ${-0.72}_{\pm{}0.05}$  &                      ${-0.97}_{\pm{}0.07}$  &                      ${-1.1}_{\pm{}0.06}$   &                      20  &                      100  &                                     ${2.2}_{\pm{}0.3}$    &                      ${-1.1}_{\pm{}0.1}$   &                     ${1.9}_{\pm{}0.2}$    \\  
\\
MB09Q          &               12                         &   15                         &   23                         &    20                         &    ${-0.9}_{\pm{}0.2}$    &                      ${-2.1}_{\pm{}0.1}$    &                      ${-0.64}_{\pm{}0.01}$  &                      6                      &   18                     &    ${0.7}_{\pm{}0.2}$                    &                     ${1.6}_{\pm{}0.3}$     &                     ${1.3}_{\pm{}0.06}$   \\                    
\\
A-0.9N100      &               12                         &   30                         &   110                        &    110                        &    ${-0.52}_{\pm{}0.05}$  &                      ${-1.5}_{\pm{}0.05}$   &                      ${-0.98}_{\pm{}0.03}$  &                      20                     &   73                     &    ${1.4}_{\pm{}0.1}$                    &                     ${0.99}_{\pm{}0.03}$   &                     ${1.6}_{\pm{}0.09}$   \\                    
A-0.5N100      &               12                         &   30                         &   100                        &    100                        &    ${-0.6}_{\pm{}0.1}$    &                      ${-1.7}_{\pm{}0.09}$   &                      ${-0.96}_{\pm{}0.03}$  &                      20                     &   64                     &    ${1.9}_{\pm{}0.2}$                    &                     ${1.1}_{\pm{}0.02}$    &                     ${2.2}_{\pm{}0.1}$    \\                    
A-0.2N100      &               12                         &   30                         &   100                        &    100                        &    ${-0.65}_{\pm{}0.02}$  &                      ${-1.7}_{\pm{}0.03}$   &                      ${-1}_{\pm{}0.05}$     &                      20                     &   65                     &    ${1.8}_{\pm{}0.2}$                    &                     ${1.2}_{\pm{}0.04}$    &                     ${2.1}_{\pm{}0.08}$   \\                    
A0.0N100       &               12                         &   30                         &   110                        &    110                        &    ${-0.94}_{\pm{}0.05}$  &                      ${-2}_{\pm{}0.06}$     &                      ${-0.98}_{\pm{}0.05}$  &                      20                     &   69                     &    ${1.5}_{\pm{}0.2}$                    &                     ${1.2}_{\pm{}0.09}$    &                     ${1.8}_{\pm{}0.1}$    \\                    
A0.1N100       &               12                         &   30                         &   110                        &    110                        &    ${-0.78}_{\pm{}0.04}$  &                      ${-1.9}_{\pm{}0.04}$   &                      ${-1}_{\pm{}0.04}$     &                      20                     &   65                     &    ${1.5}_{\pm{}0.2}$                    &                     ${1.1}_{\pm{}0.07}$    &                     ${1.9}_{\pm{}0.2}$    \\                    
A0.2N100       &               12                         &   30                         &   110                        &    100                        &    ${-0.71}_{\pm{}0.05}$  &                      ${-1.8}_{\pm{}0.05}$   &                      ${-1}_{\pm{}0.03}$     &                      20                     &   66                     &    ${1.7}_{\pm{}0.2}$                    &                     ${1.2}_{\pm{}0.06}$    &                     ${2}_{\pm{}0.1}$      \\                    
A0.5N100       &               12                         &   30                         &   100                        &    100                        &    ${-0.86}_{\pm{}0.02}$  &                      ${-1.9}_{\pm{}0.06}$   &                      ${-1.2}_{\pm{}0.03}$   &                      20                     &   61                     &    ${1.1}_{\pm{}0.07}$                   &                     ${0.98}_{\pm{}0.03}$   &                     ${1.4}_{\pm{}0.04}$   \\                    
A0.9N25        &               12                         &   30                         &   120                        &    120                        &    ${-1.4}_{\pm{}0.1}$    &                      ${-2.3}_{\pm{}0.1}$    &                      ${-1.1}_{\pm{}0.02}$   &                      20                     &   90                     &    ${0.66}_{\pm{}0.02}$                  &                     ${1}_{\pm{}0.06}$      &                     ${0.84}_{\pm{}0.02}$  \\                    
A0.9N50        &               12                         &   30                         &   110                        &    110                        &    ${-1}_{\pm{}0.1}$      &                      ${-2.1}_{\pm{}0.1}$    &                      ${-1.3}_{\pm{}0.02}$   &                      20                     &   81                     &    ${0.51}_{\pm{}0.08}$                  &                     ${0.55}_{\pm{}0.04}$   &                     ${0.76}_{\pm{}0.07}$  \\                    
A0.9N100       &               12                         &   30                         &   97                         &    97                         &    ${-0.89}_{\pm{}0.09}$  &                      ${-2}_{\pm{}0.09}$     &                      ${-1.1}_{\pm{}0.02}$   &                      20                     &   60                     &    ${1.1}_{\pm{}0.1}$                    &                     ${0.79}_{\pm{}0.02}$   &                     ${1.4}_{\pm{}0.1}$    \\                    
A0.9N200       &               12                         &   30                         &   86                         &    84                         &    ${-0.88}_{\pm{}0.04}$  &                      ${-1.9}_{\pm{}0.05}$   &                      ${-1.2}_{\pm{}0.009}$  &                      20                     &   66                     &    ${0.87}_{\pm{}0.07}$                  &                     ${0.85}_{\pm{}0.008}$  &                     ${1}_{\pm{}0.05}$     \\                    
{\bf           A0.99N100}      &                          12  &                          30  &                          120  &                          110  &                      ${-0.99}_{\pm{}0.07}$  &                      ${-2.1}_{\pm{}0.08}$   &                      ${-1.2}_{\pm{}0.01}$   &                      20  &                      77   &                                     ${0.65}_{\pm{}0.04}$  &                      ${0.68}_{\pm{}0.04}$  &                     ${0.87}_{\pm{}0.05}$  \\  
\hline
\hline
\end{tabular}
\end{center}
\label{tbl14}
\end{table*}

We now consider radial power-law fits (to the form $f=f_0(r/r_0)^n$)
for various quantities.  Table~\ref{tbl14} shows some of our results.
The 5-$\sigma$ errors in the least squares fits for the power-law
index are shown as a subscript, where systematic errors (not
statistical errors) dominate the lack of a fit.  We also by-eye looked
at all fits and confirmed that they are reasonable.  This is how we
tuned the time, number of inflow times, and number of $\sigma$ that
best represent something about the systematic uncertainties.  A ``-''
is shown if the 5-$\sigma$ error permits $n$ to pass through zero,
except for $|n|<0.25$ in which case a ``-'' is shown if the 5-$\sigma$
error is larger than $0.25$.  MB09D and 2D models are not shown
because of their small inflow equilibrium radii. The resulting
power-law can change at larger radii because the flow is not strongly
self-similar where fits are obtained.

Table~\ref{tbl14} shows fits for ``disk quantities'': $\rho_0$, $p_g$,
and $|b|$ (where $|b_\phi|\sim |b|$).  We also consider (not shown)
$|v_r|$, $|v_{\rm rot}|$, $|b_r|$.  These ``disk quantities'' have
been averaged using the ``dcden'' averaging (i.e. density-weighted
average).  Power-law fits for cases ``f'', ``fdc'', ``dc'',
``$\theta^d$'', ``eq'', and ``jet'' are also considered but not shown
(see section~\ref{integrations} for definitions of these other cases
and more details).

These ``disk quantities'' are fit for radii $r^{\rm dcden}_i$ to
$r^{\rm dcden}_o$, where $r^{\rm dcden}_i=12$ is typically set and
$r^{\rm dcden}_o$ corresponds to $N=3$ inflow times for time $T^a_i$
computed via Eq.~(\ref{tieofrie}) except limited to $r^{\rm
  dcden}_o\le 30r_g$ to get an accurate fit across all models.  A
single inflow for time $T^a_i$ is given as $r^{\rm dcden}_f$.  The
stagnation radius ($r_s$), used in section~\ref{sec:magneticflux}, is
used to restrict $r^{\rm dcden}_o<r_s$.  For some models (e.g. MB09D),
if $r_s<12r_g$ or $r^{\rm dcden}_o<12r_g$, then we force $r^{\rm
  dcden}_i=12$ and $r^{\rm dcden}_o=16$.

Table~\ref{tbl14} also shows quantities $\dot{M}_{\rm in}-\dot{M}_{\rm
  H}$, $\dot{M}_{\rm mw}$, and $\dot{M}_{\rm w}$.  For these wind
quantities, fits are from $r=20r_g$ up to $r_o^w=r^{\rm dcden}_o$
defined before, but $r_o^w>40r_g$ and $r_o^w<100r_g$ are enforced.

For thick poloidal field models, roughly: $\rho_0\propto r^{-0.6}$,
$p_g\propto r^{-0.9}$, $|v_r|\propto r^{-0.8}$, $|v_{\rm rot}|\propto
r^{-0.5}$, $|b_r|,|b_\phi|,|b|\propto r^{-1}$, and $\dot{M}_{\rm
  in},\dot{M}_{\rm w}\propto r^{1.3}$.  However, our A0.94BfN40 and
A0.94BpN100 models have $|v_{\rm rot}|\propto r^{-1.0}$.  This occurs
because our fiducial model (A0.94BfN40) was run for a longer time than
any other models, while A0.94BpN100 has no polarity inversions so also
has accumulated magnetic flux over a longer time.  So these models are
choked out to a larger radial range.  The flows are sub-Keplerian, with
$v_{\rm rot}\sim 0.2 v_{\rm K}$ at $r\sim 10r_g$ for the fiducial model
(similar to in \citealt{pmw03}).

For thick toroidal field models, roughly: $\rho_0\propto r^{-0.6}$,
$p_g\propto r^{-0.8}$, $|v_r|\propto r^{-0.7}$, $|v_{\rm rot}|\propto
r^{-0.3}$, $|b_r|\propto r^{-1}$, and $|b_\phi|,|b|\propto r^{-1}$,
$\dot{M}_{\rm in},\dot{M}_{\rm w}\propto r^{1.3}$--$r^{2}$ with
secular spin dependence.  The flow is slightly sub-Keplerian in
magnitude with $v_{\rm rot}\sim 0.5 v_{\rm K}$ at $r\sim 10r_g$.

For our thinner (TNM11) models with poloidal magnetic flux, roughly:
$\rho_0\propto r^{-0.7}$ (depending upon initial $\beta$), $p_g\propto
r^{-1.9}$, $|v_r|\propto r^{-1}$ (but depends upon spin),
$|v_{\rm rot}|\propto r^{-0.3}$, $|b_r|\propto r^{-1.5}$,
$|b_\phi|,|b|\propto r^{-1}$, and $\dot{M}_{\rm in},\dot{M}_{\rm
  w}\propto r^{1}$--$r^{2}$ with secular spin dependence.  These flows
are mildly sub-Keplerian with $v_{\rm rot}\sim 0.6 v_{\rm K}$ at $r\sim
10r_g$ for $a/M=0.99$.

The MB09 models are close to Keplerian both in profile ($v_{\rm
  rot}\propto r^{-0.5}$) and value, as expected for weakly magnetized
thinner disks.

Consider an example usage of these fits.  Table~\ref{tbl5} gives
$\dot{M}_{\rm j+mw}(r=50r_g) \sim \dot{M}_{\rm j} + \dot{M}_{\rm
  mw,o}$, and $\dot{M}_{\rm w}$ dominates.  Table~\ref{tbl14} gives
the radial power-law index, so the total mass ejected to large radii
is $\dot{M}_{\rm j+mw} \sim \dot{M}_{\rm j+mw}(r=50r_g) (r/50r_g)^{n}$
where $n$ is from the second to last column in Table~\ref{tbl14}.  For
thick toroidal field models, the larger errors in $\dot{M}_{\rm w}(r)$
is because it flattens-out at larger radii.  Perhaps steady-state has
not been fully achieved by $r\sim 100r_g$, or perhaps the self-similar
region actually has a much shallower power-law index (consistent with
$\dot{M}_{\rm w}\propto r^{0.5}$ in \citealt{ppmgl10}).

In summary, our thicker disk MHD simulations lead to: flatter density
profiles than the ADAF model, quite sub-Keplerian motion, and flat
pressure profiles at large radii due to the hot material reservoir.
Our thinner disk models have flatter density
profiles than the ADAF model and somewhat sub-Keplerian motion.
Our thick poloidal/toroidal power-law results are stable
to the chosen time-averaging interval, while some TNM11 models evolve
to slightly steeper density profiles.

\subsection{Resolution}\label{sec:resolution}

In this section, we determine whether our models are numerically
converged.  We consider both explicit convergence testing and
so-called ``convergence quality factors'' that (e.g.) measure how many
grid cells resolve various critical length scales
\citep{sano04,hgk11,sdgn11,2011arXiv1106.4019S}.  Note that MRI
quality factors like $\Qone,\Qthree$ are not too useful for the new
poloidal field models for which the MRI is suppressed.

First, consider our overall resolution, box size, and numerical method
choices.  We knew that $|m|=1$ modes would dominate and even be
required for (non-axisymmetric) accretion once the magnetic barrier
formed in poloidal field models. So, we chose to treat the poloidal
and toroidal dimensions equally by ensuring the grid has a uniform
grid cell aspect ratio, which also allows us to use less poloidal
resolution to accurately capture the MRI \citep{hgk11}.  Also, because
$m=1$ modes generally dominate, $\Delta\phi=2\pi$ is required, as also
found by \citet{sdgn11} and \citet{heni09}.  In addition, other
HARM-based GRMHD codes use the diffusive T{\'o}th scheme, while our
code uses a staggered field scheme that treats field quantities more
accurately.

One indicator that an MHD simulation is unresolved is that MHD
turbulence decays, magnetic field strengths drop, and fluxes secularly
tend towards Novikov-Thorne values.  These were used by \citet{sdgn11}
and \citet{nkh10} as evidence that certain resolutions were
insufficient to resolve the MRI.  Except for the MB09D model, we see
no evidence for such behavior.

Just because turbulence does not decay does not mean the saturated
state is converged.  Convergence can only be ensured by performing
explicit convergence testing and by resolving critical length scales.
$\alpha_{\rm mag}\gtrsim 0.4$, $Q_{nlm,\rm cor}\gtrsim 6$,
$\Qone\gtrsim 10$, and $\Qthree\gtrsim 20$ may be required to achieve
tens of percent level convergence for the MRI \citep{hgk11,sdgn11}.
These quantities are given in Table~\ref{tbl4} as described in
section~\ref{sec:alphamri}.

Consider how changing $N_\phi$ changes the results for models
A0.94BfN?c? and A0.94BfN30(r).  Although initial $\beta$ varies in
those models, we already discussed how the magnetic flux near the BH
has saturated independently from the initial magnetic flux.  All
tabulated results for these 3D models are similar to within $\sim
30\%$ fractional differences.  Also, different realizations
(A0.94BfN30(r)) and different $\beta$ (A0.94BfN100c1) are within $\sim
10\%$ fractional differences in $\eta_{\rm H}$. Statistically, our
averages are only accurate to $\sim 20\%$ fractional differences, so
values within this range are considered similar.  However, we already
saw from section~\ref{sec:timephidep} and section~\ref{sec:phidep}
that there is significant power at higher $|m|$.  While these higher
$|m|$ modes only moderately affect the $\theta-\phi$ integrated
quantities on the horizon or other radii, they significantly affect
the time dependence of the solution.  A plot (not shown) of (e.g.)
efficiency vs. time shows more violent oscillations at lower $N_\phi$.
This suggests that the temporal behavior is qualitatively affected by
how well-resolved the $|m|$ modes are, and only the $N_\phi=128$ model
shows temporal variability similar to the fiducial model.

Consider how changing $N_r,N_\theta,N_\phi$ by a factor of two changes
the results for models A0.94BfN40 (fiducial model) vs. A0.94BfN100c1
and A0.94BfN30(r).  Changes in measured quantities are order tens of
percent as discussed before.  In addition, consider the azimuthal
correlation length's $m$ mode ($m_{\rm cor}$, via Eq.~(\ref{mcor}))
for quantities $\rho_0,u_g,b^2,\uvec^t,b_i,B_i,F_M,F_E^{\rm
  EM},F_E^{\rm MA}$ (and their absolute value versions) both in the
``Disk'' and ``Jet''.  For A0.94BfN40, across all quantities, $m_{\rm
  cor}\sim 6$--$14$, except in the ``Disk'' we found $m_{\rm cor}\sim
20$ for $b_r$ and $m_{\rm cor}\sim 15$ for $b^2$ at $r/r_g=4,8,30$.
On the horizon itself, where the disk is quite geometrically thin and
the flow is causally disconnected from the rest of the solution,
magnetic field components in the disk+corona have $m_{\rm cor}\le 45$.
Even for field components, $Q_{m,\rm{}cor}\ge 14$ (via
Eq.~(\ref{Qmcor}), grid cells per correlation length) outside the
horizon and $Q_{m,\rm{}cor}\ge 6$ on the horizon.  Beyond the horizon,
the jet is always even better resolved than the disk.  So, our
lower-resolution choice of $N_\phi=128$ is sufficient to resolve most
structures beyond the horizon.

Consider the vertical and radial correlation lengths in the ``Disk''.
At $r=r_{\rm H},4r_g,8r_g,30r_g$, respectively, we find $l_{\rho_0,\rm
  cor}\approx 95,58,58,57$ giving $Q_{l,\rm cor,\rho_0}\approx
3,6,6,5$ grid cells per vertical correlation length.  Also,
$l_{b^2,\rm cor}\approx 108,37,27,17$ giving $Q_{l,\rm cor,b^2}\approx
3,8,12,18$. Also, the grid cells per radial correlation lengths are
$Q_{n,\rm cor,\rho_0}\approx 7,18,22,22$ and $Q_{n,\rm cor,b^2}\approx
7,16,21,22$.  Note that taking the spectrum of the averaged flow
rather than the averaged spectrum leads to about twice higher {\it
  apparent} mode resolution for the vertical and radial correlation
lengths.  With $\theta^d\approx 0.06,0.13,0.29,0.59$, this gives
$\lambda_{\theta,\rm cor,\rho_0}/\theta^d\approx 0.6,0.4,0.2,0.1$ and
$\lambda_{\theta,\rm cor,b^2}/\theta^d\approx 0.5,0.7,0.4,0.3$.
Beyond the horizon, the jet is always even better resolved than the
disk.  Summarizing, this shows that the narrow density filaments are
fairly resolved despite the strong magnetic Rayleigh-Taylor
instabilities, while the magnetic field that fills-in the region
between the dense filaments is well-resolved beyond the horizon.
Across our poloidal field models, $N_\theta=128$ is optimal to resolve
the compressed dense magnetic Rayleigh-Taylor filaments outside the
horizon, while the magnetic field is marginally resolved even at
$N_\theta=64$ (used for sweeping over spin).

Consider the A0.94BtN10(HR, i.e. high-resolution) toroidal field
models.  The HR model gives $\alpha_b$ and all $\eta$'s as quite
similar.  So, our lower-resolution toroidal field models are probably
quantitatively converged.  Indeed, all our toroidal models have
$\Qone\gtrsim 10$, $\Qoneweak\gtrsim 10$, $\Qthree\gtrsim 20$,
$\Qthreeweak\gtrsim 20$, and $\alpha_{\rm mag}\approx 0.4$ as required to
well-resolve the MRI \citep{hgk11}.  The A0.94BtN10 model has
$\Qone\ge10$, $\Qoneweak\ge 10$, $\Qthree\ge58$, $\Qthreeweak\ge32$,
and $\alpha_{\rm mag}\approx 0.34$.  The A0.94BtN10HR model has
$\Qone\ge50$, $\Qoneweak\ge30$, $\Qthree\ge170$, $\Qthreeweak\ge80$,
and $\alpha_{\rm mag}\approx 0.38$.  (The stated $Q$'s are limited by
flow at $r=r_o$ where $3$ inflow times have passed.) So the MRI is
probably well-converged.  In addition, for A0.94BtN10HR, across all
quantities (see list in previous paragraph) and locations (disk+corona
and jet), the azimuthal correlation's $m_{\rm cor}\approx 6$--$22$
(typically $\sim 10$) at all radii $r=r_{\rm H},4r_g,8r_g,30r_g$
corresponding to $Q_{m,\rm cor}>12$ (typically $\sim 20$) grid cells
per correlation length.  Also, this corresponds to a typical azimuthal
correlation length $d\phi_{\rm cor}\sim 0.9\theta^d$, so that the
largest correlated azimuthal structures are about as extended as the
half-vertical disk extent.  These facts suggest that $N_\phi=128$ (all
our toroidal field models have $N_\phi\ge 128$) is sufficient to
well-resolve azimuthal structures.  The vertical correlation lengths
are resolved with $Q_{l,\rm cor}\sim 18$--$30$ (typically $\sim 25$)
cells across all quantities and all radii, while the radial
correlation length is resolved with $Q_{n,\rm cor}\approx 7,20,25,25$
grid cells at $r=r_{\rm H},4r_g,8r_g,30r_g$, respectively, for both
$\rho_0$ and $b^2$.  In summary, our A0.94BtN10HR model is among the
highest-resolved global MHD simulations.

The 2D axisymmetric simulations are inappropriate for studying MCAFs
(see also \citealt{igu09}).  Once magnetic flux has accumulated up to
a saturation point even beyond the BH, accretion cannot occur in
axisymmetry except through reconnection.  Once so much magnetic flux
has accumulated just beyond the BH, the entire flow rebounds backwards
leading to low $\eta_{\rm H}$.  Eventually, mass builds up and forces
magnetic flux back onto the BH leading to high $\eta_{\rm H}$.  Also,
of course, 2D axisymmetric simulations cannot resolve the
non-axisymmetric MRI or sustain a magnetic dynamo.

We now compare our resolutions with prior simulations of magnetic flux
accumulation.  \citet{ss01} used a resolution up to $N_r\times
N_\phi=156\times 128$ per decade in radius.  So their simulations are
roughly equally resolved to our fiducial models that have about half
of the resolution per decade.  They used van Leer interpolation, which
has less than half the accuracy of our PPM-type interpolation. The 3D
pseudo-Newtonian (PN) MHD simulations by \citet{ina03} used a
Cartesian grid with $\Delta\phi=\pi/2$.  Their inner-most cell size is
$0.5r_g$ at $R_{\rm in}=4r_g$. (a quite large $R_{\rm in}$, see
\citealt{mck02}.)  Our fiducial model has $dr\sim 0.1r_g$, $dz\sim
0.037r_g$, and $r\sin\theta d\phi\sim 0.097r_g$ at $r=4r_g$ and has
$dr\sim 0.035r_g$, $dz\sim 0.0076r_g$, and $r\sin\theta d\phi\sim
0.033r_g$ at $r=r_{\rm H}$, so our $z$-resolution is about $10\times$
higher. The 3D PNMHD energy-conserving PPM-type simulations by
\citet{igumenshchev08,pih09} are full $\Delta\phi=2\pi$ with
$N_r\times N_\theta\times N_\phi=182\times 84\times 240$ for a
comparable resolution per radii as our fiducial model, except very
close to the BH where we have about $4\times$ the
$\theta$-resolution.

\section{Discussion}
\label{sec:discussion}

Our simulations show that the accumulation of poloidal magnetic flux
leads to a two-phase-like magnetospheric accretion flow that is
dramatically different than the standard MRI-driven MHD turbulent
accretion flow.  The flow that develops in our simulations is
conceptually similar to the ``magnetically arrested disk'' (MAD) flow
\citep{nia03}.  While the standard weakly magnetized MRI-driven MHD
turbulent flow has gas and magnetic pressures in force balance near
the hole, the MAD state develops as magnetic flux accumulates and
magnetic forces balance the inflow's ram or gravitational forces.  The
originally-conceived MAD flow has a sharp magnetospheric boundary
layer with a large density contrast at some radius, as confirmed by
low-resolution 3D MHD simulations (e.g., fig.~13 in \citealt{ina03} ;
also seen in our 2D axisymmetric simulations).  In these pioneering
studies, accretion occurs primarily via diffusive reconnection events.

Our high-resolution fully 3D simulations show that efficient
non-axisymmetric magnetic Rayleigh-Taylor (RT) instabilities prevent
the formation of the MAD's sharp magnetospheric barrier.  Any
additional magnetic flux that tries to accrete onto the BH is
redistributed out in the disk by these instabilities.  Also, we found
that the magnetosphere geometrically compresses the dense inflow.  We
call this fully non-linear MAD flow a ``magnetically choked accretion
flow'' (MCAF), referring to the magnetic flux compressing the dense
inflow leading to enhancement of the magnetization over much of the
horizon.  Such a magnetic choke is analogous to chokes in man-made
engines, within which it enriches the fuel mixture by partially
shutting off the air intake.

Our simulations confirm the brief 3D pseudo-Newtonian MHD simulations
by \citet{igumenshchev08} that also show MCAF formation.  We also
roughly confirm the magnetospheric QPO mechanism by \citet{ln04},
which in their model drives some disk-based frequency at a vertical
magnetic barrier.  However, our 3D simulations show that the disk
inflow interacts with the polar magnetic flux threading the rotating
BH, which leads to a new ``jet-disk'' QPO (JD-QPO) mechanism based
upon the BH rotation frequency.  We also reaffirm that the BZ
mechanism operates efficiently, except our low-spin thick-disk models
do not form relativistic jets due to mass infall.  Otherwise, the BZ
mechanism leads to powerful jets directly from the BH for our poloidal
field models.  We confirm the 3D GRMHD simulations by \citet{tnm11},
who showed that outflow efficiencies of $\eta\gtrsim 100\%$ are
possible once the BH with $|a/M|\gtrsim 0.9$ has reached poloidal
magnetic flux saturation.  As in other MHD simulations, the entire
wind's outflow rate is roughly $\dot{M}_{\rm w}\propto r$.

We also confirm the results of \citet{igumenshchev08} that $a/M=0$
MCAF models have low heat+outflow efficiencies of $\eta\sim$ few
percent.  One would expect heat+outflow efficiencies of $\eta\sim
100\%$ even for $a/M=0$ if even the dense inflow were significantly
arrested \citep{nia03}.  In particular, for $a/M=0$, our thinner disk
models have $\eta_{\rm H}\sim 5\%$, while the NT efficiency is
$\eta_{\rm NT}\sim 6\%$.  For the thick disk models, $\eta_{\rm
  H}<0\%$ and $\eta^{\rm EM}_{\rm H}\sim 0\%$. In the simulations, the
heavy disk inflow is relatively unmagnetized and not sufficiently
slowed to achieve high $\eta$.

Radiatively efficient MCAF states (not studied in this paper) might
still be hyper-NT efficient even for $a/M=0$ \citep{nia03}.  Our
thinner disk models have a high specific enthalpy such that
$\eta^{PAKE}_{\rm H}\sim 34\%$ and $\eta^{EN}_{\rm H}\sim -28\%$ for
$a/M=0$ and $\eta^{PAKE}_{\rm H}\sim 64\%$ and $\eta^{EN}_{\rm H}\sim
-62\%$ for $a/M=0.99$.  Emission of that free thermal energy would
give a radiative efficiency of up to $\eta_{\rm rad}\sim 28\%$ for
$a/M=0$ and $\eta_{\rm rad}\sim 62\%$ for $a/M=0.99$.  However, the
trend with thickness is roughly $\eta^{\rm EN}\propto \theta^t$ across
our models, so we would predict that thin disks with $\theta^t\lesssim
0.03$ (as relevant for BH x-ray binaries in the thermal state ;
\citealt{2011MNRAS.414.1183K}) would have no enhanced radiative
efficiency.  Thin radiatively efficient MCAFs should be studied with
GRMHD simulations to check.

We confirm the suggestion by \citet{ina03} that most prior MHD disk
simulations used initial conditions that limited the available
magnetic flux.  For example, a relatively radially-narrow torus can
only have a relatively small amount of magnetic flux inserted if also
keeping $\beta\gg 1$ to allow for the MRI.  Also, much of the matter
and field can be ejected or remain beyond the torus pressure maximum
rather than being accreted.  After $\Psi_{\rm H}$ reaches a
quasi-steady-state, one can test whether an MHD simulation is limited
by such initial conditions.  One computes $\Psi_{\rm H}/\Psi_a$
(i.e. [flux on hole] per unit [flux on the hole plus just outside the
hole of the same polarity]) and also $\Psi_{\rm H}/\Psi_s$ ([flux on
hole] per unit [flux on the hole plus available within the stagnation
radius for radial inflow]).  One must compute both because there may
be plenty of same-polarity magnetic flux beyond the hole, but it may
not be accreting.  One could also compute how much flux is available
within the inflow equilibrium region.  Both $\Psi_{\rm H}/\Psi_a\sim 1$ and $\Psi_{\rm
  H}/\Psi_s\sim 1$ for MB09D, so the initial conditions artificially
limited the magnetic flux that can reach the hole.  Also, it appears
that $\Psi_{\rm H}/\Psi_s\sim 1$ in the simulations by
\citet{bhk09trans}, who show much magnetic flux
is ejected to (or remains at) large radii.

A local $\alpha$-viscosity leads to a poor model of angular
momentum transport for the simulations.
The effective $\alpha$ is different
than predicted by local stresses divided by either total or magnetic
pressure for all our thick ($H/R\sim 1$) disk and poloidal field thinner ($H/R\sim
0.3$) disk models.  The $\alpha$-disk theory only works well for our
weak poloidal field models of either very thin ($H/R\sim
0.05$) or relatively thin ($H/R\sim 0.3$) disks.
Large-scale magnetic confinement forces, rather than local stresses, may be acting on the dense inflow.
Also, for our toroidal thick disk models, convection
may be important because Reynolds stress dominates Maxwell stress.

Interestingly, $\Qtwo\sim 1$ (or $\Qtwo\sim 0.25$--$0.5$)
gives the disk's saturated vertical field strength in our toroidal
(or poloidal) field models.  Also, the effective
viscosity is $\alpha_{\rm eff}\sim 0.1$--$1$
($\alpha_{\rm eff2}$ is at most $2\times$ smaller for the new thinner
models and at most $4\times$ smaller for the new thick disk models),
which is sufficiently consistent with observations
\citep{2007MNRAS.376.1740K}.

MCAFs might explain observations of apparently highly efficient jets
\citep{1983MNRAS.204..151L,1999MNRAS.309.1017W,2004ApJ...607..800B,2006ApJ...647..161O,2006ApJS..166..470R,2010MNRAS.405..387G,2011ApJ...727...39M,2011MNRAS.411.1909F,2011ApJ...728L..17P}.
However, more work is required to ensure the observations are
accurately modelled.  For example, the jet models in
\citet{2011MNRAS.411.1909F} have factors (e.g. $f$, see
\citealt{1999MNRAS.309.1017W}) that can significantly change depending
upon estimates of the invisible work done by the jet.  Also, there are
similar uncertainties in translating to the jet power in
\citet{2011ApJ...727...39M} as related to work done by bubble
expansion \citep{2004ApJ...607..800B}.  Also, the Blazar estimates of
jet power by \citet{2010MNRAS.405..387G} can be affected by models of
the Doppler factor.  178Mhz observations by
\citet{2006ApJ...647..161O} are affected by a lack of simultaneity
between optical and radio emission, and the systems they observe could
be low-luminosity radiatively inefficient systems that lowers their
jet efficiency requirements.  Our simulations may help constrain such
jet power estimates.

MCAFs may also help explain the variety of spectral states seen in BH
x-ray binaries \citep{1999ApJS..124..265R,dl05,igumenshchev08,igu09}.
The accumulation of magnetic flux might lead to the low-hard,
bright-hard, and hard:steep-power-law intermediate states with LFQPOs
whose frequency is controlled by the ``magnetospheric radius'' ($r_m$)
and with a persistent (mostly invisible) radio jet.  The steep power
law (very high) state could be due to reconnection as a field polarity
reversal makes its way to the hole, where the reconnection with the
jet on slightly larger scales makes the HFQPOs visible near the BH.
Dissipation of the accumulated magnetic flux on the hole could cause
the jet to light-up as a radio bright outgoing reconnecting plasmoid.
The remaining disk without accumulated flux can lead back to the
thermal (high-soft) state that has no jet and no (or very weak) QPOs.
Notice that we distinguish between QPOs and jets being present
vs. visible.  Future work can test these speculations.

\section{Conclusions}
\label{sec:conclusions}

Black hole systems may have plenty of coherent magnetic flux available
at large radii that can feed the accretion flow down to the black
hole.  Using fully 3D GRMHD simulations of extended radiatively
inefficient accretion flows, we found that poloidal magnetic flux
readily accretes from large radii and builds-up to a natural
saturation point near the black hole independently from the initial
poloidal magnetic flux.  The accumulated poloidal magnetic flux
naturally leads to a highly non-axisymmetric ``magnetically choked
accretion flow'' (MCAF), within which the MRI is suppressed.

In such a choked state, the polar magnetic flux forces accretion to
occur through magnetic Rayleigh-Taylor instabilities and geometric
compression.  Near the black hole, the inflow is highly compressed,
which leads to a highly magnetized state over most of the horizon.  So
the accumulated flux acts as an inflow choke (analogous to air chokes
in engineering that lead to fuel enrichment), which allows for greater
than $100\%$ jet efficiencies for $|a/M|\gtrsim 0.9$.

Once the horizon's magnetic flux reaches a saturated state, the inflow
pushes its way through the jet's bulging magnetosphere.  This leads to
a new jet-disk high-frequency quasi-periodic oscillation (JD-HFQPO)
mechanism driving spherical harmonic $|m|$ mode oscillations with a
frequency set by the black hole's field line angular rotation
frequency at the jet-disk interface. So these HFQPOs may allow one to
measure black hole spin.  The coherence quality factor is highest in
the jet and at the jet-disk interface (harboring large magnetic
energy) rather than in the disk plane, so these HFQPOs could be
dominated by non-thermal emission.  More work is required to test
their observability.

In contrast to our initially poloidally-dominated magnetic field
models that develop the uncommonly-found MCAF state, our initially
toroidally-dominated magnetic field simulation results are similar to
as seen in prior works.  A significant new result is that such models
do show large-scale dipolar magnetic flux formation near the horizon.
However, only weak transient highly magnetized relativistic jets
emerge.  Higher resolutions or spins could promote this emergent
dipolar field to launch persistent relativistic jets.

As with any numerical study, more convergence testing is required.
Our toroidal field models satisfy the convergence quality factors of
$\Qone\gtrsim 10$, $\Qthree\gtrsim 20$, and $\alpha_{\rm mag}\approx 0.4$
in \citet{hgk11} and $r,\theta,\phi$ correlation lengths are
well-resolved.  Our fiducial poloidal model well-resolves
all correlation lengths.  We explicitly tested convergence with
$r,\theta,\phi$-resolutions for $a/M=0.9375$ poloidal/toroidal
models, and most values are converged to $30\%$ and some
values (e.g. $\Upsilon$ for poloidal models) are converged to
$10\%$.  We found that the convergence criteria of \citet{hgk11} are
not generally sufficient (e.g. jet efficiency continues
to change significantly in toroidal models) or applicable (e.g. MRI can be suppressed).
Our code conserves energy during large-scale field reconnection (as in \citealt{lm11}),
occurring in magnetic Rayleigh-Taylor disruptions and field polarity inversions,
but the reconnection rate may be set by grid diffusivity.

In summary, the MCAF state and the spontaneous large-scale dipolar
field generation seen in toroidal field models should be accounted for
when seeking to understand SgrA*, M87, Blazars, high-powered quasars,
black hole x-ray binaries, and other systems.  In future work, we will
consider the spin dependence of the jet/wind power
\citep{2012arXiv1201.4385T} and of the JD-QPO, characterize the
``magnetospheric radius'' and determine its spin dependence, perform
radiative transfer to identify the observational signatures of MCAFs,
and study the source of angular momentum transport (e.g. turbulence
vs. large-scale stresses).

\section*{Acknowledgments}

We thank Ramesh Narayan, Serguei Komissarov, Robert Penna, Eliot
Quataert, Ioannis Contopoulos, Julian Krolik, Charles Gammie, Jonathan
Arons, Vasily Beskin, and Kris Beckwith for useful discussions.  This
work was supported by a NASA Chandra Fellowship PF7-80048 (JCM),
Princeton Center for Theoretical Science Fellowship (AT), and
NSF/XSEDE resources provided by TACC (Lonestar/Ranch) and NICS
(Kraken) under awards TG-AST080025N (JCM) and TG-AST100040 (AT).

\section*{SUPPORTING INFORMATION}

Additional Supporting Information may be found in the online version
of this article: Movie file. Movie of the fiducial model A0.94BfN40
showing the animated version of Fig.~\ref{evolvedmovie} (see caption
alongside movie file for more detail).

\appendix

\section{Numerical Methods}
\label{sec:nummethods}

Our simulations use the GRMHD code HARM based upon a conservative
shock-capturing Godunov scheme with 2nd order Runge-Kutta
time-stepping, Courant factor $0.8$, LAXF fluxes, simplified wave
speeds, PPM-type interpolation (with no contact steepener, but with
shock flattener based upon rest-mass flux density and total pressure)
for primitive quantities ($\{\rho_0,u_g,\reluvec^i,\Bvec^i\}$), a
staggered magnetic field representation, and any regular grid warping
\citep{gam03,nob06,tch_wham07}.

One key feature of our HARM is its handling of grid cells with large
magnetic energy per unit rest-mass energy.  The code first tries to
convert conserved to primitive quantities using the ideal MHD
equations \citep{mm07}.  If that fails, the entropy conservation
equations are tried.  If that fails, the cold ideal MHD equations
(with the failed grid cell having its internal energy averaged over
its immediate neighbors) are tried.  If that fails, the primitives are
averaged over non-failed neighbors.  The inversions are sought to
machine accuracy.  Density floors are used to avoid too low densities
that lead to inversion failures.  $u_g$ is chosen as to enforce
$u_g/\rho_0\le \uorhomax$, then $\rho_0$ is chosen as to enforce
$b^2/\rho_0\le \bsqorhomax$ and $b^2/u_g\le \bsqoumax$, and then these
implied comoving changes are added to the stress-energy tensor in the
ZAMO frame.  Using a fixed (rather than comoving) frame avoids
run-away energy-momentum gains.

These inversion reductions and density floors mostly occur in the cold
highly-magnetized jet with ordered magnetic field along which the flow
moves.  So, we set $\rho_0=u_g=0$ in calculations (except in color
images) if, e.g. for the ``thick disk'' simulations, $b^2/\rho_0\ge
\bsqorhomaxdiaghigh$ to $b^2/\rho_0\ge \bsqorhomaxdiaglow$ from
$r_{\rm H}$ to $r=9r_g$, respectively, as interpolated linearly with
radius (tracing injected mass-energy, as in \citealt{tnm11}, gives a
bit more accuracy).  Floors affect $\reluvec^\phi,\Omega_F$ only near
the axes (not in the disk or collimated jets).

Another key feature of our HARM is how the 3D polar axis is treated
\citep{mb09}.  To resolve the staggered field component $\Bvec^{(2)}$,
a machine error polar cut-out of size $d\theta=10^{-13}$ is used.  The
values of $\Bvec^{(2)}$ on the polar cut-out are copied asymmetrically
across the pole.  All other primitives are interpolated from active
cells to ghost cells at the same physical locations and also to the
polar cut-out.  All fluxes are set to vanish at the pole, except for
the induction update to $\detg \Bvec^{(2)}$. The small $\theta$
factors in $\detg$ cancel on both sides of the induction equation
allowing one to recover $\Bvec^{(2)}$ on the polar cut-out.  This
procedure introduces a complex phase to the otherwise singular polar
axis, where for general discontinuous flows the only constraint is
from $\nabla\cdot B=0$ that forces $\Bvec^{(2)}$ to be continuous
along $\theta$ across the pole for each $\phi$.  This transmissive
polar boundary condition allows the flow to pass through the polar
region with only additional dissipation due to the polar triangular
mesh as compared to a more uniform mesh.

Another key feature is the connection coefficients are corrected to
obtain zero body forces to machine error for constant
pressure. Nominally, constant pressure flux and source terms do not
cancel, and this secular differencing error sits at the polar axes.
For constant pressures, the relevant equation of motion is
\begin{equation}
\partial_t (\detg \delta^t_\nu) = -\partial_j(\detg \delta^j_\nu) + \detg \delta^k_\lambda  \Gamma^\lambda_{\nu\kappa} .
\end{equation}
Assuming $\partial_t(\detg)\sim 0$ and $\partial_j(p_{\rm tot})\sim
0$, then $0 = -\partial_\nu(\detg) + \detg \Gamma^\kappa_{\nu\kappa}$
or $\Gamma^\kappa_{\nu\kappa} = \partial_\nu(\detg) / \detg$.  We need
to subtract off face-related flux-positioned values and add
center-related conservative-positioned values. Let a grid cell indexed
by $i,j,k$ have grid face position $x_-^\kappa \equiv
x^\kappa_f(i,j,k)$ and upper grid face position of $x_+^\kappa \equiv
x^\kappa_f(i+dx^\kappa \delta^1_\kappa,j+dx^\kappa
\delta^2_\kappa,k+dx^\kappa \delta^3_\kappa)$ ($\kappa$ chosen, not
summed) such that the cell size is $\Delta x^\kappa=x_+^\kappa -
x_-^\kappa$ and cell center position is $x_c^\kappa = (x_+^\kappa +
x_-^\kappa)/2$.  First, compute $C_\kappa = \Gamma^\mu_{\mu\kappa}$ as
using a {\it continuous} approximation to the derivatives at $x_c$.
These are computed using a modified dfridr() in
\citet{1986nras.book.....P} to achieve a uniformly low error. The
conversion from internal to physical coordinates is generated from a
simplified continuous approximation to the derivative with a
difference size of $10^{-5}dx^\kappa$, which depends upon range in
$x^\kappa$ as necessary to ensure the differencing scales with the
grid.  Second, compute $D_\kappa = \Gamma^\mu_{\mu\kappa}$ using {\it
  discrete} differences across the actual grid cell via $D_\kappa =
(\detg[x_+^\kappa] - \detg[x_-^\kappa])/(\Delta x^\kappa \detg[x_c])$.
Now, changing each $\kappa$ term independently can lead to too large
changes in some limits.  Instead, we construct the weights $S_\kappa =
10^{-300} + |\Gamma^\mu_{\kappa\mu}|$ (sum over $\mu$) and let
$dS_\kappa = D_\kappa - C_\kappa$, $W_{\kappa\mu} =
|\Gamma^\mu_{\kappa\mu}|/S_\kappa$ (no sum over $\kappa$ or
$\mu$). Finally, for each $\mu$ and each $\kappa$, the correction is
\begin{equation}
\Delta \Gamma^\mu_{\kappa\mu} = dS_\kappa W_{\kappa\mu}
\end{equation}
(no sum over $\kappa$ or $\mu$).  The new connection gives machine
error cancellation between source and flux differencing of pressure.

{\small

}

\label{lastpage}
\end{document}